\documentclass[a4paper,fleqn,usenatbib,useAMS]{mnras}

\usepackage{graphicx}	
\usepackage{amsmath}	
\usepackage{multicol}   
\usepackage{pdflscape}	

\usepackage[T1]{fontenc}
\usepackage{ae,aecompl}

\usepackage{newtxtext,newtxmath}

\title[MNRAS A New Planet in the Kepler-159 System From Transit Timing Variations a]{\textit{A New Planet in the Kepler-159 System From Transit Timing Variations} \\}

\author[Chris Fox, Paul Wiegert]{Chris Fox, Paul Wiegert$^{1}$
\thanks{Contact e-mail: \href{mailto:cfox53@uwo.ca}{cfox53@uwo.ca}}
\author[Fox, Wiegert]{Fox, Wiegert$^{1}$}
\\
$^{1}$The University of Western Ontario, Department of Physics \& Astronomy, London, Ontario, Canada}

\date{Last updated 2018 October 2\\ Accepted for publication by the Monthly Notices of the Astronomical Society}

\pubyear{2018}

\begin{document}
\label{firstpage}
\pagerange{\pageref{firstpage}--\pageref{lastpage}}
\maketitle

\begin{abstract}
The Kepler Space Telescope has discovered thousands of planets via the transit method.  The transit timing variations of these planets allows us not only to infer the existence of other planets, transiting or not, but to characterize a number of parameters of the system.
Using the transit timing variations of the planets Kepler-159b and 159c, the transit simulator TTVFast, and the Bayesian Inference tool MultiNest, we predict a new non-transiting planet, Kepler-159d, in a resonant 2:1 orbit with Kepler-159c. This configuration is dynamically stable on at least 10~Myr time scales, though we note that other less stable, higher-order resonances could also produce similar TTVs during the three-year window Kepler was in operation.
\end{abstract}




\section{Introduction}

While the Kepler Space Telescope primarily found planets via the transit method \citep{bor2010}, additional planets have been discovered by examining the changes in the transit timings.  These transit timing variations (TTVs) result from gravitational perturbations due to other planets, which can hurry or delay the time that a planet transits its parent star \citep{agol2005, holmur2005}.  Here we look at the case of the Kepler-159 system (also known as  KIC 5640085, KOI-448, 2MASS J19481684+4052076, WISE J194816.85+405207.5).  This star has two confirmed transiting planets, Kepler-159b and Kepler-159c (hereafter named 159b and 159c).  The latter shows strong transit timing variations, as high as 5 hours, suggesting a significant gravitational influence acting upon it.

To assess the possible existence of a third planet, we used two publicly available pieces of software.  The first is TTVFast \citep{da2014}.  This program simulates the orbits of planets around a star and outputs calculated TTVs.  The second piece of software used was MultiNest \citep{feroz2009}, a Bayesian Inference tool.  

Under the assumption that the observed TTVs are produced by an unseen planet,  these two programs allow us to determine the best set of planetary parameters that reproduce the observed TTVs.  We find a degenerate set of parameters, though with our new planet consistently in (or near) an orbital resonance with 159c.  We will refer to this inferred planet as Kepler-159d.

\section{Target Details}
The relevant stellar information of Kepler-159 comes from \citet{muir2012} and \citet{math2017}.  Transit Timing data is taken from \citet{hm2016}.  Planetary radii were taken from \citet{rowe2015}.  The host star is a M0V star with an estimated mass of 0.52 $M_{\sun}$, radius of 0.50 $R_{\sun}$, and effective temperature of 3893$\pm$80 K \citep{muir2012, math2017}.  There are not any direct measurements of the mass, but we used initial mass estimates from \citet{ck2017} (using their 2$\sigma$ values) and the mass-radius relation for sub-Neptunes from \citet{wolfgang2016}.

\begin{figure}
 \includegraphics[width=\columnwidth]{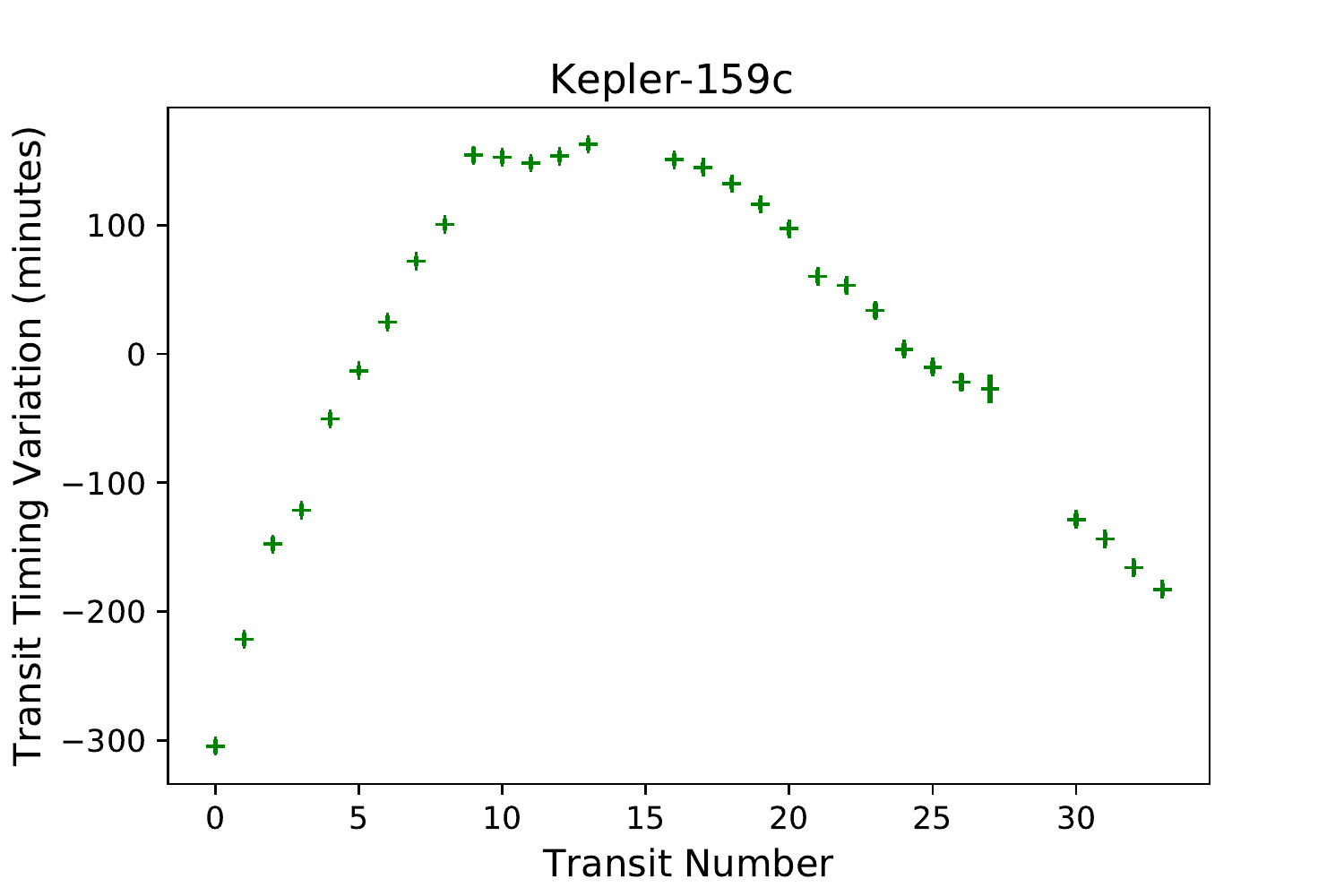}
 \includegraphics[width=\columnwidth]{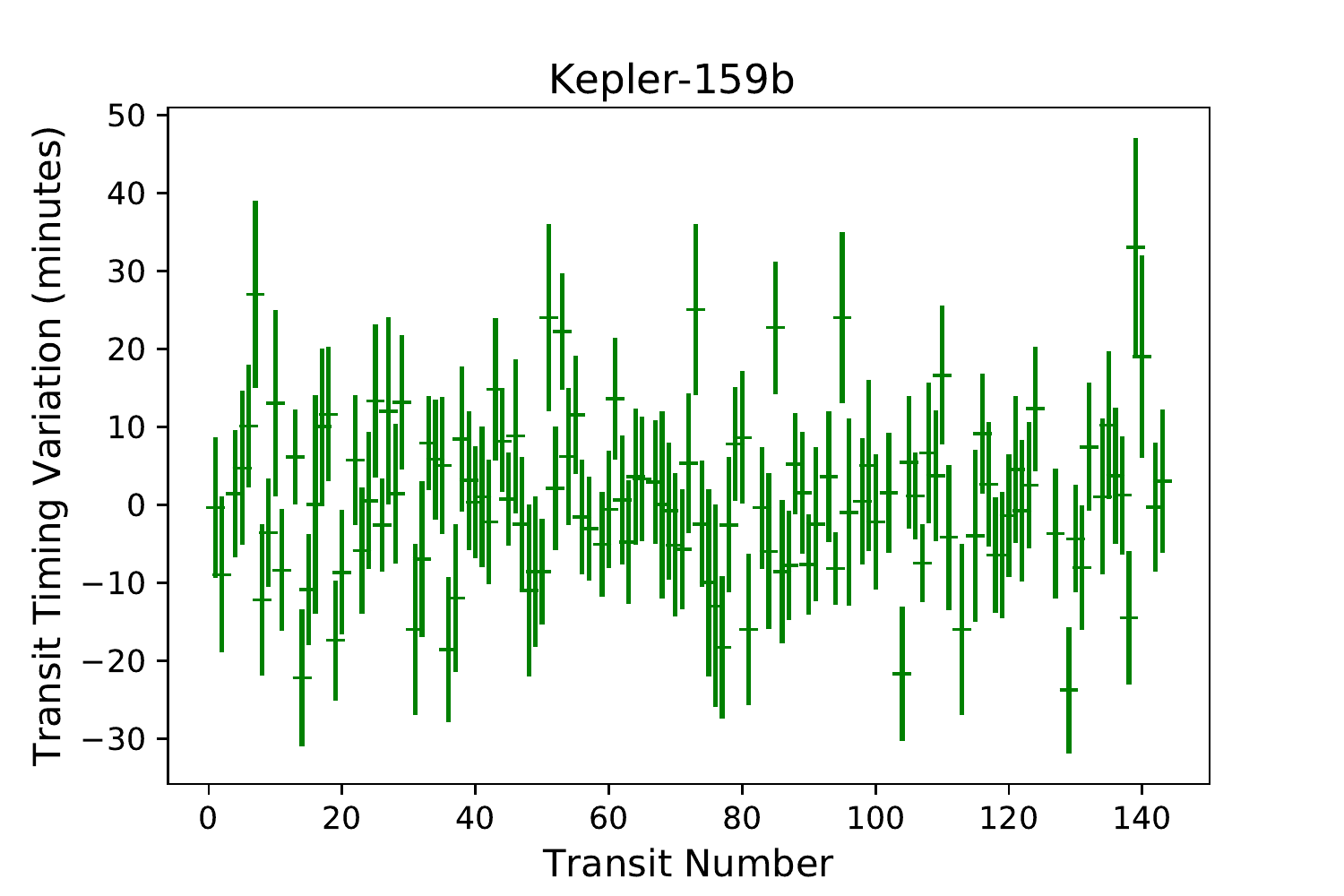}
 \caption{Transit Timing Variations of Kepler-159b and Kepler-159c, created using data from \citet{hm2016}  \label{fig:bcdat} }
\end{figure}

The two transiting planets have estimated radii based on the observed transit depths.  Their radii suggest both are likely gaseous \citep{rogers2015}.  Figure \ref{fig:bcdat} shows the transit data for both planets.  The inner planet, 159b, has a period of 10.14 days. All TTVs are within 30 minutes of the expected time, and the majority are consistent with zero.  Kepler 159b is periodic to within measurement error.  However 159c has an average period of 43.58 days and shows significant TTVs over nearly all of its 30 transits, all well outside the measurement error.  This planet's behaviour is the primary reason for predicting the existence of a third body.

\begin{table}
 \caption{Properties of Kepler-159 Transiting Planets} 
 \label{tab:bcprops}
 \begin{tabular}{lll}
  \hline
  Parameter & Kepler-159b & Kepler-159c \\
  \hline
  Period (days) & 10.139624$\pm$0.000022 & 43.582588$\pm$0.000044 \\[2pt] Semi-major axis (AU) & 0.0737$\pm$0.0028 & 0.1949$\pm$0.0075 \\
  Radius ($R_{\earth}$) & 1.870$\pm$0.230 & 2.640$\pm$0.320 \\
  Mass (Chen) & $4.2\substack{+9.0\\-2.5}$ & $7.1\substack{+15.3\\-4.7}$ \\
  Mass (Wolfgang) & $4.0\substack{+3.4\\-1.6}$ & $9.0\substack{+5.0\\-2.8}$ \\ 
  \hline
 \end{tabular}
  Semi-major axis computed using stellar mass and planetary periods from \citet{hm2016}.  Planetary radii from \citet{rowe2015}.  Mass estimates are from \citet{wolfgang2016} and \citet{ck2017}. 
\end{table}

The inclinations of 159b and 159c must be near 90$\degr$ for a transit to occur.  Note that when we use the term inclination, we are referring exclusively to the value as measured from the plane of the sky, not the mutual inclination of the planetary orbits.  The impact parameters of -b and -c are $0.72\pm0.27$ and $0.78\pm0.3$ respectively \citep{rowe2015}.  These would give best-fit inclinations of $90\degr\pm1.8\degr$ and $90\degr\pm0.7\degr$. 

Transit simulations were set to start at the first transit of the inner planet, providing the starting position for 159b.  This corresponds to a BJD of 2454970.890506.  The timing of the first transit of 159c then gives its relative location at the start of the simulation.

We adopt the TTV measurements of \citet{hm2016}. Of the 30 observed data points for 159c, there are 4 that are considered outliers  by \citet{hm2016}.  These data points are transit numbers 9, 19, 26 and 27.  Similarly, 159b had 10 points excluded.  All such outliers were removed from the runs reported on here.  We also ran simulations which included these points, and while they have an effect on the reported log-Evidence values, the conclusions are unchanged if the outliers are included.

\section{Methods and Setup}

\subsection{Parameters}
Each planet has 6 orbital elements plus a mass for a total of 7 parameters.  Of the 14 total parameters of the two transiting planets, only six are known to high precision.  These are the periods, inclinations, and relative starting positions (mean anomalies). 

Periods of both 159b and 159c were kept fixed for all simulations.  It was expected that slight deviations from 90$\degr$ inclination (with respect to plane of the sky) would have little impact on the observed TTVs \citep{agol2005}.  Some test simulations allowing inclination to vary by $\pm$2$\degr$ from 90$\degr$ (chosen to maintain the transit condition) were performed and confirmed this expectation. Thus, to reduce the parameter space, for all simulations the inclination was set to 90$\degr$ for both known transiting planets.  

Because the mean anomaly is measured from the argument of periastron, the mean anomaly for each planet was not fixed but rather predetermined from the eccentricity, argument of periastron, and timing of the first transit.  As such, the mean anomaly of 159c was different for each simulation but not a free parameter.

This leaves 8 unknown parameters, plus 7 for the presumed third planet.  This is a 15-dimensional parameter space to search.  To further reduce computational requirements, we reduced this to 11 parameters.  This reduction was accomplished by presuming 159b was in a circular orbit, using its orbital plane as the reference for the system, and using a constant estimate of its mass.  Thus, 159b's eccentricity, argument of periastron, longitude of the ascending node, and mass are removed from the priors. This is discussed in more detail in section \ref{priorsdisc}. 

\subsection{TTVFast and MultiNest}
We used TTVFast \citep{da2014}, a symplectic numerical integrator, to compute the transit times for each hypothesized set of orbital parameters.  These times were converted to TTVs, then compared to the observed TTV data from \citet{hm2016}.  Another set of parameters would then be chosen by MultiNest, and the values compared again.  To search the parameter space  efficiently, we used the Bayesian Inference algorithm MultiNest \citep{feroz2009, feroz2013}.  This iterative process can find solutions far more quickly than searching the entire parameter space.  We used the Python interface for MultiNest, PyMultiNest by \citet{buch2014} for this project.

Due to the size and complexity of the parameter space, finding the best fits required an iterative process and dozens of sets of runs.  Each "run" produces a single result from MultiNest, estimated from millions of samples of the parameter space and as many simulations from TTVFast.  Repeated runs, even with the same priors, would often return different best-fits. This is indicative of multiple maxima in our parameter space, particularly associated with different resonances between 159d and 159c. It is known that resonances produce strong TTVs and that there can be degeneracy between resonances in this regard \citep{boue2012}.  To examine this effect, we performed additional runs with periods of 159d constrained to be near the resonances (periods within 10 days of resonance) of 159c.  This narrowing of ranges allowed us to determine the best-fits associated with each resonance independent of the others.  
We used the default parameters of PyMultiNest with exception of the evidence tolerance, which we set to 0.15 (from default of 0.5).  This reduced tolerance value allows for a more fine-tuned search across the peaks of this rapidly varying landscape.

\subsection{Likelihood}
The Likelihood, $L$ we used is based on the usual $\chi^2$ statistic and is given by:
\begin{equation}
    L = \prod_{i=0}^{N} \frac{1}{\sigma_i\sqrt{2\pi}} \exp\Big(-\frac{(x_i-\mu_i)^2}{2\sigma_i^2} \Big)
\end{equation}
where $x_{i}$ is the calculated TTV,  $\mu_{i}$ is the observed TTV, and $\sigma_{i}$ is the error in the observed TTV for point $i$, respectively \citep{icvg2014}.  The log-likelihood is the value we aim to maximize in the simulation.
\begin{equation}
    \ln{L} = \sum_{i=0}^{N} \Bigg( \Big(\ln{\frac{1}{\sigma_i\sqrt{2\pi}}\Big)} -\frac{(x_i-\mu_i)^2}{2\sigma_i^2} \Bigg)
\end{equation}
In this case, summation includes transit data from both 159b and 159c.  

\subsection{Priors}
The priors used for the search were constrained based on physical grounds.  First we tested the two-planet case, then we moved to the three-planet scenario.  Several priors differed between these two cases.  In this section, we describe the ranges of parameter space explored in our scenarios.

\subsubsection{Testing the Two-Planet case}
First, simulations were performed to ensure that the observed TTVs were not mutually induced.  159b was used as the reference planet, with longitude of the ascending node of 0\degr.  The timing of the first transit of 159b serves as the start time for all simulations.  This time along with the timing of the first transit of 159c thus determines the position of 159c at the start of each simulation.  Masses of both planets were allowed to vary from Earth-mass to 10 $M_{J}$.  Most orbital parameters were allowed to vary freely so long as the initial conditions (initial transit times) were maintained.  The periods were known and thus fixed, and the eccentricity was constrained such that the two orbits would not cross.  The stellar mass was allowed to vary from 0.3 $3M_{\sun}$ to 1.0 $M_{\sun}$. 

\subsubsection{Adding the Third Planet}\label{priorsdisc}
After it was determined that the two planets could not mutually induce the observed TTVs, we looked at adding a third planet.  This required new priors for 159d, as well as modifications to the priors of 159b and 159c.

\textit{Kepler 159b}:  The inner planet, 159b shows no trends in its TTVs.  It has a smaller radius from 159c suggesting it is likely a less massive planet.  We use it as a constraint for the system; any good solution must maintain its lack of TTVs.  159b has a radius estimate of 1.87 $R_{\earth}$ (Table \ref{tab:bcprops}).  Based on the work of \citet{ck2017}, and supported by the radius-mass relationship for sub-Neptune-radius planets established by \citet{wolfgang2016}, we estimate the mass of 159b to be 4.0-6.0 $M_{\earth}$, though both of these sources allow for a wide range of mass values.  Because we did not expect 159b to be a major player dynamically, we kept the mass of 159b to a fixed value of 4.0 $M_{\earth}$.  We set the longitude of the ascending node of 159b to 0$\degr$ as the orbital reference plane for the system.  The time of first transit is used as the starting point of each simulation, and thus not computed by TTVFast.  We expect this planet to be in a circular orbit due to its proximity to the host star, so eccentricity was fixed to 0.

\textit{Kepler 159c}: To encompass the mass possibilities for the outer transiting planet, 159c, its mass was allowed to vary from 2.0 $M_{\earth}$ to 35 $M_{\earth}$, covering the likely ranges for sub-Neptune planets estimated by \citet{wolfgang2016} for the planetary radius of 2.64 $R_{\earth}$ (see Table \ref{tab:bcprops}).  Eccentricity was limited to a maximum of 0.66 to prevent crossing with the orbit of 159b.  Initially, the full range of values of the longitude of the ascending node were explored.  This resulted in some solutions with ascending node values of 159c and 159d near 180$\degr$ from each other.  The inclination values for the planets were always within 15$\degr$ of 90$\degr$. This combination of out of phase ascending nodes and near-90$\degr$ inclinations correspond to the two planets orbiting in opposite directions. Retrograde planets appear to be valid solutions, but we deem it more likely that the planets all orbit in the same sense.  So, we then restricted the ascending node prior to -45$\degr$ to +45$\degr$ for the solutions reported here.  This ensured the orbital axes of the planets were not anti-aligned.

\textit{Kepler 159d}: The parameters for the hypothesized planet, 159d, were completely unknown.  Like 159c, the entire range of ascending node values was initially explored, but ultimately limited to -45$\degr$ to +45$\degr$ to prevent retrograde motion.  Similarly, inclination was constrained to 45$\degr$ to 135$\degr$.  We limited the eccentricity to no higher than 0.7 to prevent orbital crossing with 159c for periods as high as 250 days.  Its period was explored from 50 days (just outside 159c's orbit) to 300 days, and its mass allowed to range up to 30 Jupiter masses.  We found that 159d's period always settled near a resonance with 159c, and ultimately we did individual runs with the period prior limited to $\pm$10 days around each resonance.

The stellar mass was set to the value from \citet{muir2012} and \citet{rowe2015} of 0.520 $M_{\sun}$ for all reported runs.  

In total we varied 11 parameters.  The mass of 159b was static and its orbit assumed circular.  The three static values of 159c were the period, inclination, and (effectively) mean anomaly.  The new planet had no static parameters.  We used uniform priors in all tests, summarized in Table \ref{tab:paramranges}.

\begin{table}
 \caption{Planetary Parameter Priors for Testing}
 \label{tab:paramranges}
 \begin{tabular}{llll}
  \hline
  Parameter & Kepler-159b & Kepler-159c & Kepler-159d \\
  \hline
  Mass & 4.0 $M_{\earth}$ & [2.0, 35.0] $M_{\earth}$ & [0.05, 20.0] $M_{J}$ \\[1pt]
  Period & 10.1396236 d* & 43.58258676 d* & [50.0, 300.0] d \\[1pt] 
  Eccentricity & 0.0 & [0.0, 0.66] & [0.0, 0.7] \\[1pt]
  Inclination & 90.0\degr* & 90.0\degr* & [45.0, 135.0]\degr \\[1pt]
  Asc. Node & 0.0\degr & [-45.0,45.0]\degr & [-45.0,45.0]\degr \\[1pt]
  Arg. of Peri. & 0.0\degr & [0.0, 360.0]\degr & [0.0, 360.0]\degr \\[1pt]
  Mean Anom. & 90.0\degr* & ** & [0.0, 360.0]\degr \\[1pt]
  \hline
 \end{tabular}
 Parameters for 159b were always static, assuming a simple circular orbit, and used as a simulation constraint.\\
 *These parameters are known, and used as starting values for the start of all simulations.\\
 **The mean anomaly of 159c for each simulation is dependent on its first transit time, eccentricity and argument of periastron.
\end{table}

\section{Results} \label{results}

\subsection{Two Planet Scenario}  \label{sec:2planres}
\begin{figure}
 \includegraphics[scale=0.61]{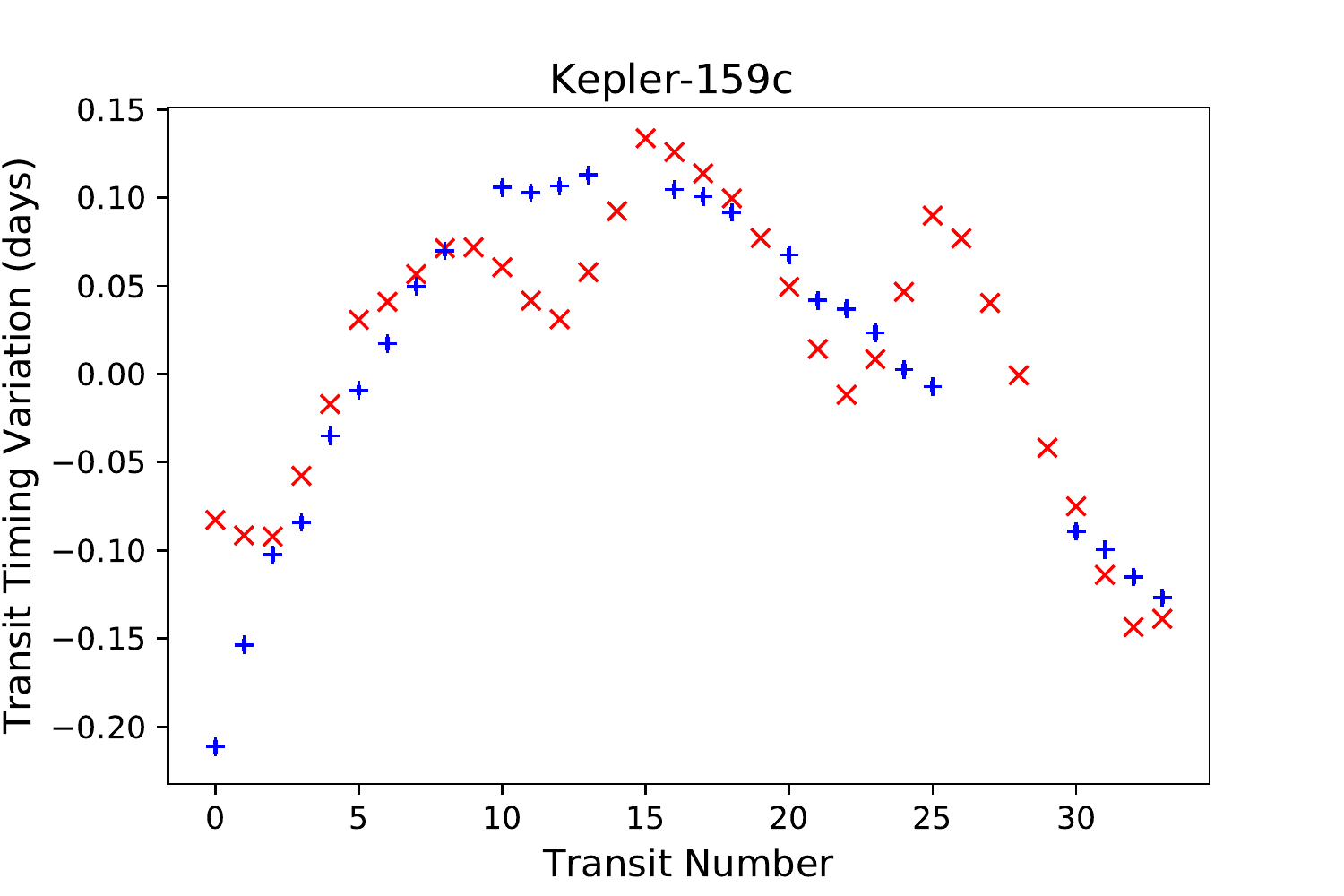}
 \includegraphics[scale=0.61]{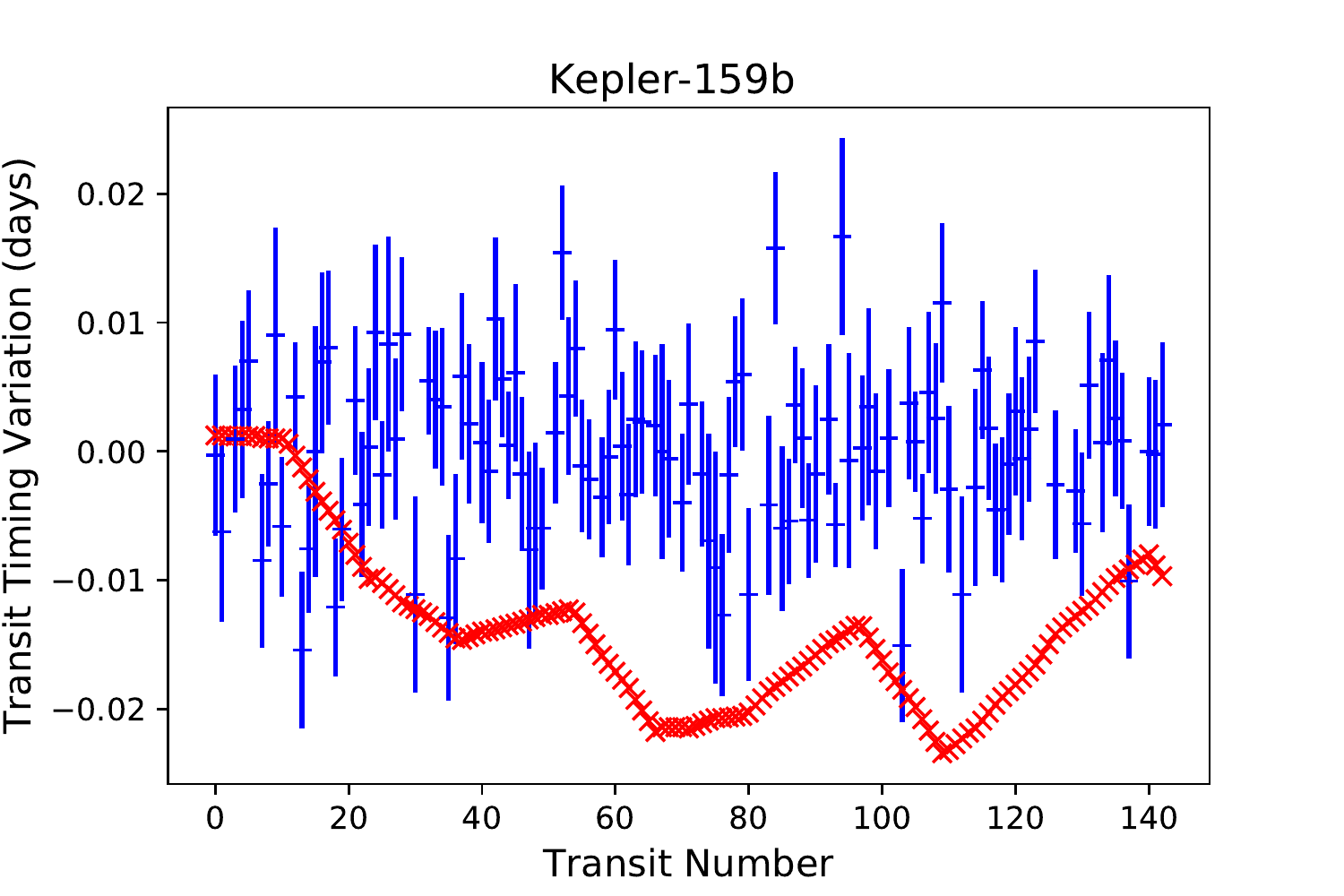}
 \caption{Simulated TTVs of 2-Planet scenario.  \label{fig:2planttv}  A blue + symbol is the ttv computed using \citet{hm2016} data, (uncertainty included), and a red x represents the simulated TTV.}
\end{figure}

The best fit scenario for the 2 planet case, where 159b and 159c mutually induce TTVs, is shown in Figure \ref{fig:2planttv}.  The posterior results are in Table \ref{tab:post2plan}.  The reported Log-Evidence value is -2508.  Not only do the TTVs from 159c not match, TTVs are induced upon the inner planet, inconsistent with the observations from Kepler.  Further, the best-fit eccentricity of 159c is 0.55, which puts its closest approach very close to the orbit of 159b.  For these reasons, we conclude that the two planets do not mutually induce the observed TTVs and the existence of a third planet was then assumed. 

\begin{table}
 \caption{Best Fit Inputs for 2-Planet Simulations}
 \label{tab:post2plan}
 \begin{tabular}{lll}
  \hline
  Parameter & Kepler-159b & Kepler-159c \\
  \hline
  Mass &  0.1049244768 $M_{J}$ & 6.2583441849 $M_{\earth}$ \\[1pt]
  Eccentricity & 0.0077796061 & 0.5542004366 \\[1pt]
  Asc. Node & 0.0\degr &  91.9345810745\degr \\[1pt]
  Arg. of Peri. & 0.3672973405\degr & 255.727757573\degr \\[1pt]
  \hline
 \end{tabular}
\end{table}

\subsection{Three Planet Scenario} \label{sec:3planres}
We found several configurations that successfully reproduced the observed TTVs with a small degree of error.  The best configurations correspond to the 2:1, 3:1, 4:1, 5:1 and 6:1 resonances of 159d with 159c.  Though most of these are only near and not actually in resonance, we will refer to each as N:1 case (or resonance) for the sake of convenience.

\begin{figure}
 \includegraphics[scale=0.61]{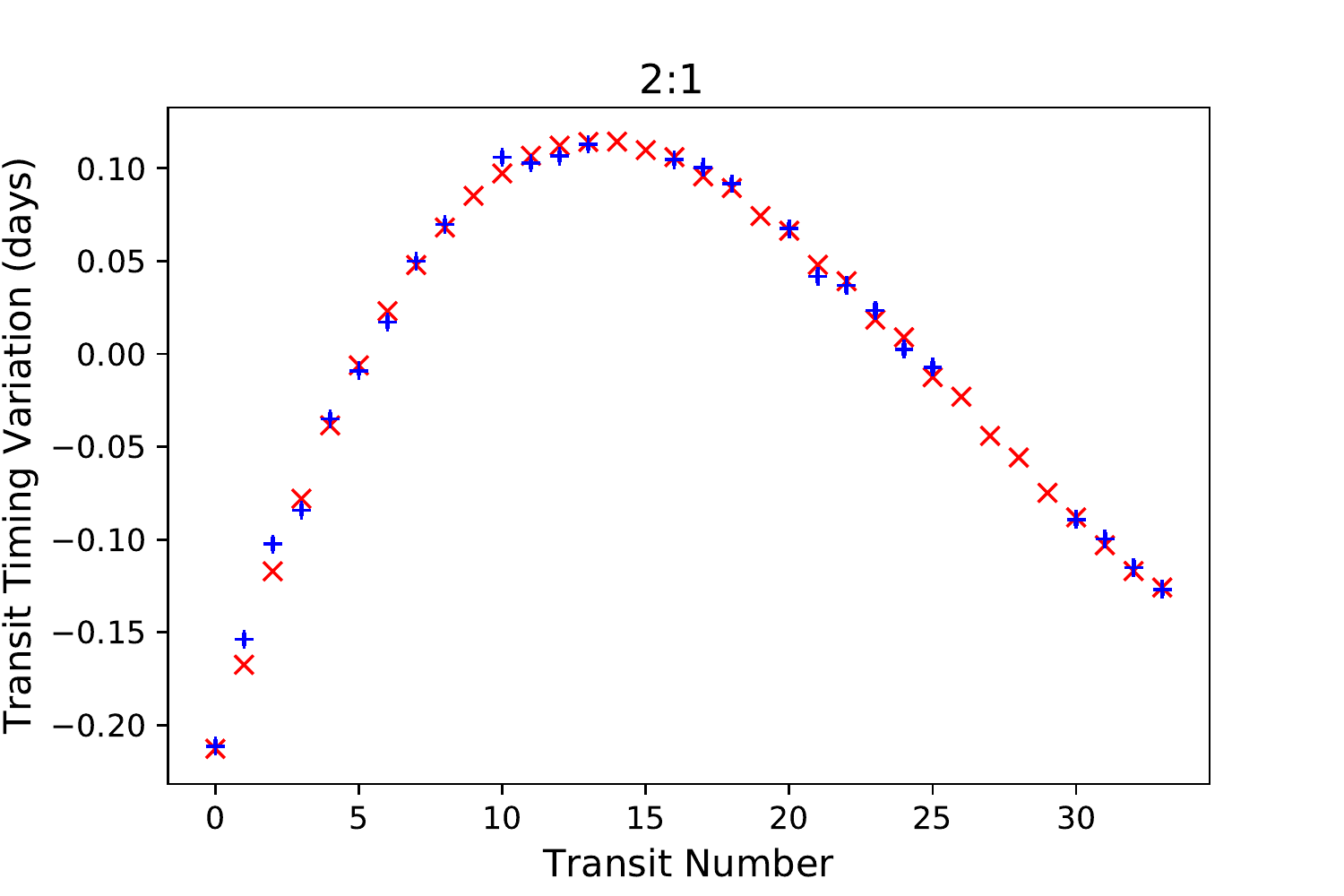}
 \includegraphics[scale=0.61]{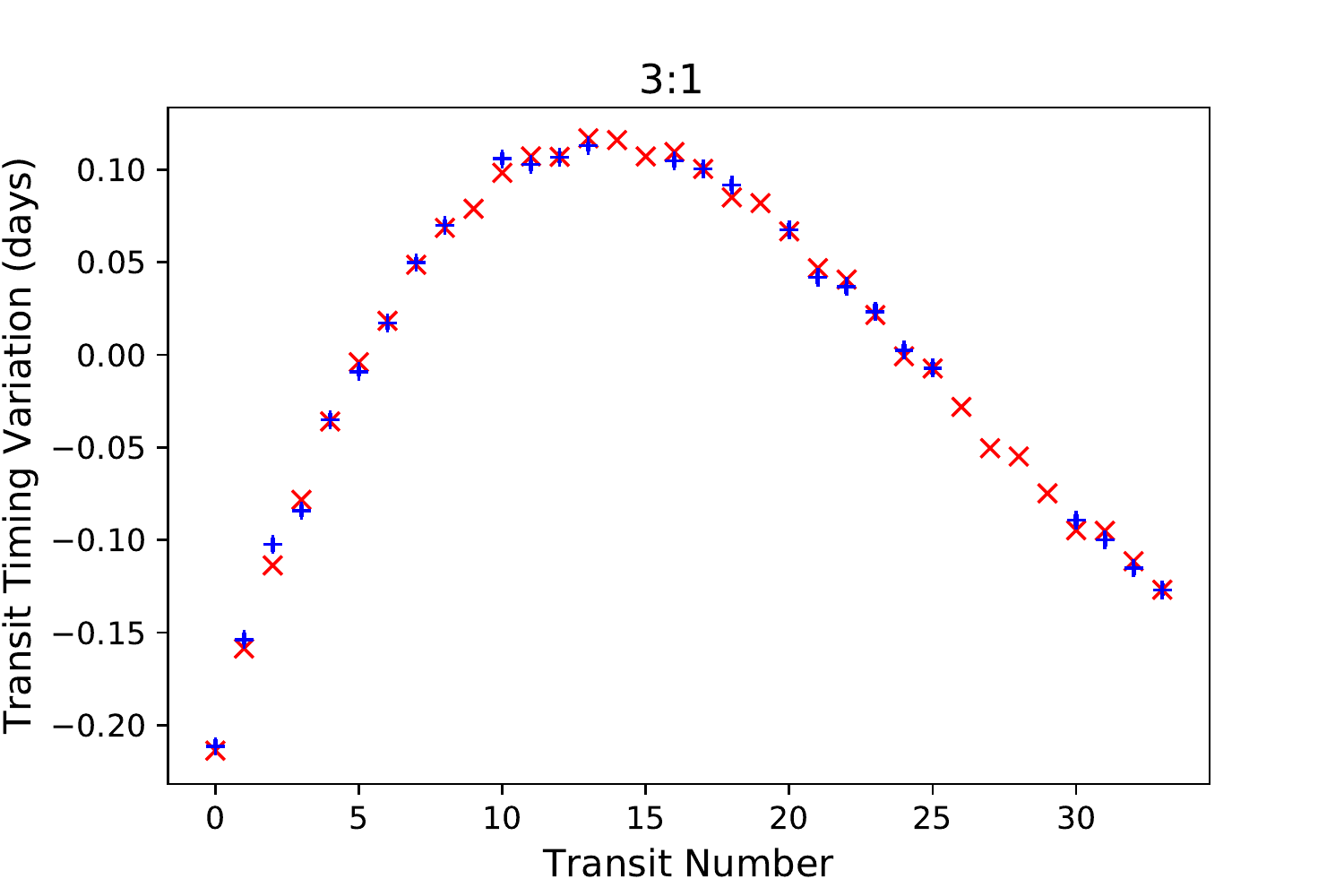}
 \includegraphics[scale=0.61]{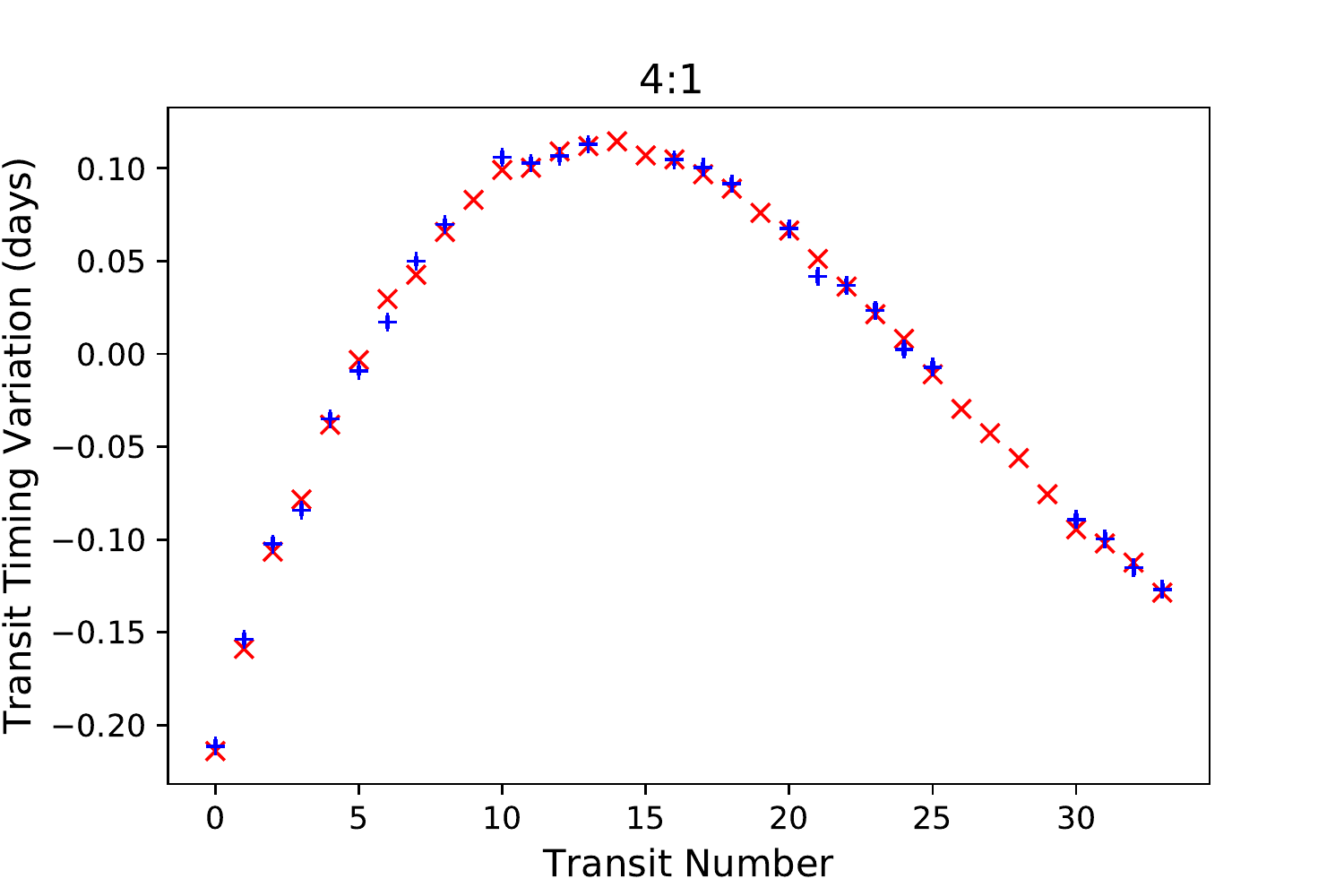}
 \includegraphics[scale=0.61]{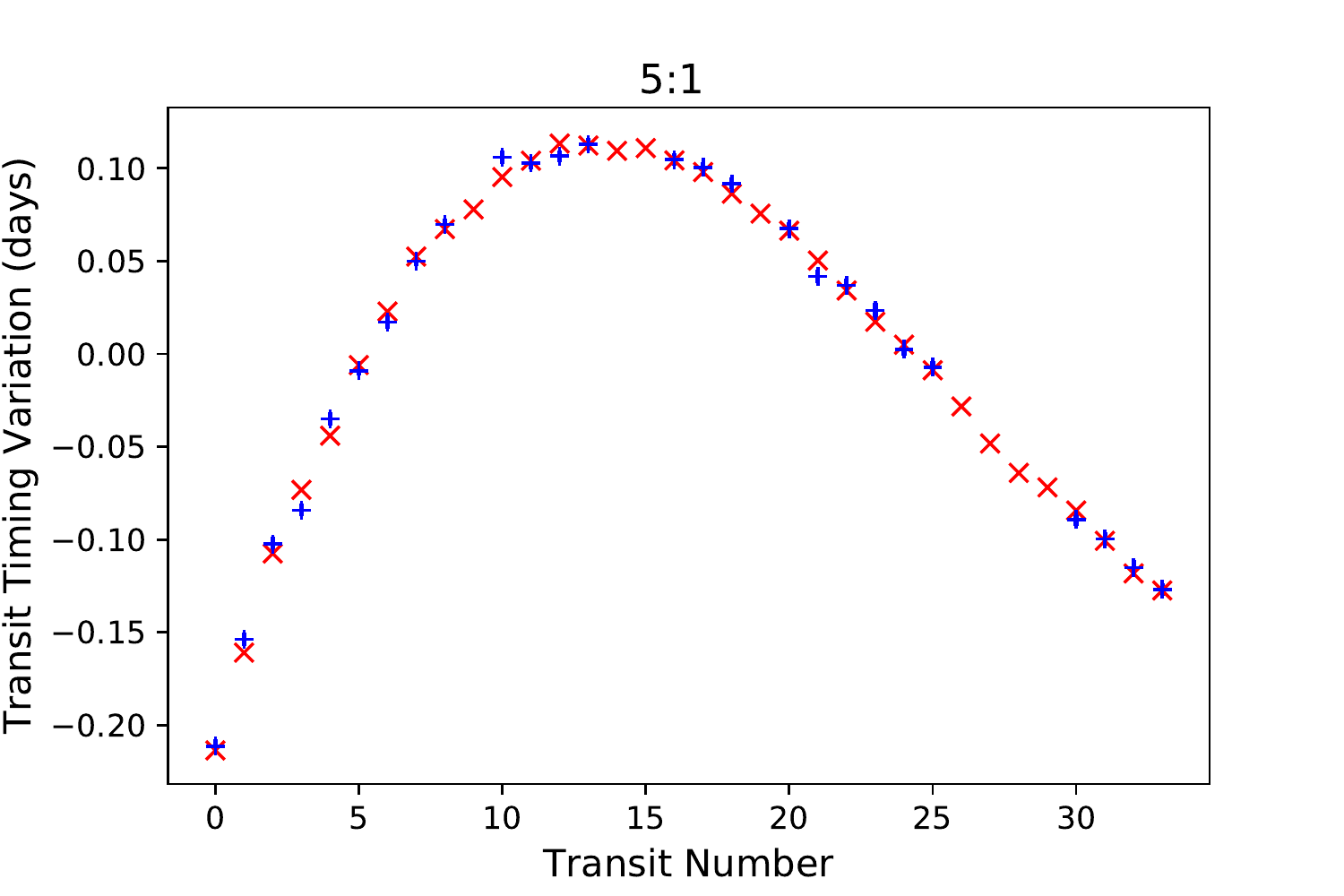}
 \includegraphics[scale=0.61]{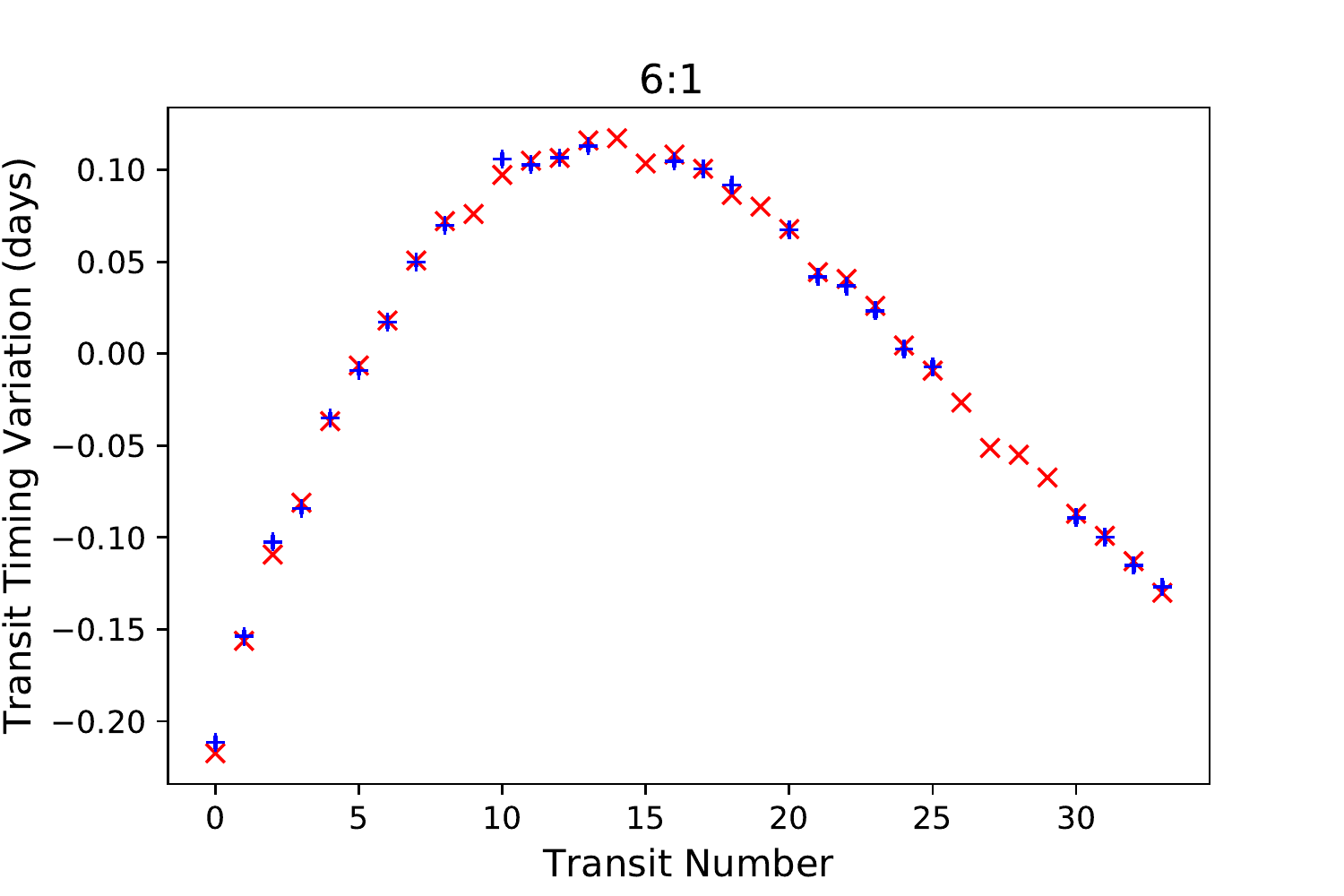}
 \caption{Best-Fit Simulated TTVs for 159c for each N:1 Resonant Case.  \label{fig:simvsobs}  A blue + symbol is the observed data with error bar, and a red x represents the simulated TTV. }
\end{figure}

\begin{figure}
 \includegraphics[scale=0.61]{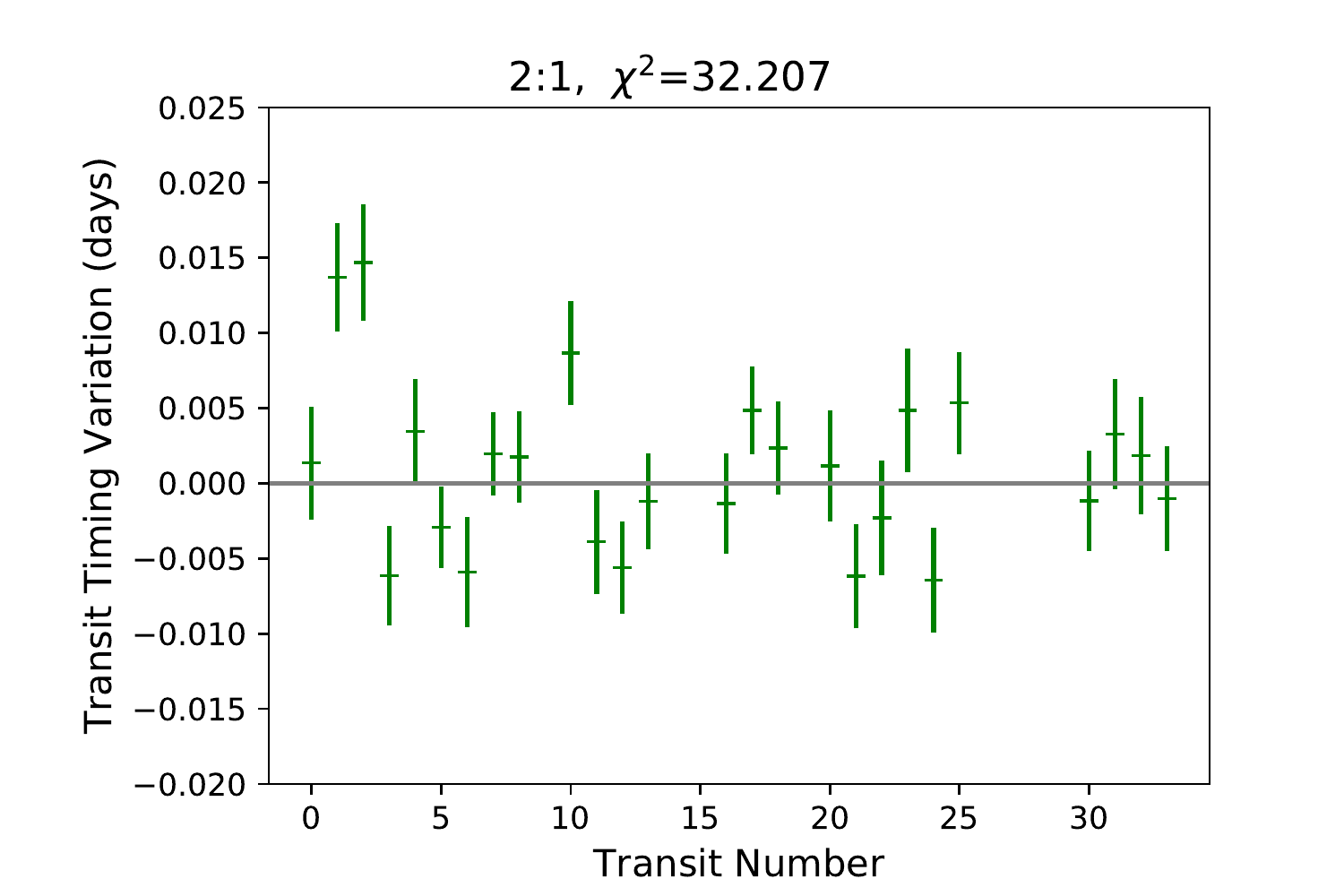}
 \includegraphics[scale=0.61]{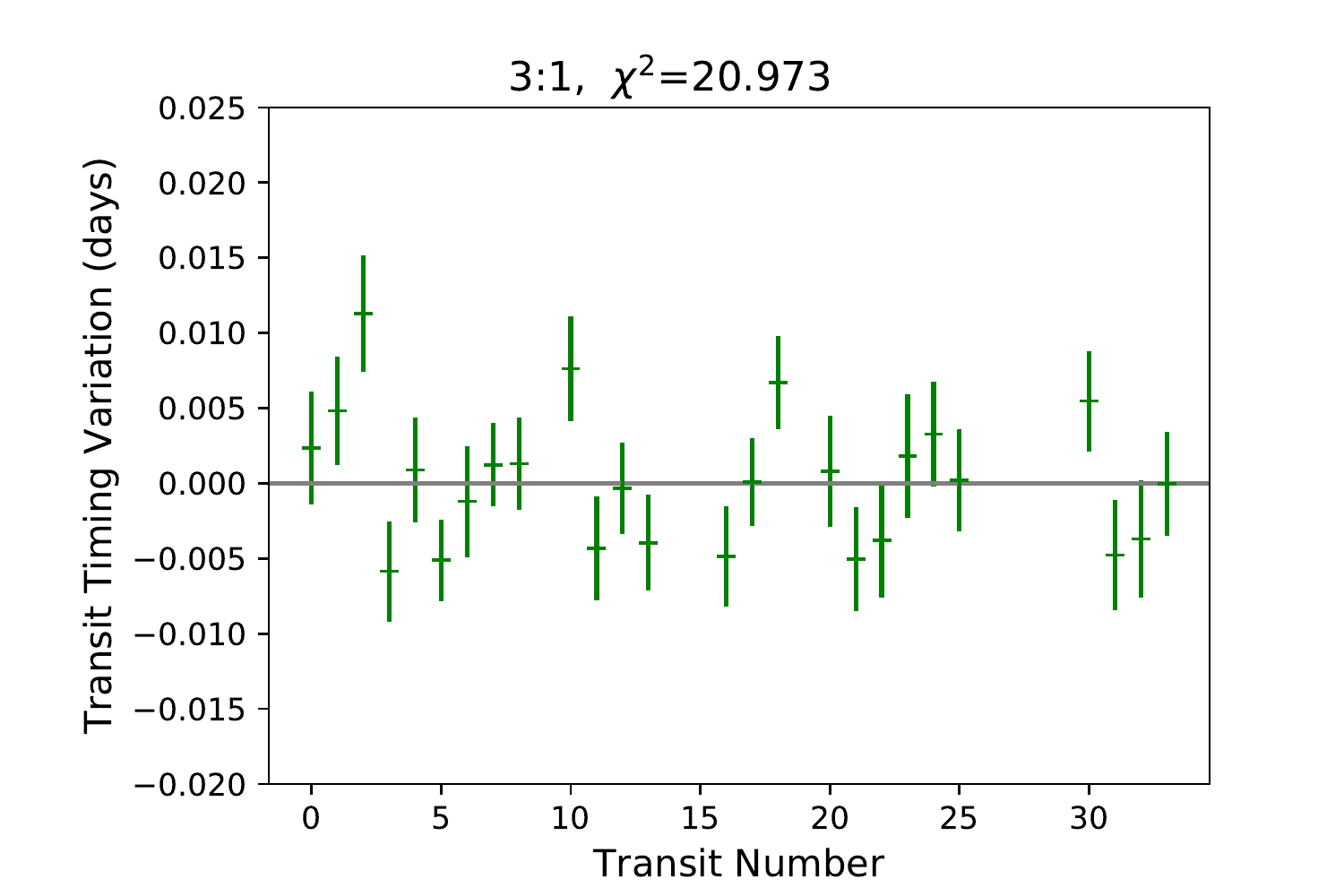}
 \includegraphics[scale=0.61]{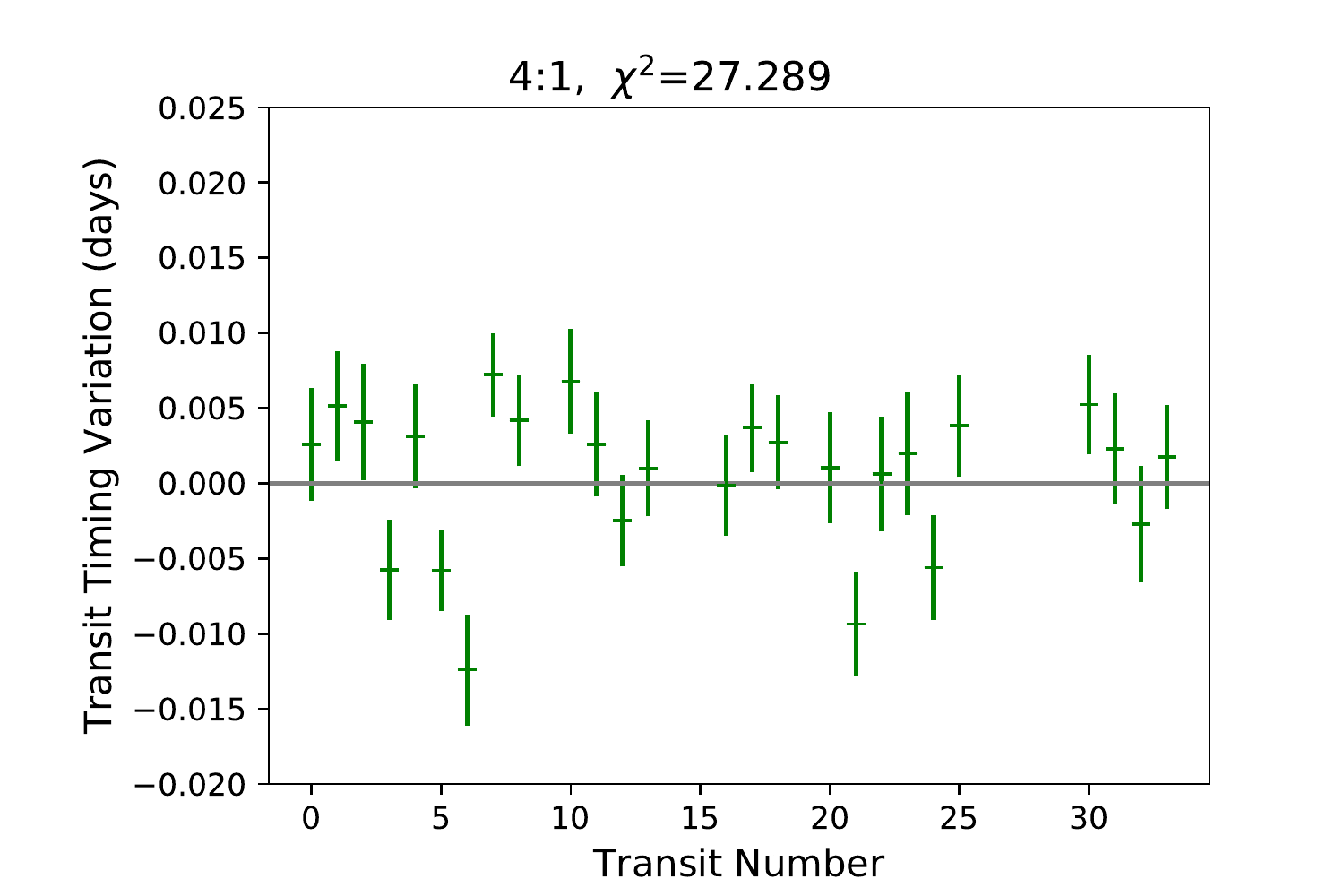}
 \includegraphics[scale=0.61]{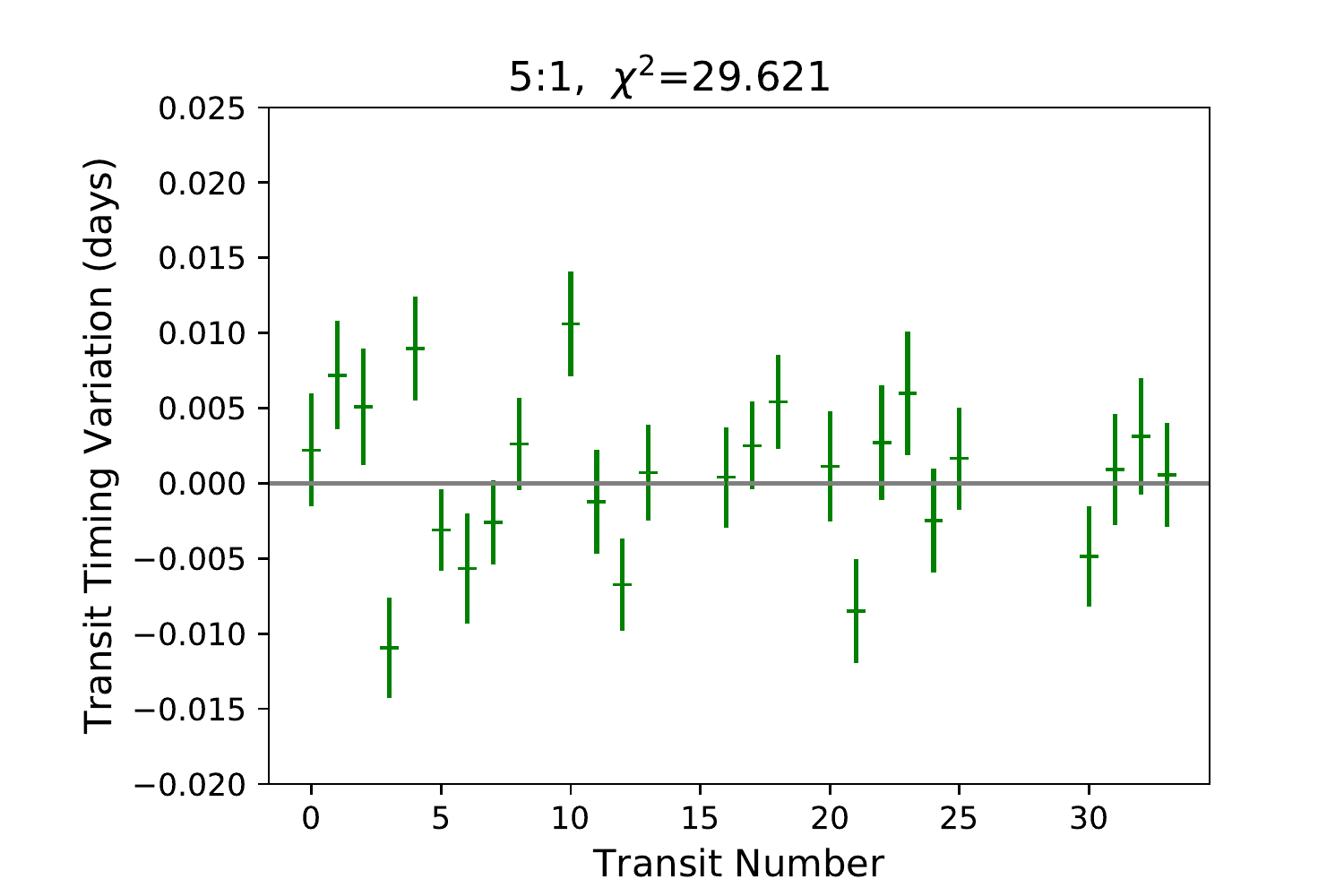}
 \includegraphics[scale=0.61]{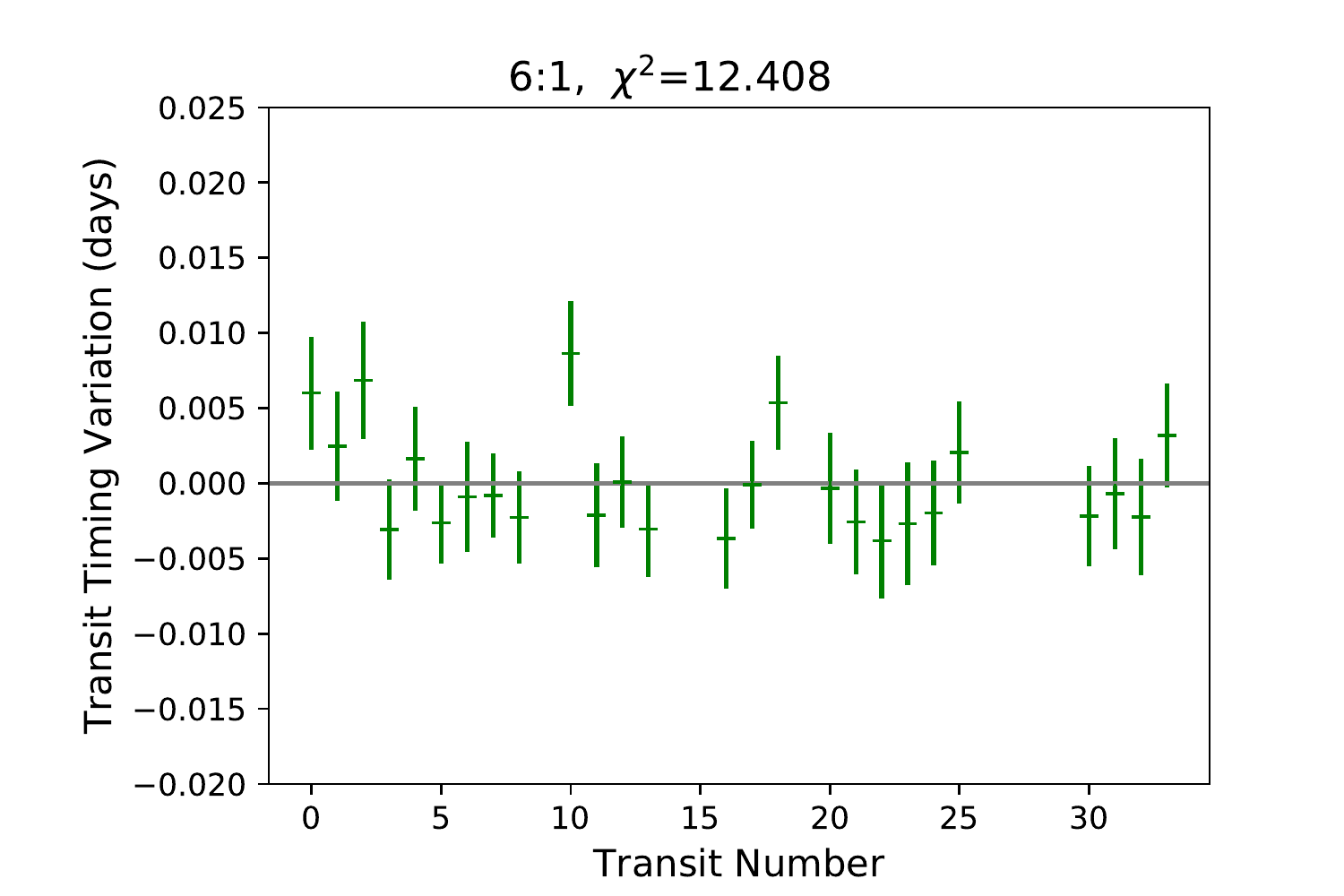}
 \caption{Best-Fit Residuals for 159c for each N:1 Resonant Case  \label{fig:resids} }
\end{figure}

The simulated TTVs are plotted on top of the observed TTVs in Figure \ref{fig:simvsobs}.  Residuals for each case are shown in Figure \ref{fig:resids} along with the $\chi^2$ fit value.  From these results, we conclude that an additional planet, Kepler-159d, exists in this system.

In all simulated best-fit solutions, the newly discovered planet d does not transit.  For the smallest orbit, the 2:1 case, the required inclination for a transit is 90$\degr\pm0.4\degr$, with even tighter constraints on the higher period solutions.  However, our results in all cases put the inclination well outside this range.  Our results are thus consistent with the lack of observations of such a transit from Kepler.

The simulated TTVs exerted on 159b were essentially non-existent in all cases reported here; never varying from the zero point by more than 0.001 days.  Thus, our solutions are consistent with the observed behavior of 159b.

The marginal values (posteriors) from MultiNest provided a range of values and errors ($\pm$1$\sigma$ (68\%) values) for each parameter.  The second set of values reported from MultiNest are the Best-Fit values, which are the set of parameters that provided the single highest Log-likelihood value sampled at any point in the algorithm.  The Best-Fit values do not have errors reported since they represent exact values.  The Best-Fit values all reside inside the errors of the posteriors. 

The results from MultiNest were fed back into TTVFast manually as a data check.  The Best-Fit values provided excellent correlation between the plots.  These are the plots shown in Figures \ref{fig:simvsobs} and \ref{fig:resids}.  However, one should use caution with the outputted posterior values.  The nominal values from Table \ref{tab:medianfits}, when plugged into TTVFast, do not provide good matches with observations; the TTV curves maintain an overall arching shape, but progressively deviate from the observed data with each successive transit.  This misfitting occurs due to the central posterior values being applied without consideration of the correlations that occur between parameters.

\onecolumn

\clearpage

 \begin{table}
  \caption{Marginalized Posterior Values with 1 Sigma Error for Each Resonant Case.  }
  \label{tab:medianfits}
  \begin{tabular}{ccccccc}
    \hline
    Parameter & 2:1 & 3:1 & 4:1 & 5:1 & 6:1 & \\
    \hline
    d mass & $0.308\substack{+0.016\\ -0.013}$ & $1.04\substack{+0.08\\ -0.05}$ & $3.09\substack{+0.07\\ -0.09}$ & $6.90\substack{+0.16\\ -0.15}$ & $13.89\substack{+0.52\\ -0.53}$ & $M_{J}$ \\
    \\
    d period & $88.73\substack{+0.60\\ -0.05}$ & $131.03\substack{+0.07\\ -0.05}$ & $173.07\substack{+0.15\\ -0.16}$ & $216.20\substack{+0.22\\ -0.21}$ & $262.95\substack{+0.22\\ -0.16}$ & days \\
    \\
    d eccentricity & $0.032\substack{+0.006\\ -0.006}$ & $0.137\substack{+0.018\\ -0.019}$ & $0.056\substack{+0.009\\ -0.007}$ & $0.078\substack{+0.005\\ -0.005}$ & $0.198\substack{+0.006\\ -0.006}$ & \\
    \\
    d inclination & $90.9\substack{+0.9\\ -0.1}$ & $77.5\substack{+3.9\\ -4.1}$ & $74.5\substack{+2.2\\ -2.5}$ & $86.5\substack{+2.7\\ -2.5}$ & $85.3\substack{+1.7\\ -1.6}$ & degrees \\
    \\
    d longitude of ascending node & $13.7\substack{+22.2\\ -24.1}$ & $-1.4\substack{+16.6\\ -15.7}$ & $12.9\substack{+16.2\\ -14.2}$ & $-19.0\substack{+19.0\\ -17.3}$ & $21.2\substack{+9.4\\ -36.8}$ & degrees \\
    \\
    d argument of periastron & $7.0\substack{+6.9\\ -4.6}$ & $222.3\substack{+12.5\\ -14.8}$ & $172.4\substack{+6.2\\ -6.1}$ & $99.6\substack{+7.9\\ -7.1}$ & $89.1\substack{+3.7\\ -4.0}$ & degrees \\
    \\
    d mean anomaly & $274.8\substack{+5.5\\ -6.9}$ & $55.9\substack{+9.2\\ -8.3}$ &$90.3\substack{+6.7\\ -7.3}$ & $170.3\substack{+6.3\\ -6.2}$ & $202.6\substack{+2.5\\ -2.3}$ & degrees \\
    \\
    c mass & $6.5\substack{+6.7\\ -2.4}$ & $5.5\substack{+3.5\\ -2.1}$ & $16.1\substack{+5.6\\ -1.8}$ & $18.2\substack{+6.7\\ -3.0}$ & $8.9\substack{+7.0\\ -4.7}$ & $M_{\earth}$ \\
    \\
    c eccentricity & $0.009\substack{+0.002\\ -0.002}$ & $0.181\substack{+0.026\\ -0.031}$ & $0.087\substack{+0.014\\ -0.013}$ & $0.096\substack{+0.003\\ -0.005}$ & $0.064\substack{+0.003\\ -0.003}$ & \\
    \\
    c longitude of ascending node & $3.6\substack{+22.6\\ -23.7}$ & $10.3\substack{+15.9\\ -15.5}$ & $-0.9\substack{+16.1\\ -14.3}$ &  $10.4\substack{+19.2\\ -17.1}$ & $6.5\substack{+10.0\\ -35.7}$ & degrees \\
    \\
    c argument of periastron & $236.4\substack{+17.8\\ -10.5}$ & $247.2\substack{+5.6\\ -7.9}$ & $281.0\substack{+5.6\\ -9.3}$ &  $261.3\substack{+2.8\\ -2.7}$ & $279.1\substack{+5.2\\ -5.2}$ & degrees \\
    \hline
    Nested Sampling Global Log-Evidence & 475.7$\pm$0.02 & 487.8$\pm$0.02 & 485.0$\pm$0.02 & 482.0$\pm$0.04 & 493.4$\pm$0.03 & \\    
  \end{tabular}
 \end{table}

 \begin{table}
  \caption{Best Fit Parameters for Each Resonant Case}
  \label{tab:bestfits}
  \begin{tabular}{ccccccc}
    \hline
    Parameter & 2:1 & 3:1 & 4:1 & 5:1 & 6:1 & \\
    \hline
    d mass & 0.3221631382 & 0.9823871944 & 3.1845969511 & 6.9502940711 & 14.9134726287 & $M_{J}$ \\
    \\
    d period & 88.6719024826 & 130.940442456 & 173.231659674 & 216.17353574 & 262.642480693 & days \\
    \\
    d eccentricity & 0.0328965561 & 0.1771459360 & 0.0491852474 & 0.0783972053 & 0.2050824733 & \\
    \\
    d inclination & 91.0428510417 & 80.000250445 & 80.2976935675 & 96.6614963008 & 85.5464215937 & degrees \\
    \\
    d longitude of ascending node & 23.9249822741 & -19.7243818881 & 13.3203978877 & 18.1524418105 & 25.9271945662 & degrees \\
    \\
    d argument of periastron & 2.7596764301 & 239.442586615 & 177.585357201 & 98.3775876453 & 93.5510409237 & degrees \\
    \\
    d mean anomaly & 276.813762079 & 44.9256736424 & 81.2210236607 & 169.477617267 & 198.029737345 & degrees \\
    \\
    c mass & 7.1875572623 & 4.4947762292 & 14.8442353074 & 16.1473341603 & 8.4575152214& $M_{\earth}$ \\
    \\
    c eccentricity & 0.0100266745 & 0.2330428640 & 0.1178657898 & 0.097737198 & 0.0673859389 & \\
    \\
    c longitude of ascending node & 14.1797256542 & 0.792090619646 & -4.41129668525 & -9.5079755857 & 7.67530949204 & degrees \\
    \\
    c argument of periastron & 223.705031533 & 252.966717978 & 287.938758322 & 260.167384522 & 278.817603374 & degrees \\
    \hline
    Log-Likelihood & 516.27 & 527.96 & 520.85 & 518.80 & 535.39 & \\
  \end{tabular}
 \end{table}

\clearpage
\begin{landscape}
  \begin{figure}
  \includegraphics[scale=0.3,trim={2cm 5cm 2cm 5cm},clip]{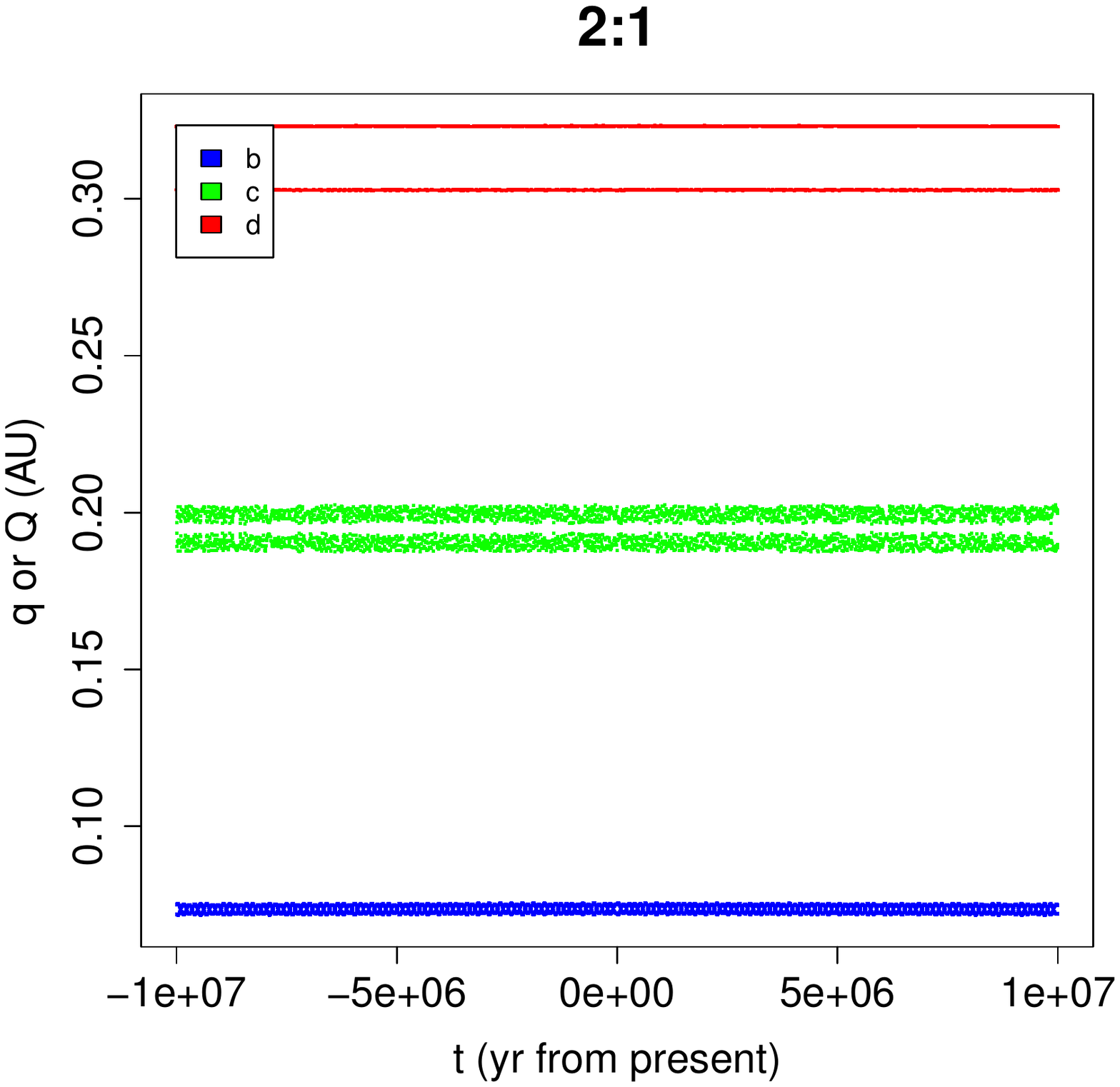}
  \includegraphics[scale=0.3,trim={2cm 5cm 2cm 5cm},clip]{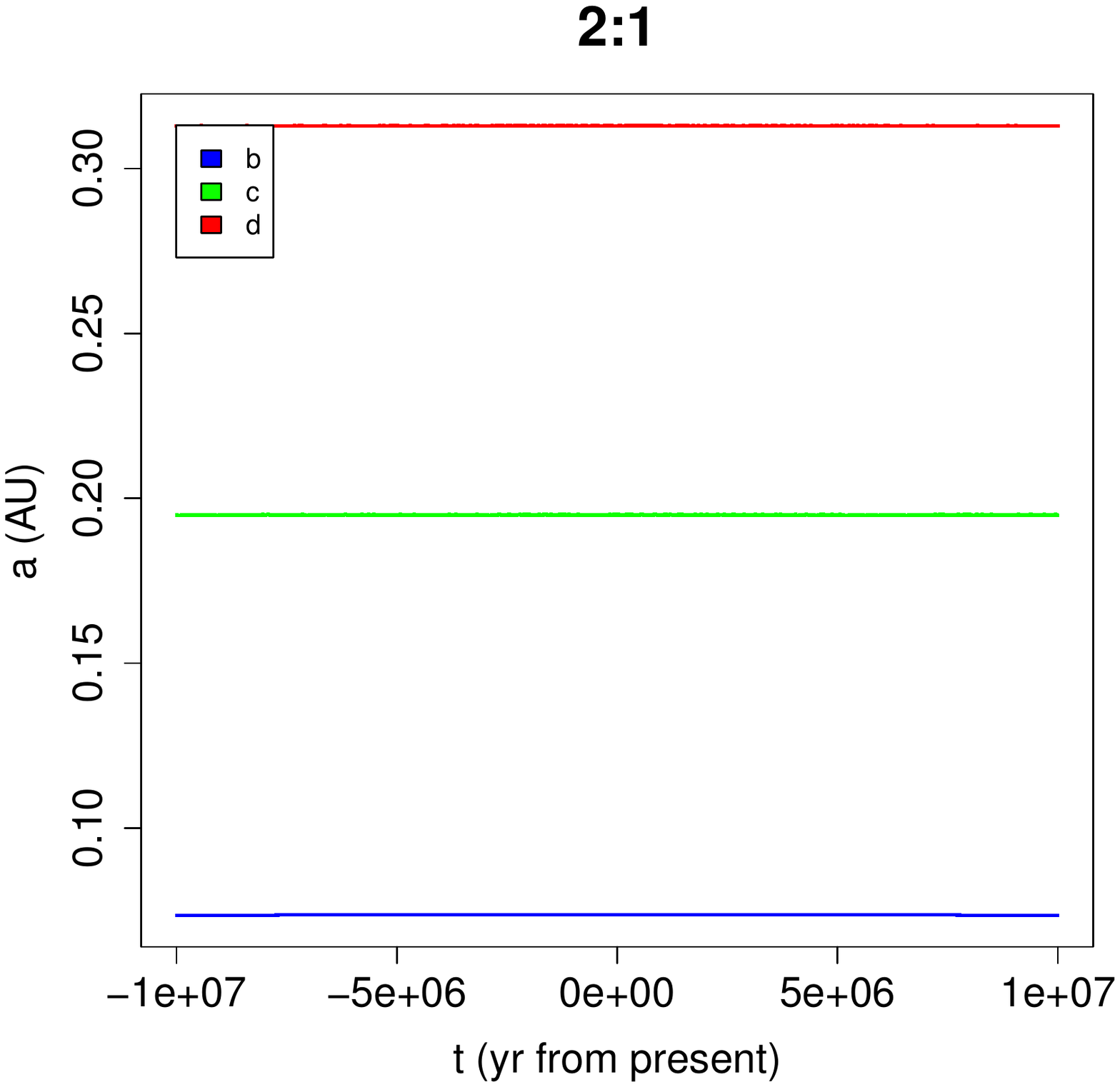}
  \includegraphics[scale=0.3,trim={2cm 5cm 2cm 5cm},clip]{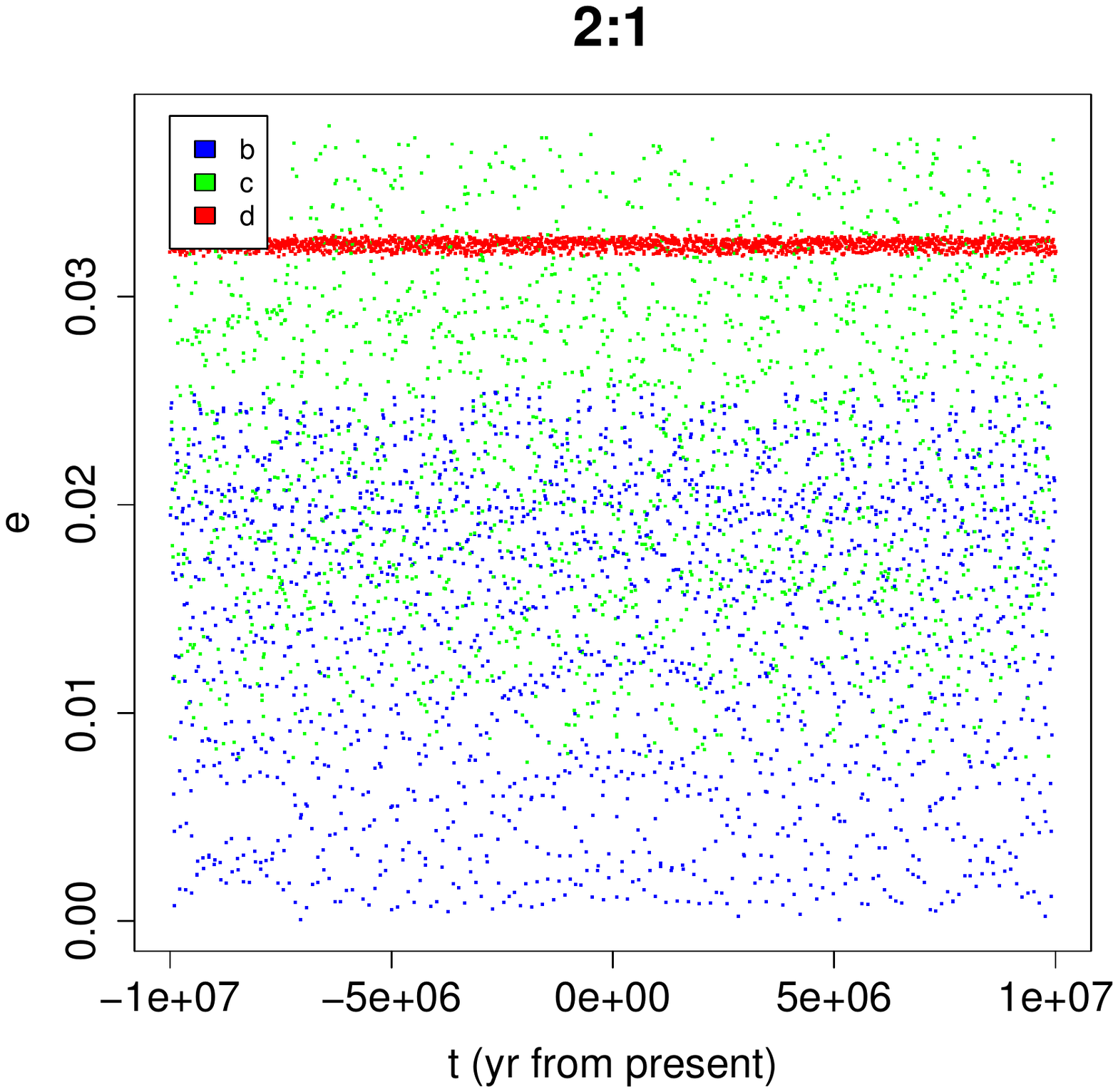}
  \includegraphics[scale=0.3,trim={2cm 5cm 2cm 5cm},clip]{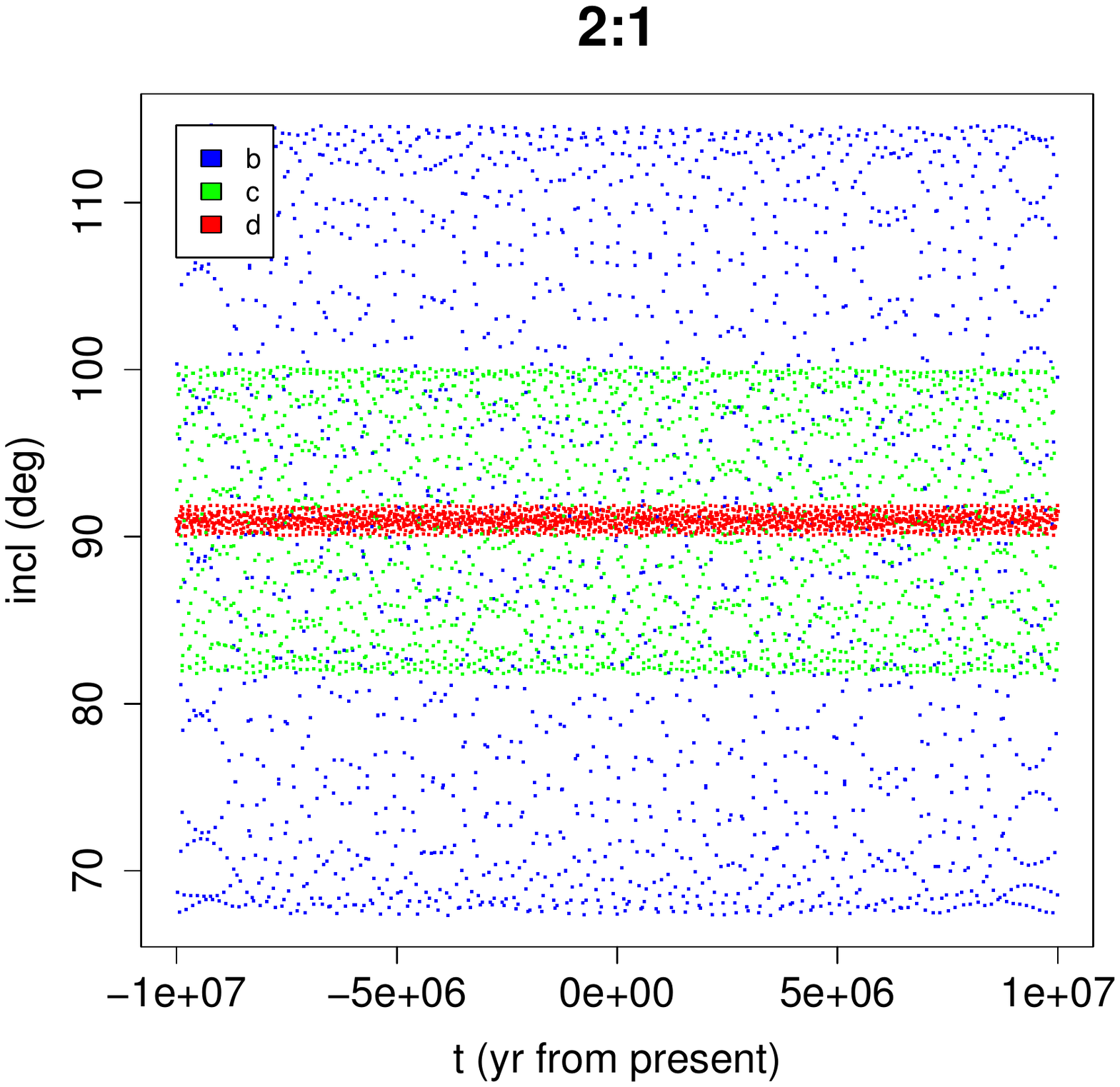}\\
  
  \includegraphics[scale=0.3,trim={2cm 5cm 2cm 5cm},clip]{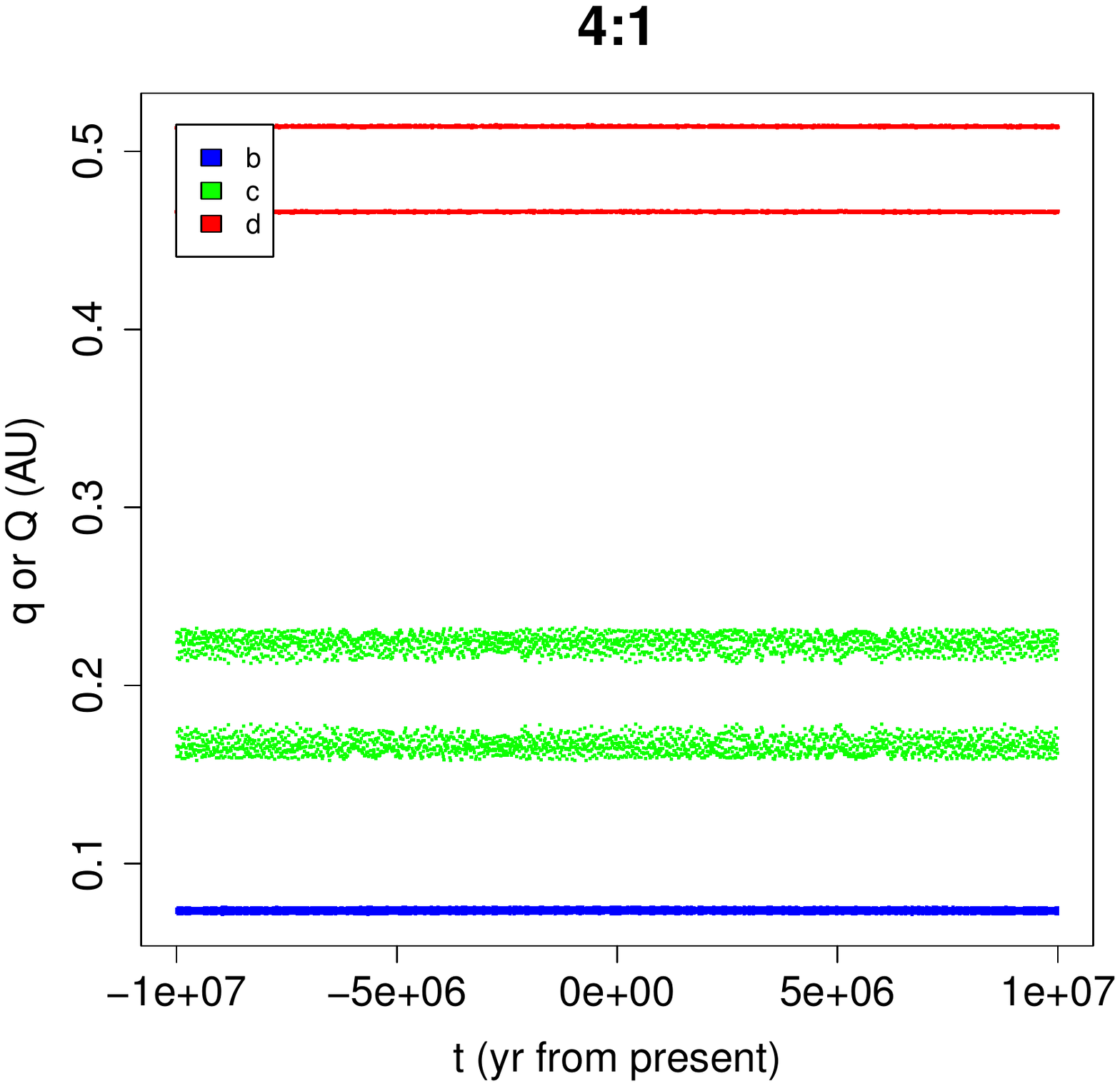}
  \includegraphics[scale=0.3,trim={2cm 5cm 2cm 5cm},clip]{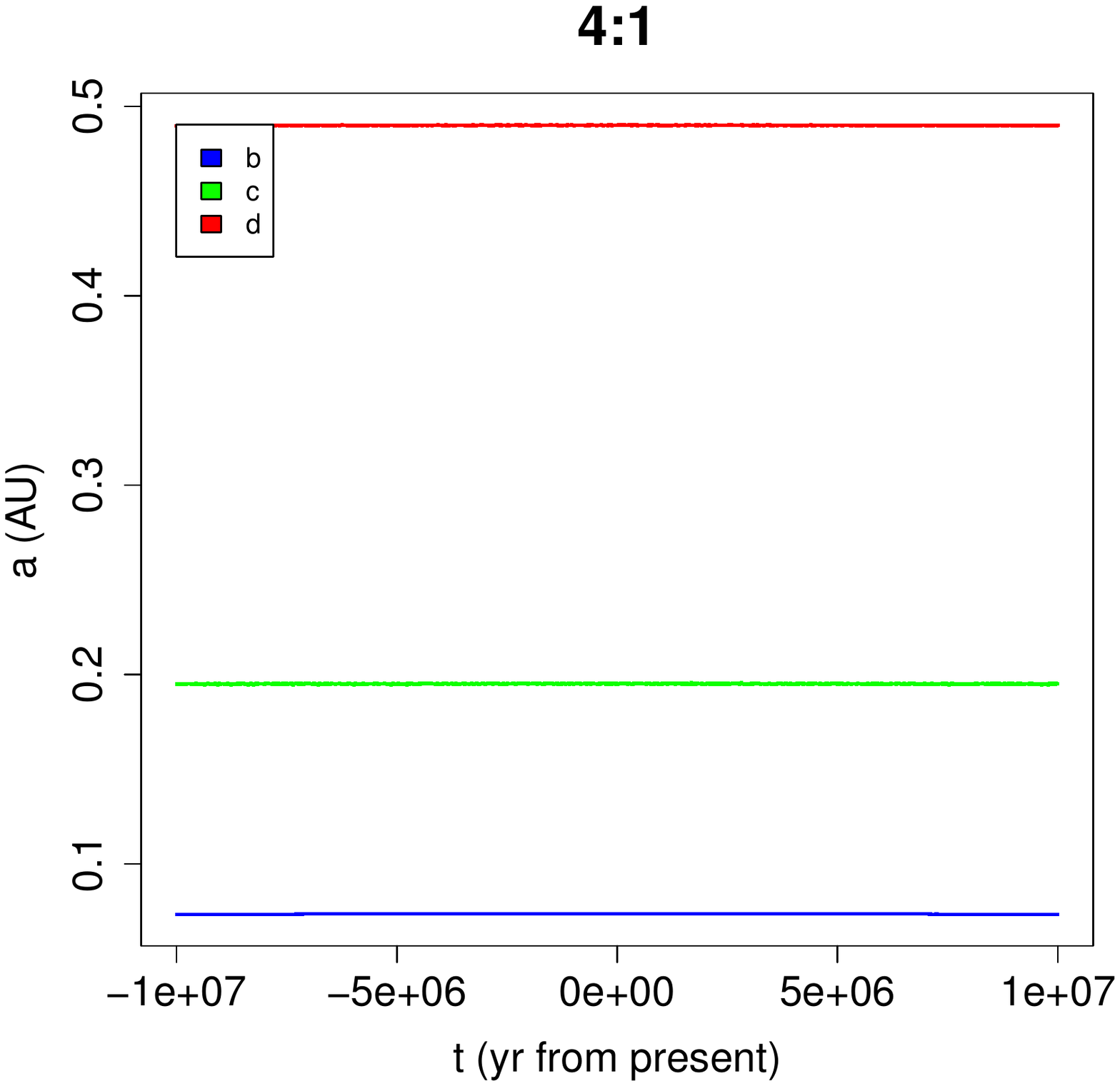}
  \includegraphics[scale=0.3,trim={2cm 5cm 2cm 5cm},clip]{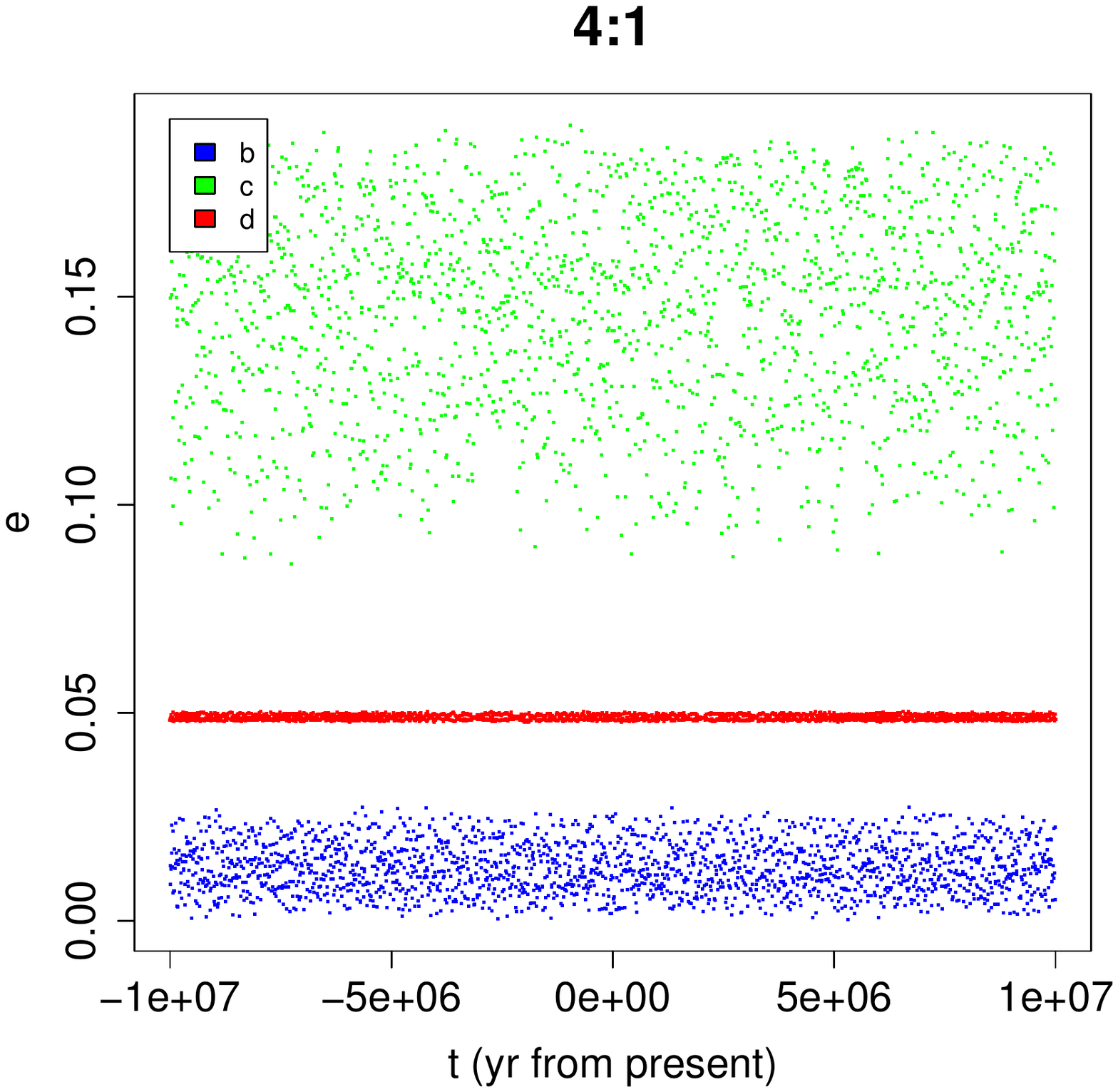}
  \includegraphics[scale=0.3,trim={2cm 5cm 2cm 5cm},clip]{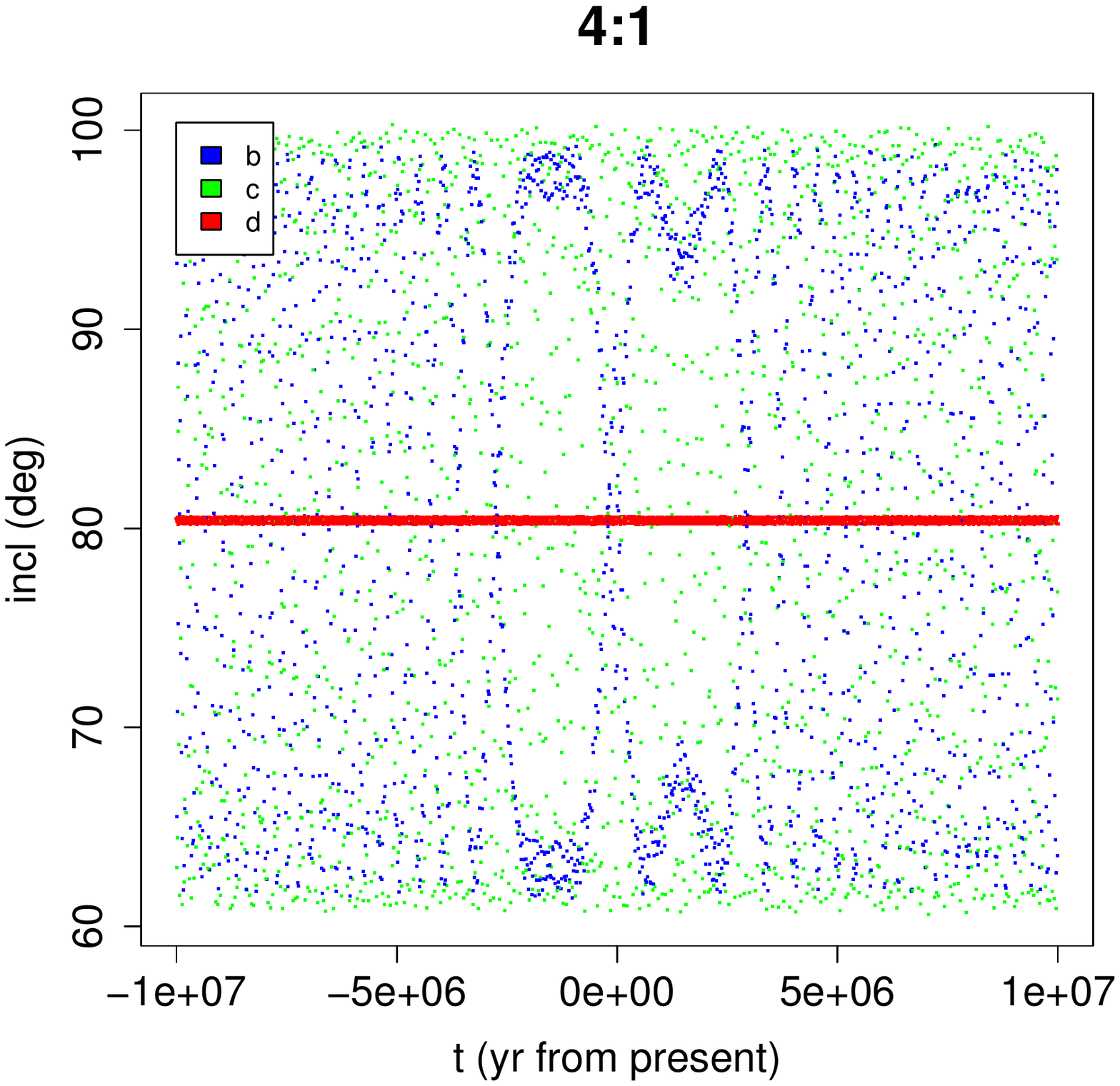}\\
  
  \includegraphics[scale=0.3,trim={2cm 5cm 2cm 5cm},clip]{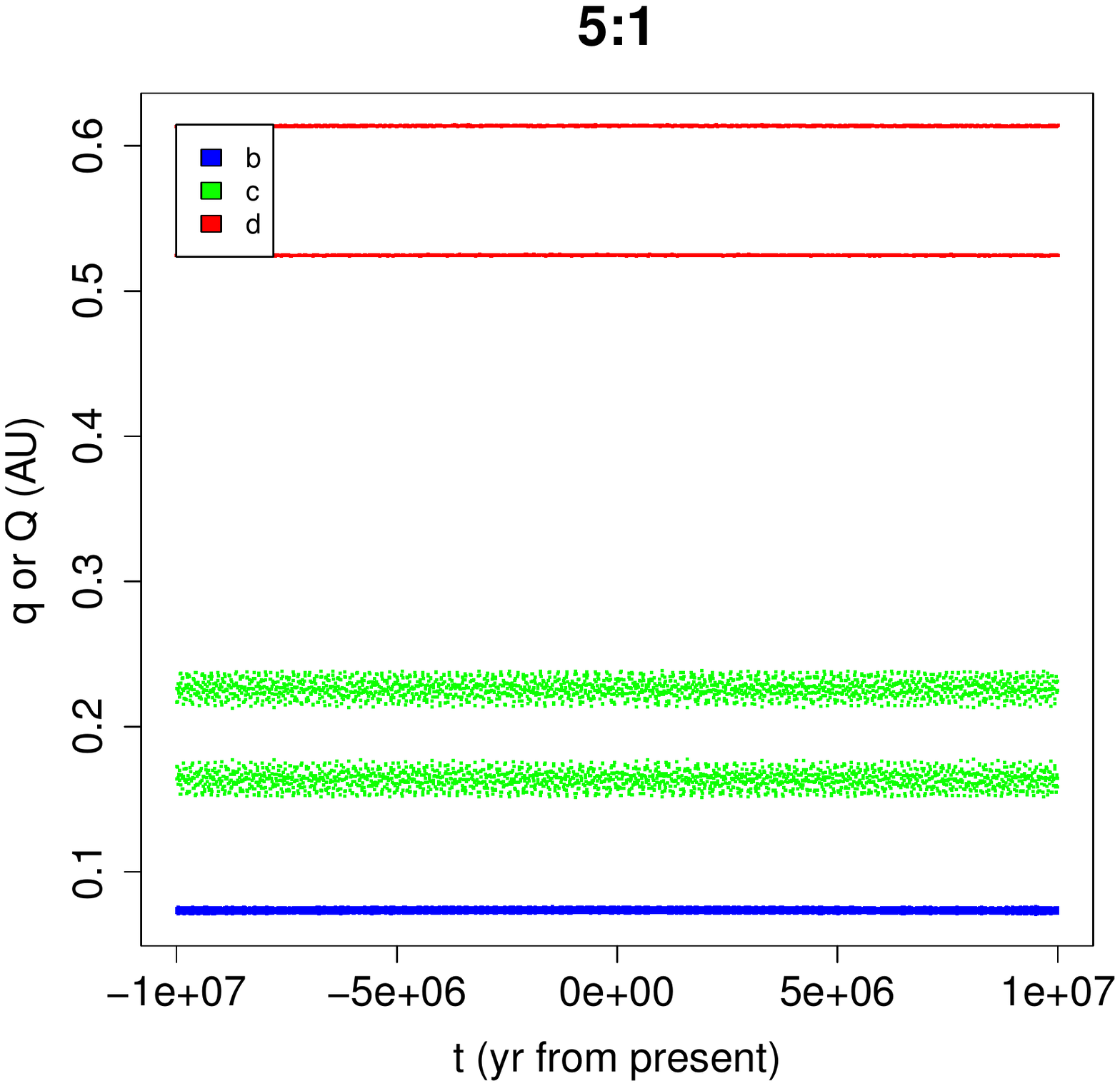}
  \includegraphics[scale=0.3,trim={2cm 5cm 2cm 5cm},clip]{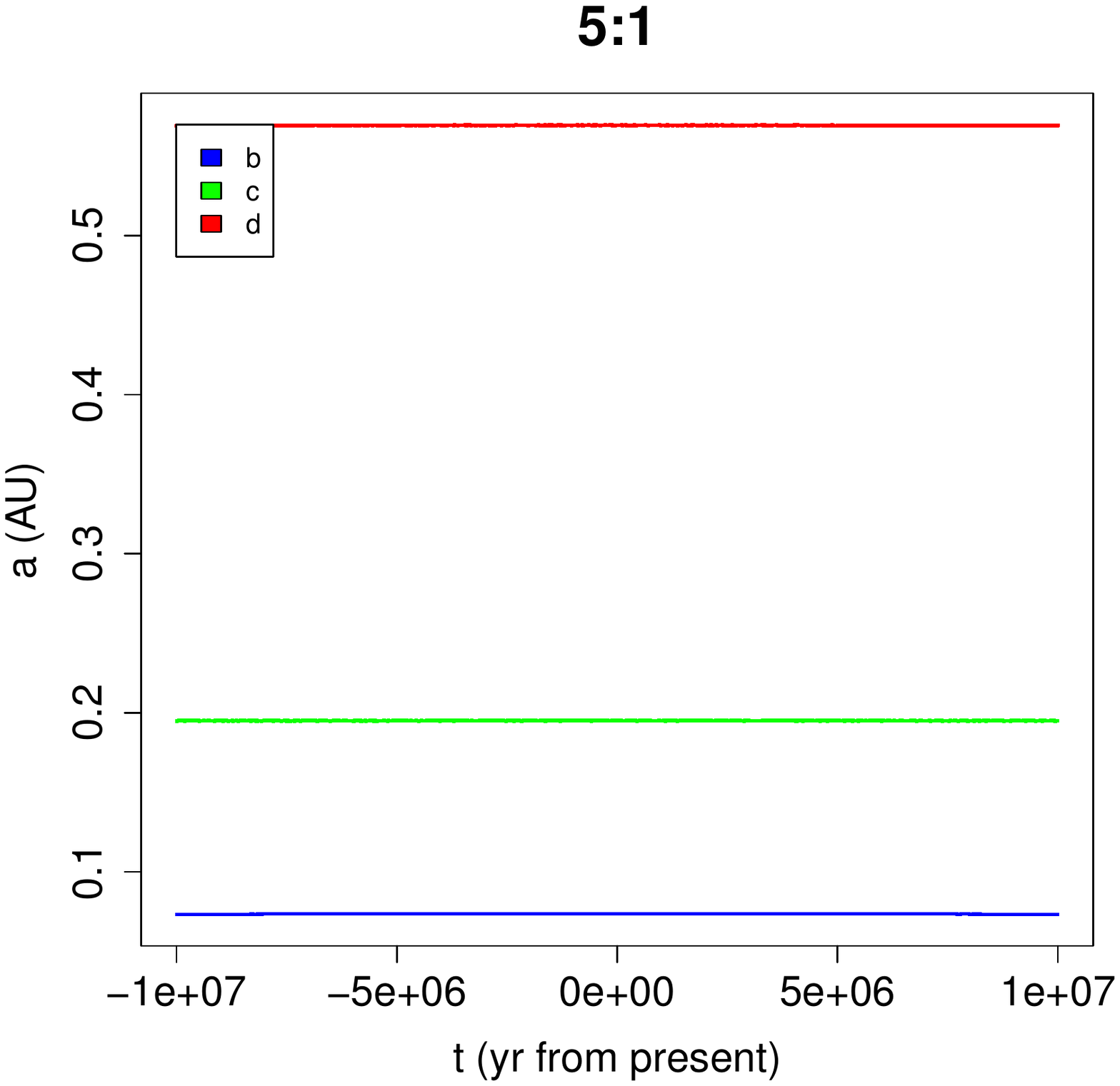}  
  \includegraphics[scale=0.3,trim={2cm 5cm 2cm 5cm},clip]{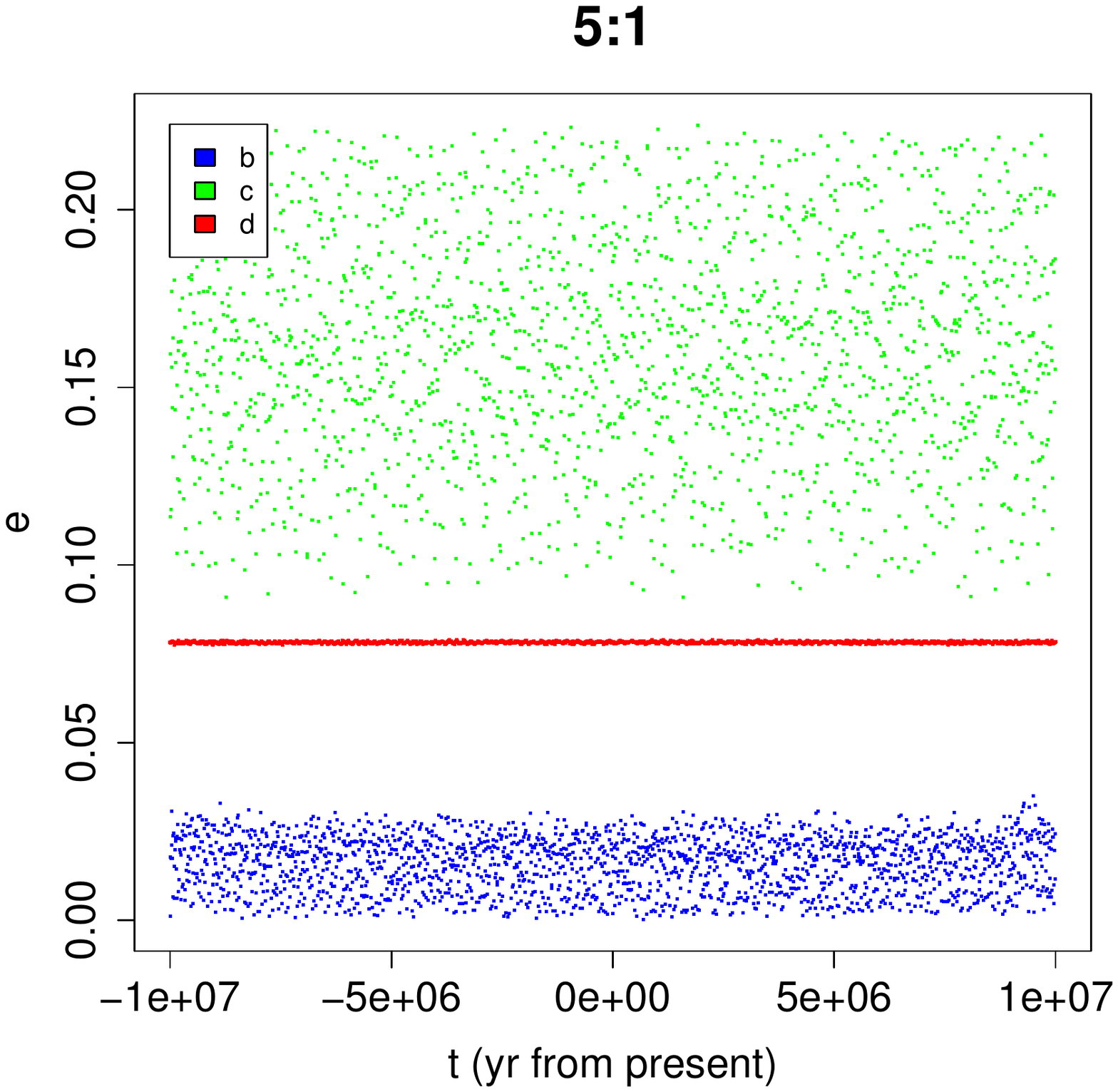}
  \includegraphics[scale=0.3,trim={2cm 5cm 2cm 5cm},clip]{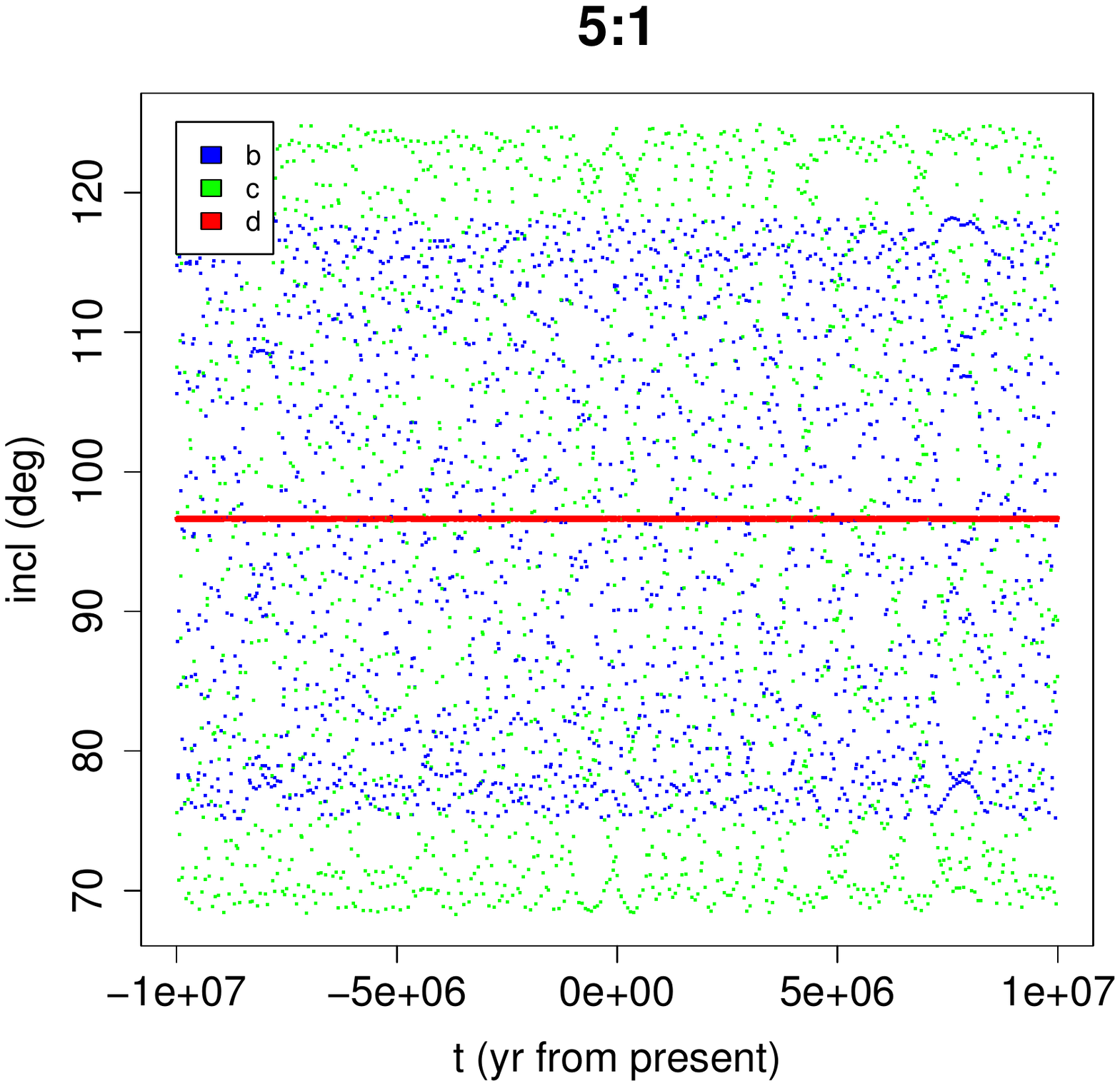}\\
  
  \caption{Long-term (10 Myr) simulation results for best-fit systems. Each row contains the results for each case: the 2:1, 4:1, and 5:1.  Each column corresponds to a different orbital element: periastron/apastron, semi-major axis, eccentricity, inclination.   \label{fig:longterm} }
 \end{figure}
\end{landscape}
\clearpage

\twocolumn

\section{Discussion} 

\subsection{Long-term stability}
The best-fit solutions provided by MultiNest represent good matches to the observed TTVs during the few year lifespan of Kepler but do not necessarily represent systems with long-term (million to billion year) stability. We make the assumption that the Kepler-159 system is stable on these longer time scales to provide an additional constraint. The solutions provided by MultiNest are subsequently simulated for 10 Myr into the past and future to test for signs of instability. The simulation code is symplectic and based on the Wisdom-Holman algorithm \citep{wh1991} using a time-step of 0.5 days, (less than 1/20th of the period of the innermost planet: 0.507 days), and includes post-Newtonian general relativistic effects. Some solutions which provide good matches in the short-term will be seen to have longer-term instability that make them unlikely to reflect the actual properties of the Kepler-159 system.

Integrations of the best-fit systems for 10 Myr into the past and the future are presented in Fig. \ref{fig:longterm}. The primary result is that only the 2:1, 4:1 and 5:1 cases are stable.  The 3:1 and 6:1 cases become unstable on time scales of only a few tens of thousands of years in both directions, resulting in a violent rearrangement of the planets. We will take this as evidence that these particular cases do not represent the actual state of Kepler-159, which is unlikely to be in such a precarious state. 

The 2:1, 4:1 and 5:1 states all remain stable over this time frame without any discernible trends in the orbital elements. The 4:1 case undergoes significant changes in the inclinations ($\pm$20$\degr$) of both 159b and 159c, but not in its semi-major axes and eccentricities.  The 5:1 case shows similar changes in 159b, but even wider variation ($\pm$30$\degr$) in the inclination of 159c.  Further, in the 5:1 case the variation of eccentricity of 159c is higher and varies rapidly between 0.1 and 0.25. The 2:1 case shows similar variations in the inclination of 159b, but less so in 159c ($\pm$10$\degr$), and all planets in the 2:1 case show less variation in the semi-major axis and eccentricities.

\label{sec:resdisc}
The 2:1 resonant case can be verified to be in resonance by calculating the resonant argument (see e.g. \citet{md1999} equation 8.8 though we recast it for an outer perturber, since 159d is more massive) where we find the resonant argument librates with an amplitude of 50$\degr$. The other higher-order cases show circulation of their resonant argument, indicating that the systems may be near but are not fully in mean motion resonance. Though it is common for exoplanetary systems to be near but not quite in resonance \citep{lrf2011,flr2014}, capture into first-order resonances like the 2:1 is preferred in models of planet formation including migration \citep{spn2011,ps2005}. 

\subsection{Other Resonances}
We found that periods near any N:1 resonance out to 6:1 could produce a quality fit.  Given the importance of the resonance in this system, we then looked at other resonances.  We ran simulations near the 7:1, 3:2, 5:2, 7:2, 4:3 and 5:3 resonances.  Of these, the 3:2 resonance is particularly important as it is a common configuration in exoplanetary systems \citep{flr2014}.  We searched several resonances of these forms with a typical range of $\pm10$ days of the resonance, so long as it didn't overlap another case.  The exception is the 4:3 case, whose range of periods also encompassed the 5:4 resonance.

\begin{table}
 \caption{Simulation Fit Quality for Other Resonances}
 \label{tab:othres}
 \begin{tabular}{lll}
  \hline
  Case & log-Evidence & $\chi^2$ \\
  \hline
  7:1 & 432.2 & 71.1 \\[1pt]
  3:2 & 432.9 & 64.2 \\[1pt] 
  5:2 & 476.4 & 34.3 \\[1pt]
  7:2 & 469.7 & 39.9 \\[1pt]
  4:3 & 135.9 & 367.8 \\[1pt]
  5:3 & 338.6 & 164.3 \\[1pt]
  \hline
 \end{tabular}
\end{table}

The quality of fits for these runs is summarized in Table \ref{tab:othres}.  The 7:1 and 3:2 simulations produced fits that did not as strongly match the earlier N:1 cases.  The 4:3 and 5:3 resulted in very poor agreement with the observed TTVs with large deviations well outside the observational error.  Only the 5:2 and 7:2 provided fits of a quality comparable to the N:1 cases.  However, creating these cases required relatively high eccentricities (0.15-0.30), unlike the N:1 cases.  Searching these regimes with forced low eccentricities resulted in far worse fits (log-Evidences near 100 and $\chi^2$ values near 400).  Long-term simulations of the 5:2 and 7:2 showed them to be highly unstable, resulting in crossing orbits after only a few thousands of years.  Thus, we conclude that the 7:1, 3:2, 5:2, 7:2, 4:3, and 5:3 cases do not represent the true nature of the system.

\subsection{Favoured Configuration}
We favour the 2:1 resonance as the most likely scenario for a number of reasons. First, we note that the quality of each N:1 fit shown here is essentially as good as any other.  The log-Evidence values are all similar, and the Best Fit $\chi^2$ values are all of order of degrees of freedom; we have 26 data points and the worst $\chi^2$ value is 32.  Second, smaller planets are more common than massive planets, particularly for lower-mass stars, and the 2:1 case requires the smallest mass for 159d.  Third, planets with periods just above an exact 2:1 resonance are among the most common seen, while those above 3:1 resonance are quite rare \citep{flr2014}.  Finally, our long-term simulations show the 2:1 to be a very stable case with orbital parameters not varying significantly over 10 million years.  We conclude that the case where 159d is in 2:1 resonance with 159c is the most probable scenario.

\subsection{Planetary Masses}
The estimated mass of the new planet, 159d, is dependent on the case chosen, but has relatively low uncertainty within each case.  The best fit masses are 0.32, 0.98, 3.18, 6.90, and 14.91 $M_{J}$ for the 2:1, 3:1, 4:1, 5:1, 6:1 cases respectively.  In the 2:1 case, near-identical TTV curve fits can be produced with a mass of 159d as high as 0.49 and as low as 0.3 $M_{J}$.  Larger masses of d also required an increase in mass of 159c, seen in the correlation on Table \ref{tab:corls}.  Our reported value of the mass of 0.32 $M_{J}$ was chosen due to this run having the highest log-Evidence value.  Regardless of resonant case, the most massive object in the system is the new planet 159d.  In all of our results, the lowest ratio of the mass of 159d to 159c in any run was 6.  In our preferred case (best fit of 2:1), the ratio of the masses of 159d to 159c is approximately 14. 

There is significant uncertainty in the mass of 159c.  In addition to the reported error, multiple runs for the same resonance could produce very different Best Fit values for the mass of 159c.  The highest value found was 30 $M_{\earth}$ with the lowest being 6.5 $M_{\earth}$, with nearly as good results in both log-Evidence and $\chi^2$ values.  There is a correlation (see Table \ref{tab:corls}) between the masses of 159c and 159d.  Our best fit of 159c's mass for the 2:1 case is 7.2 $M_{\earth}$.  This is within the initial mass estimates from Table \ref{tab:bcprops}, and would equate to a mean density about 20 percent higher than Neptune.  However, the posterior values ($6.5\substack{+6.7\\ -2.4}$) have significant uncertainty.  If the actual mass were on the upper end of this error, we obtain a density similar to that of Mars.  Given the relative insensitivity of this parameter and that other runs with similar fits could have masses of around 30 $M_{\earth}$,  we cannot say anything definitive about the planet's mass or density, nor whether it is gaseous or terrestrial.  

All simulations were performed assuming a 4.0 $M_{\earth}$ of the inner planet, 159b.  The mass was not set as a free parameters because it shows no signs of TTVs, suggesting its coupling to the other planets is not large.  To confirm this hypothesis, we re-ran each resonant case with a different (but still static) mass of 159b.  Modifying the mass by a factor of 0.25 to 3.0 (for a range of 1-12 $M_{\earth}$) resulted in only minor differences to the results for 159c.  However, we found in the 2:1 case TTVs were induced on 159b as its mass was increased above about 8 $M_{\earth}$, showing a linear increasing trend.  From these tests, we conclude that 159b has a minor impact on the reported results for 159c and 159d, and we can place an upper limit on the mass of 159b of about 8 Earth masses. 

\subsection{Orbital Parameters}
The orbit of 159b is used as the orbital reference plane (longitude of the ascending node of 0\degr).  The other planets are inclined relative to the orbital plane of 159b.  Further, 159c and 159d are consistently inclined with each other by anywhere from 10$\degr$ to 20$\degr$.  Correlation plots on Table \ref{tab:corls} show a strong relationship between the two values. The inclination with respect to 159b does not seem to matter much to this relation; it is the mutual inclination between 159c and 159d that is most relevant to the dynamics.

While the argument of periastron varies, the physical starting location of 159d relative to the other planets is quite consistent.  The starting location of 159d was found to be on the opposite side of the star (argument of periastron + true anomaly $\approx$ 270$\degr$) in all our best case fits.  That is, it starts near superior conjunction.  We did find the occasional run in which 159d started on the near side of the star, near inferior conjunction.  In the 2:1 inferior conjunction case, the masses of 159c and 159d were much higher; 29 $M_{\earth}$ and 0.5 $M_{J}$ respectively.  At no point did we ever find a result (in any N:1 resonant case) where the starting location of 159d was far from superior or inferior conjunction.

For most resonant cases, we found equally good solutions with very different eccentricities.  In most cases, the eccentricities for both planets was less than 0.1, but in some cases they were 0.15 and higher.  Long term simulations of these higher eccentricity solutions proved to be quickly unstable, regardless of the resonance case examined.  In some cases (most notably the 3:1 and 6:1), we did followup simulations forcing the eccentricity prior to less than 0.1.  While we found equally good fits with low eccentricities, long term simulations showed these systems to still be unstable.  This suggests that low eccentricity is required but not sufficient for stability in this system.

\subsection{Habitable Zone}
Using the estimated size and temperature of Kepler-159 \citep{rowe2015, math2017}, the habitable zone for this system can be computed \citep{kg2012, kop2013}.  This system has a conservative zone (runaway greenhouse to maximum greenhouse boundaries) of 0.24 AU to 0.45 AU,  while the optimistic (recent Venus to early Mars) estimate puts it at 0.18 AU to 0.47 AU.  

\begin{table}
 \caption{Relative Flux from Star for Each Planet Location}
 \label{tab:hzs}
 \begin{tabular}{lllll}
  \hline
   & Semi-major & Stellar & Equil. & \\
  Case & axis (AU) & Flux ($F_{\sun}$) & Temp. (K)* & HZ region \\
  \hline
  2:1 & 0.313 & 0.525 & 217 & conservative \\
  3:1 & 0.406 & 0.313 & 191 & conservative  \\
  4:1 & 0.489 & 0.216 & 174 & no (near edge) \\
  5:1 & 0.573 & 0.157 & 161 & no  \\
  6:1 & 0.646 & 0.124 & 151 & no \\
  \hline
 \end{tabular}
  *Equilibrium Temperatures assume a bond albedo of 0.3.
\end{table}

159b is too close to the star and is well inside the habitable zone.  159c is inside the optimistic zone.  Whether the new planet, 159d, exists in the habitable zone depends on the resonant case, as shown in Table \ref{tab:hzs}.  In the 2:1 and 3:1 case, 159d would be inside the conservative zone.  159d in the 4:1 case resides just outside the outer edge of the optimistic zone. 

\subsection{Stellar Mass}
We used a mass of 0.52 $M_{\sun}$ in all of our runs, based on \citet{muir2012}, \citet{rowe2015} and \citet{math2017}.  NASA's Exoplanet Archive lists a "standard" mass of 0.72 $M_{\sun}$, based on \citet{rowe2014}.  This value is often seen in other sources, such as exoplanets.org.  For comparison we ran some simulations with this higher mass estimate of 0.72 $M_{\sun}$. 

We found that changing the mass primarily affected the reported planetary masses with only slight differences in orbital parameters.  In effect, the masses of the planets scale with the mass of the star with each resonant case requiring a mass typically around 50 percent higher than for the 0.52 $M_{\sun}$ star.  The reported periods were identical to within 1$\sigma$.

\section{Conclusions}
We report the existence of a new planet in the Kepler-159 system, likely in a 2:1 resonant orbit, period of 88.7 days with 159c and with a mass similar to that of Saturn's.  We cannot with complete certainty rule out the 4:1 and 5:1 cases (corresponding masses of 3.1 and 6.9 $M_{J}$), but these are less likely.  We can rule out the other resonant cases as they all prove to be unstable on the order of a few ten-thousands of years or less.

Further, we have gained insights into the other aspects of the system.  All planets have low (<0.1) eccentricities.  There is a strong correlation in the orbital planes of 159c and 159d, differing by an average of 12$\degr$.  The mass of 159c and 159b are uncertain, but are less than the mass of 159d by at least a factor of 10.

The Kepler-159 system is an illustration of the degeneracy of TTVs produced by a non-transiting planet.  Firming up certain parameters will require further observations, either from more transits, radial velocity measurements (which we estimate from reflexive motion should be approximately 20 m/s in the 2:1 case, and higher for the other cases), or possibly transit duration variations.  We hope to explore this system further in the future.

\section{Acknowledgements} We thank the anonymous referee for their work which much improved the paper. We also thank Stephen Kane for his useful feedback and advice. This work was funded in part by the Natural Sciences and Engineering Research Council of Canada.

\begin{landscape}
 \begin{table}
  \caption{Correlation Plots for the 2:1 Resonance Case}
  \label{tab:corls}
  \begin{tabular}{ccccccccccc}
    \hline
    & d mass & d period & d eccentricity & d inclination & d ascending node & d periastron & d mean anomaly & c mass & c eccentricity & c ascending node \\
    d period & \includegraphics[scale=0.11]{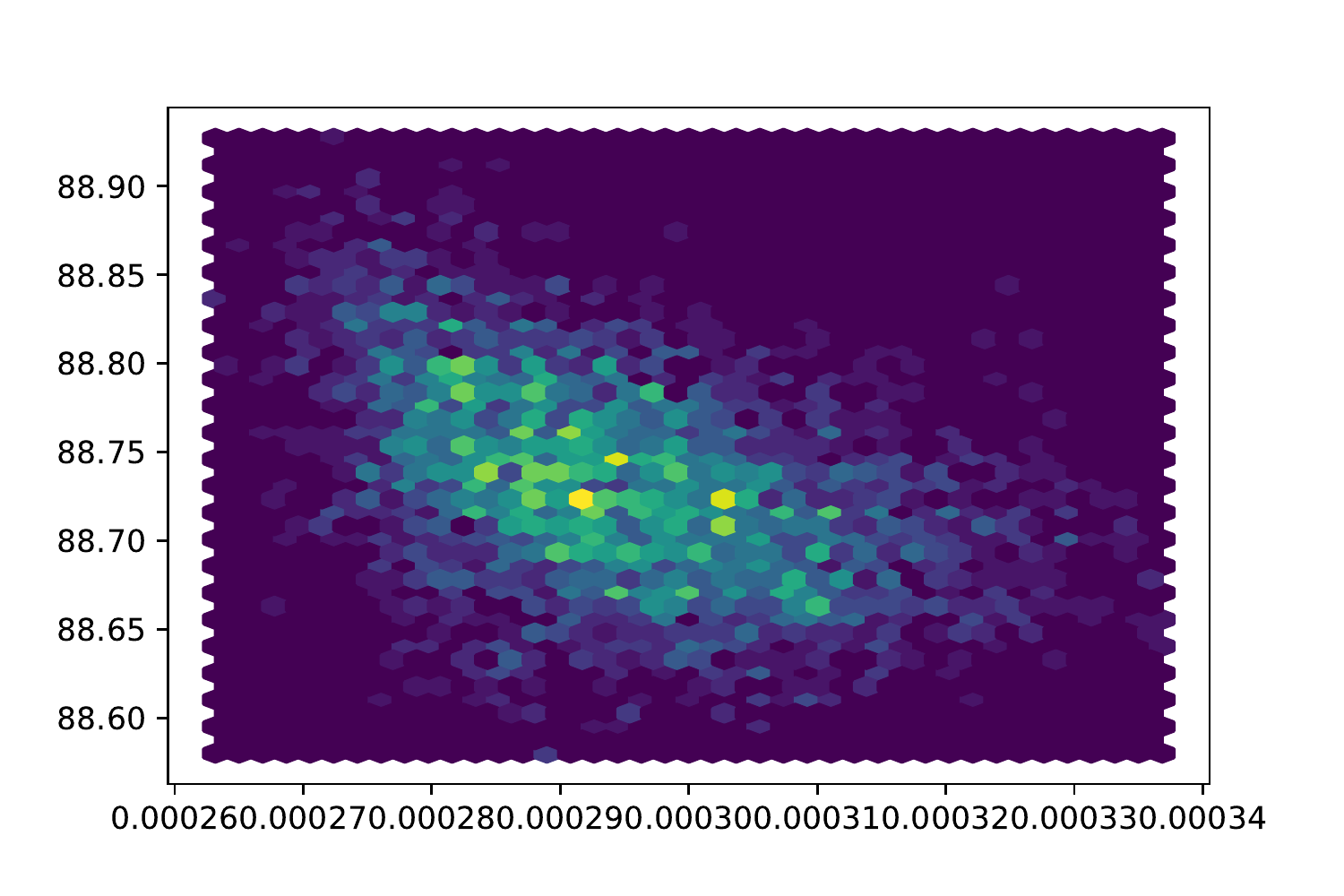} & & & & & & & & & \\
    d ecc & \includegraphics[scale=0.11]{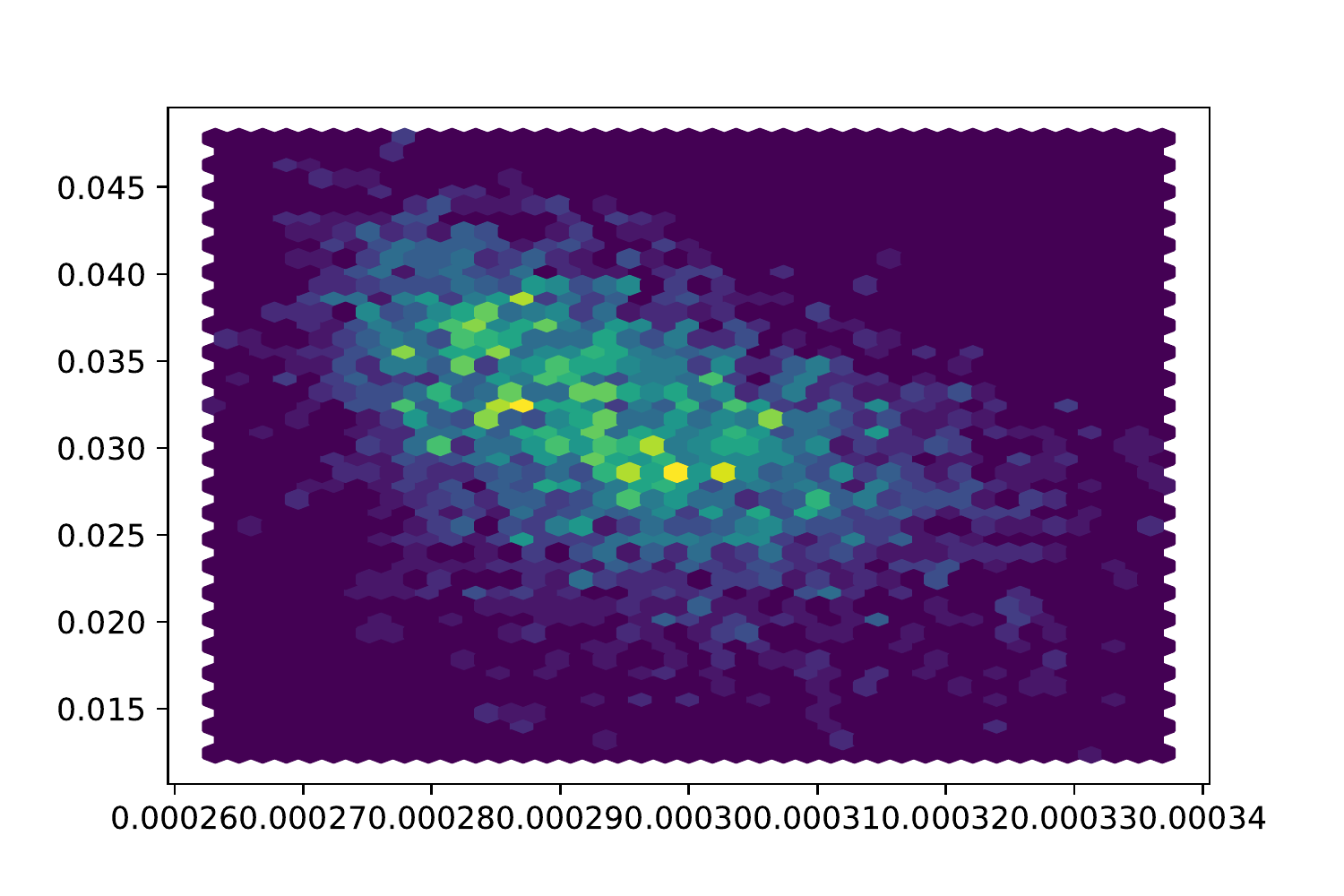} & \includegraphics[scale=0.11]{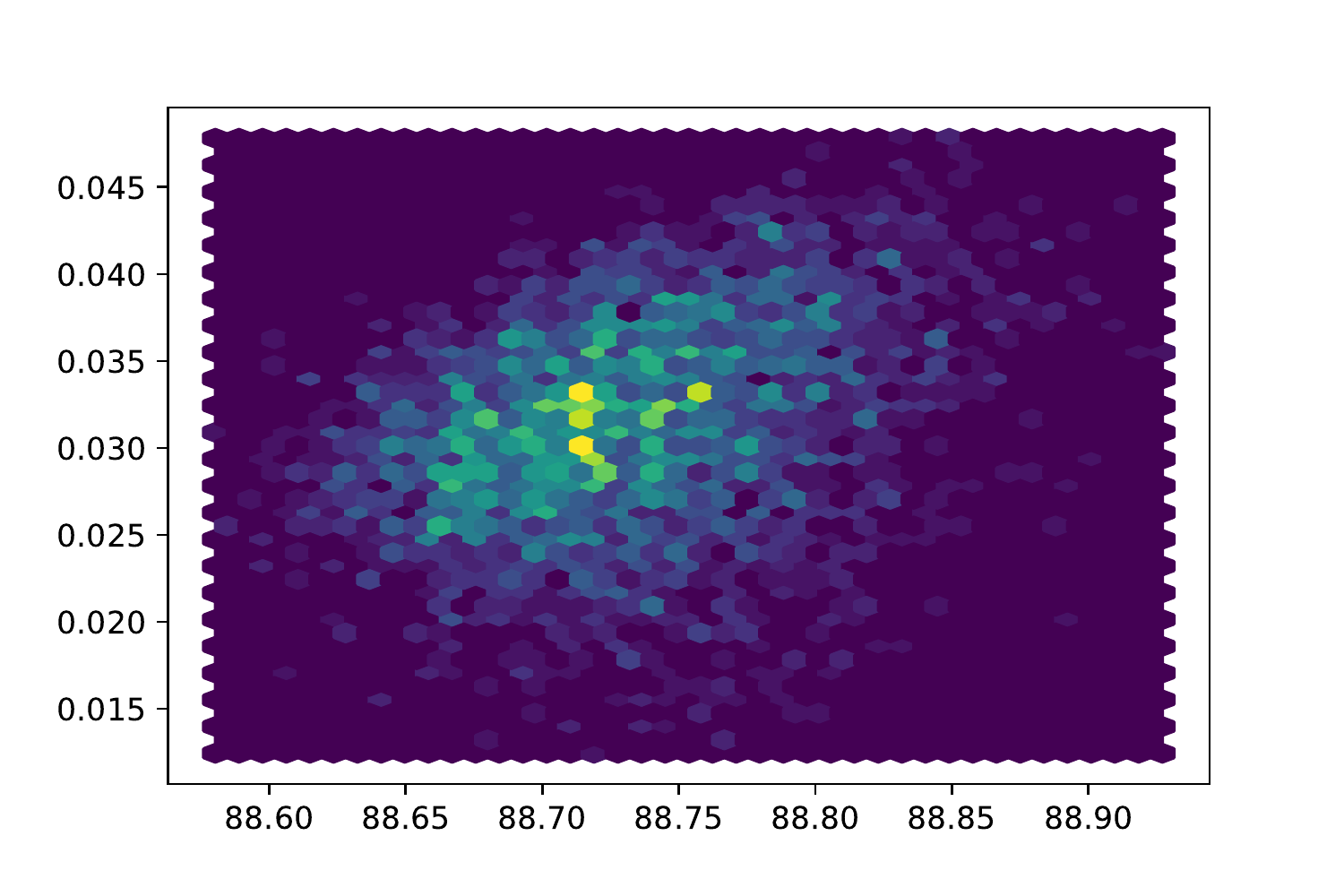} & & & & & & & & \\
    d inc &\includegraphics[scale=0.11]{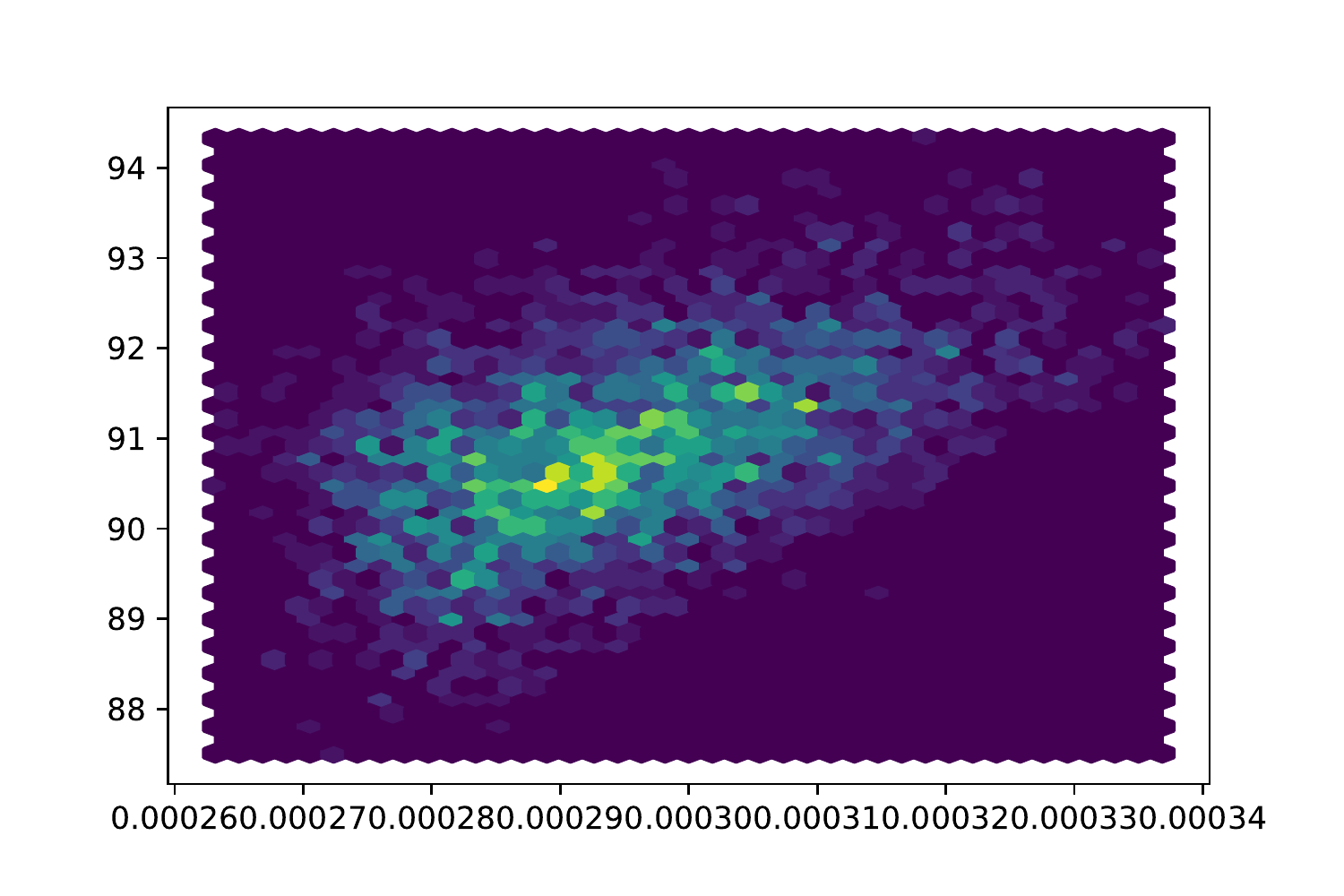} & \includegraphics[scale=0.11]{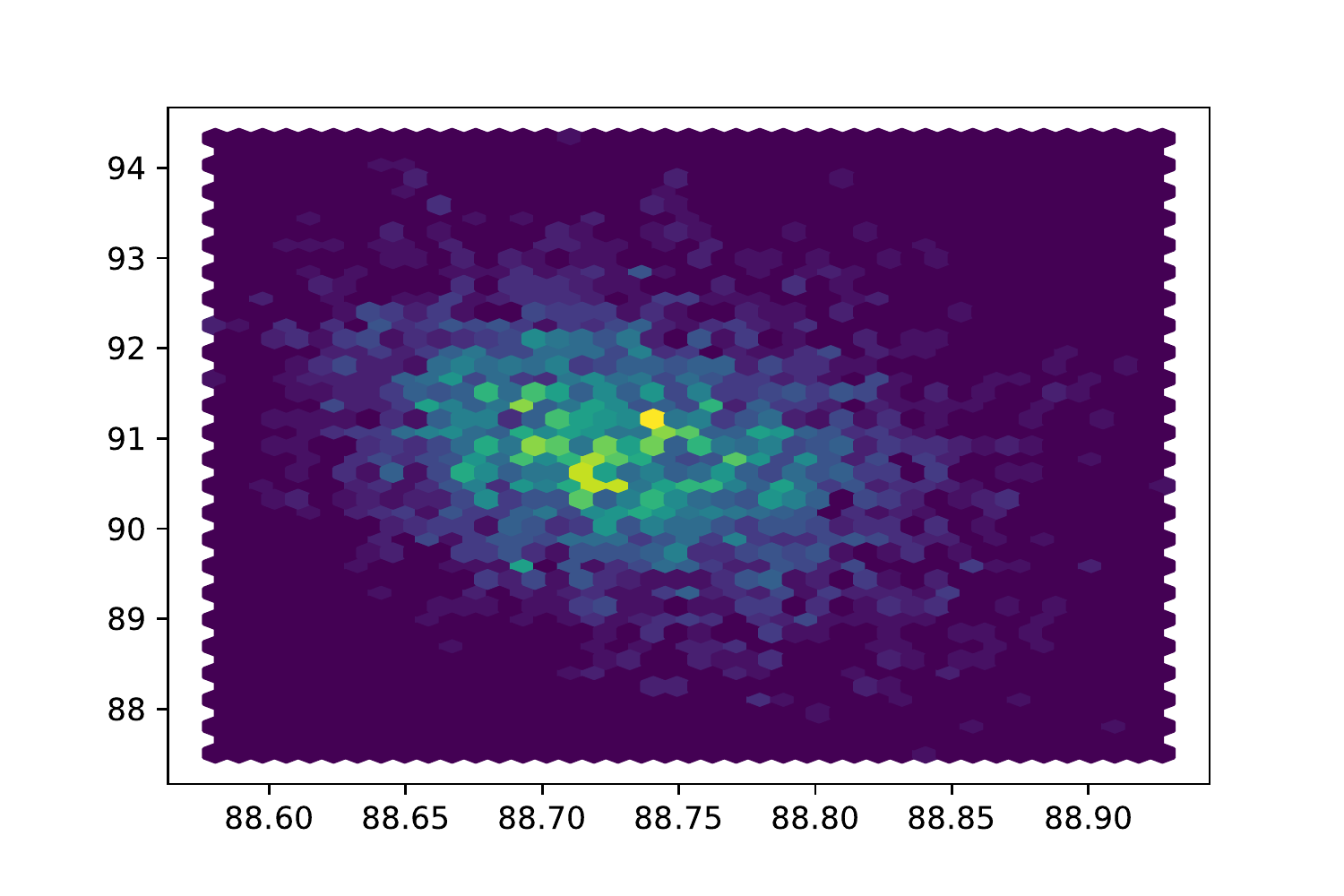} & \includegraphics[scale=0.11]{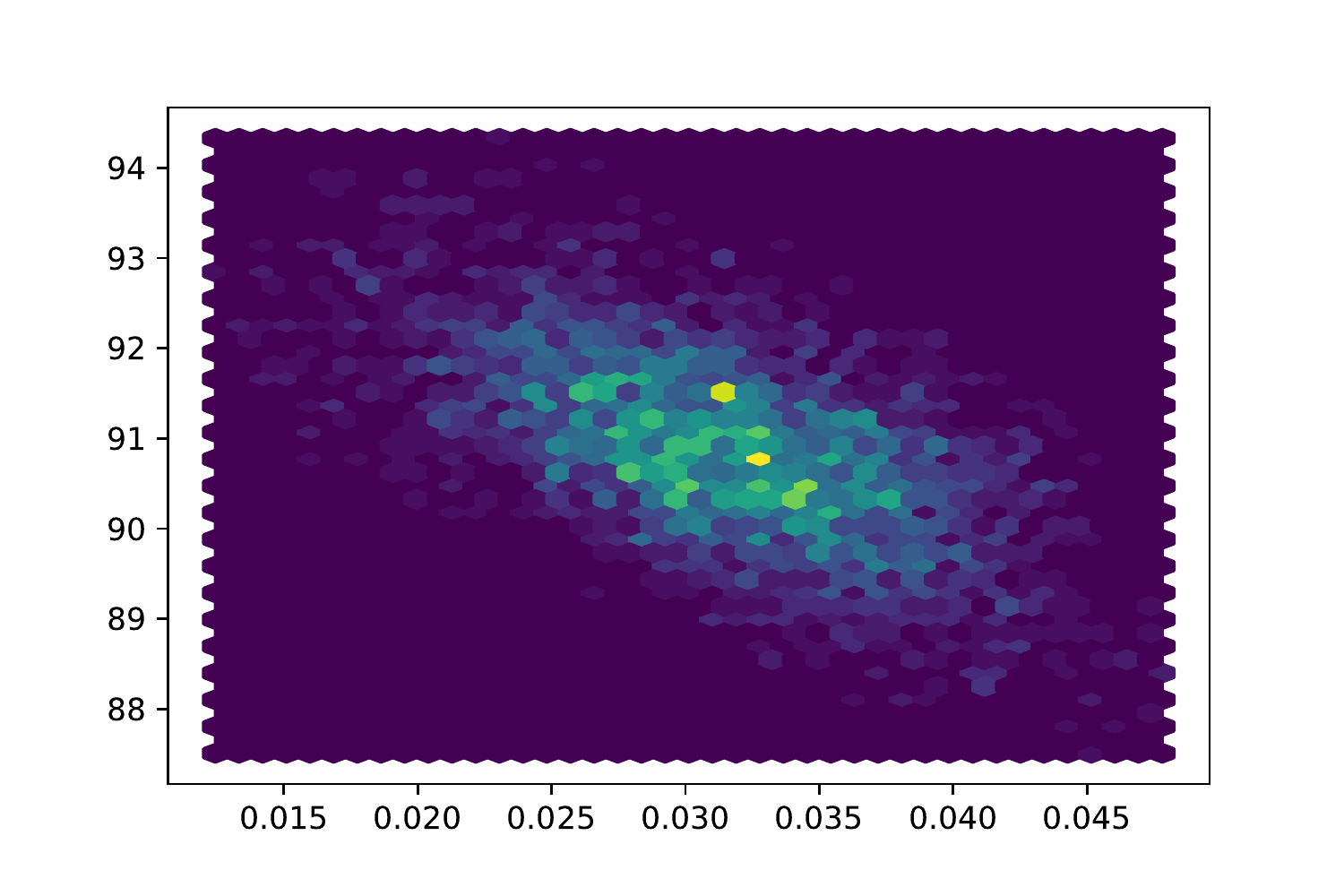} & & & & & & & \\
    d asc node & \includegraphics[scale=0.11]{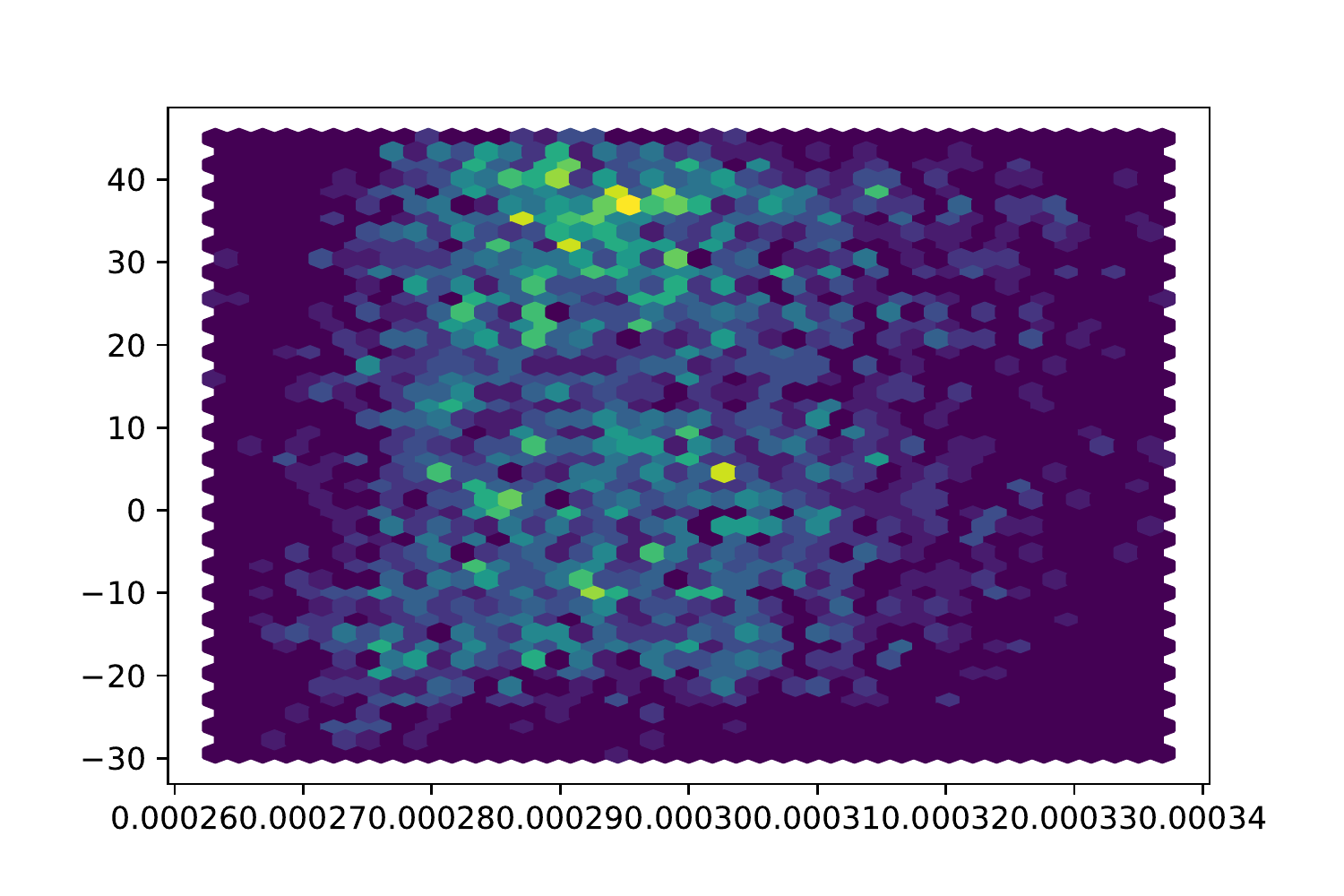} & \includegraphics[scale=0.11]{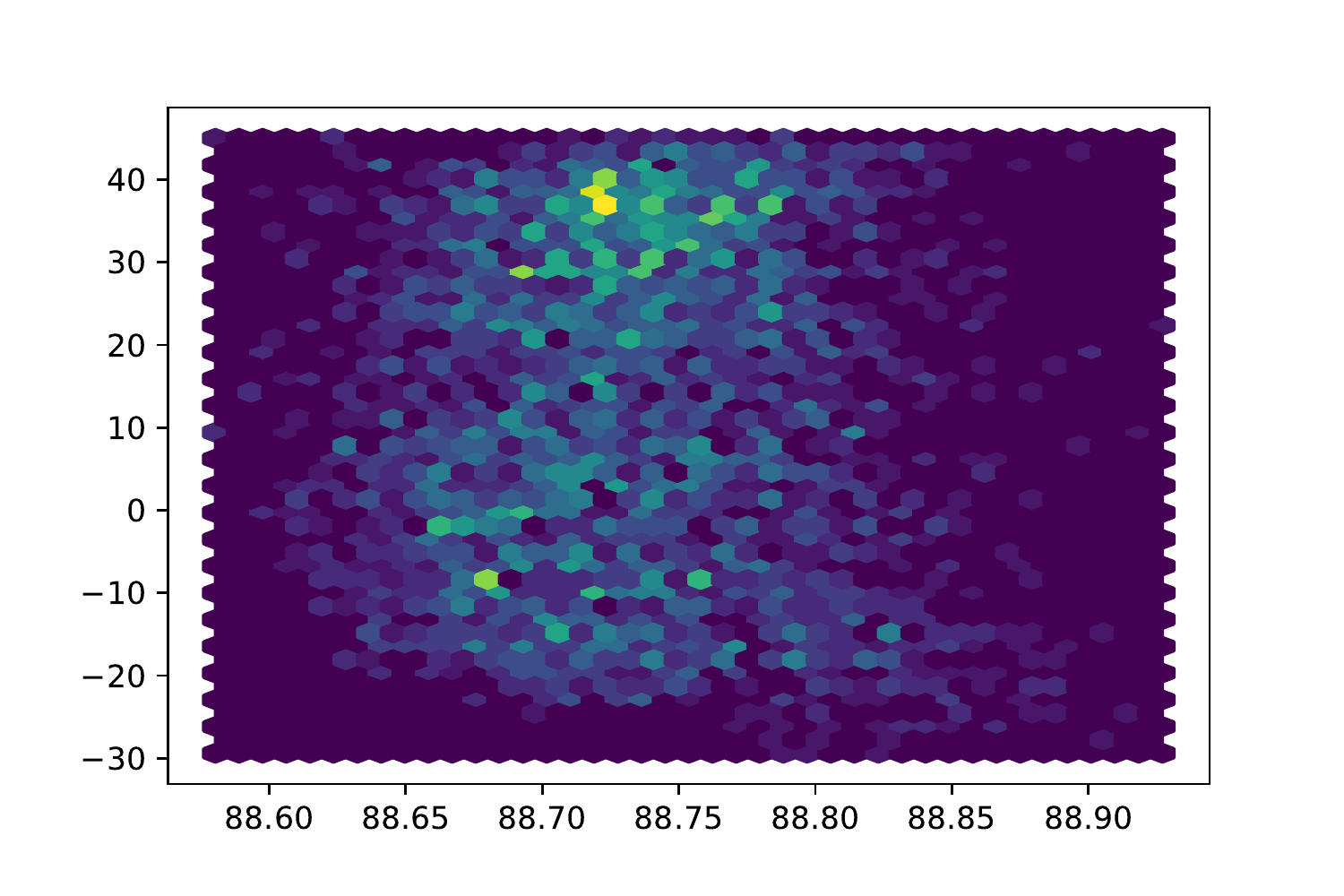} & \includegraphics[scale=0.11]{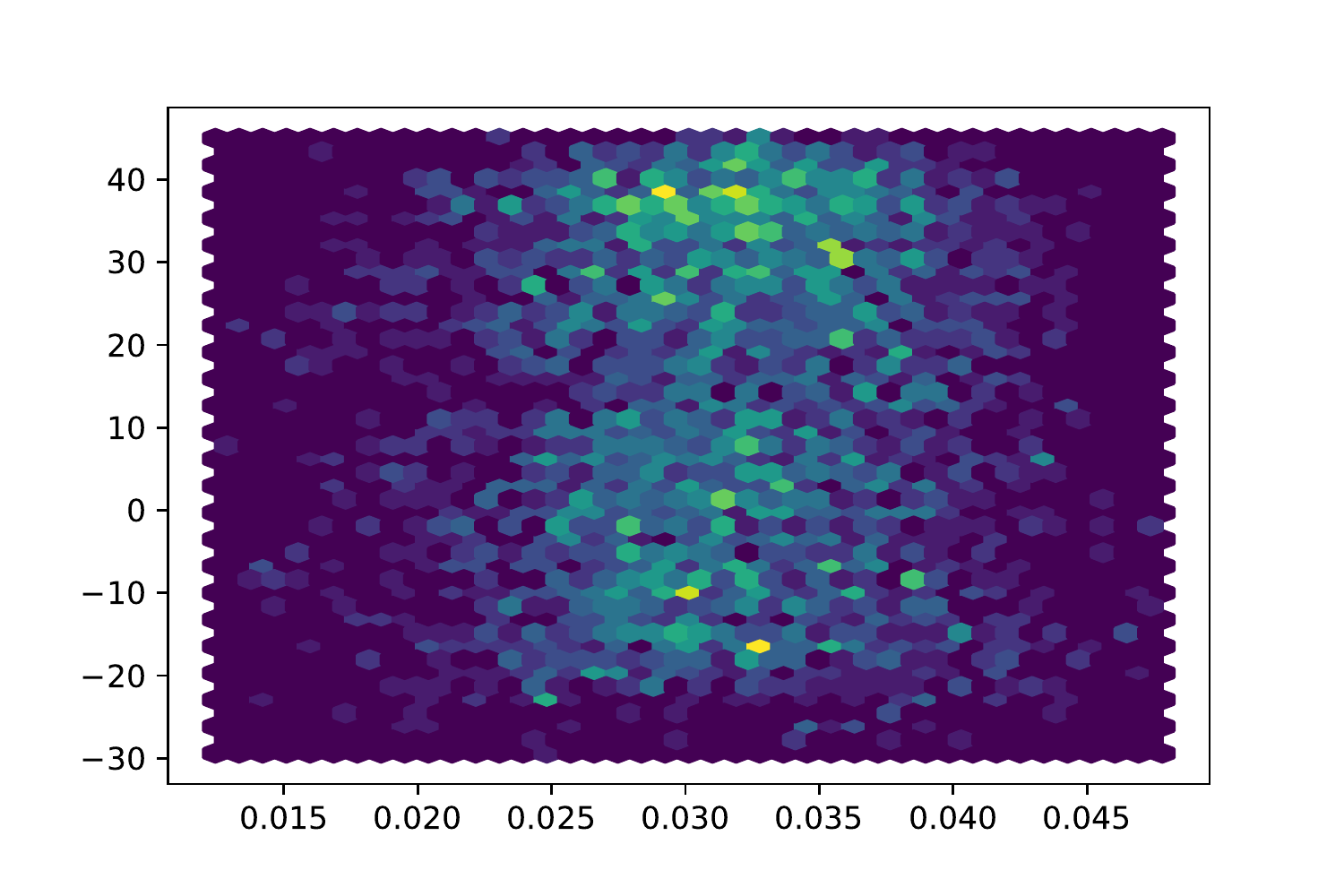} & \includegraphics[scale=0.11]{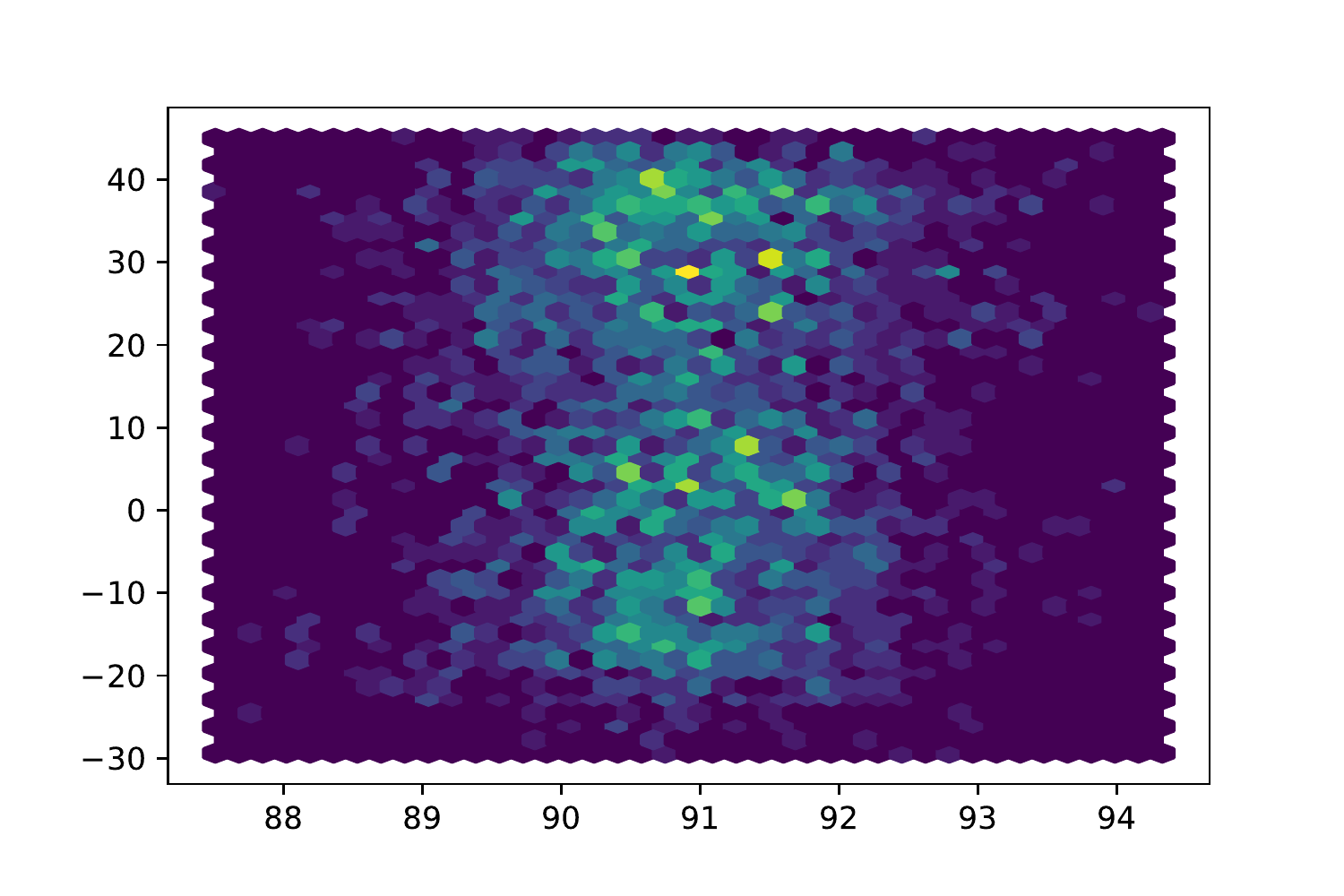} & & & & & & \\
    d peri & \includegraphics[scale=0.11]{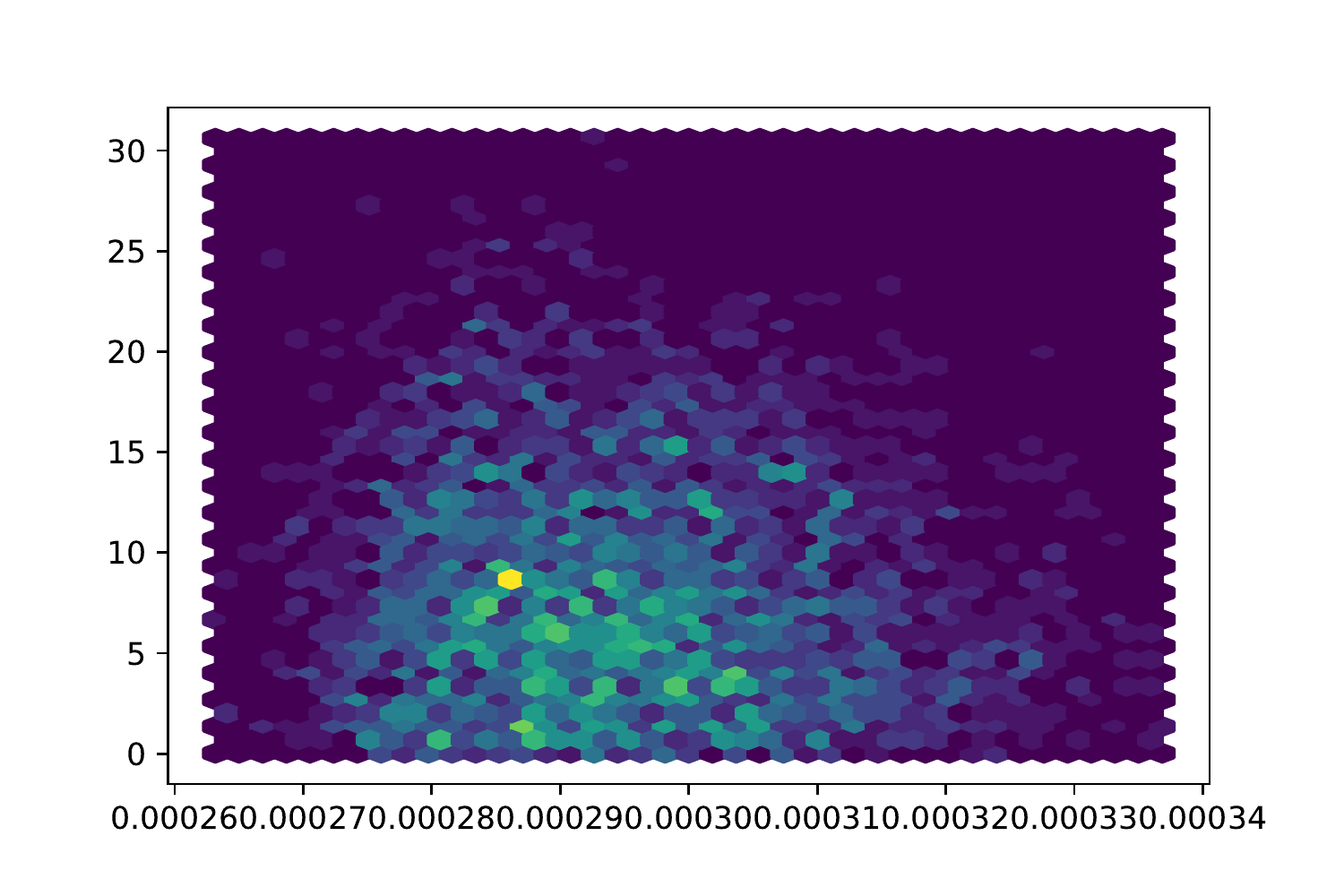} & \includegraphics[scale=0.11]{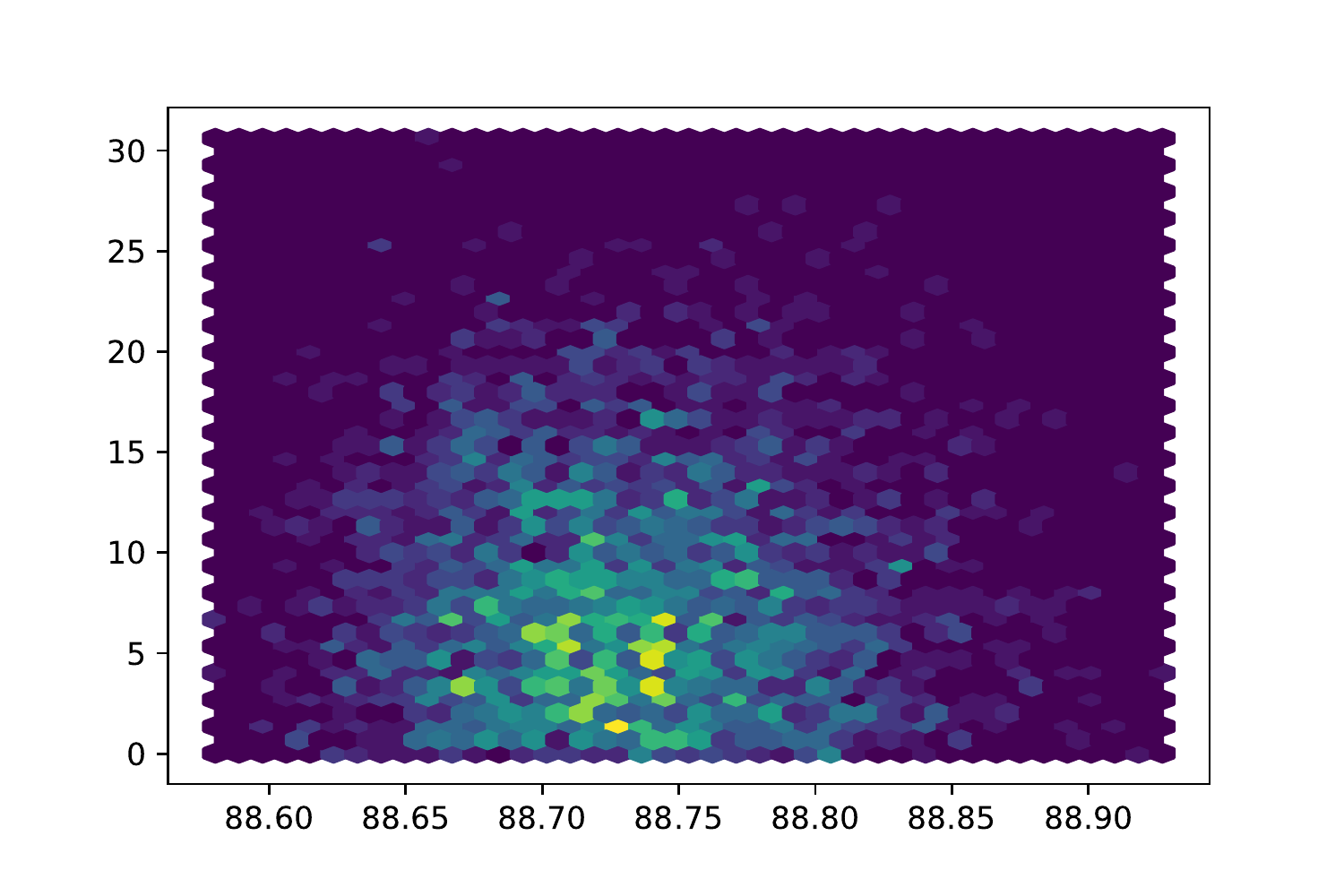} & \includegraphics[scale=0.11]{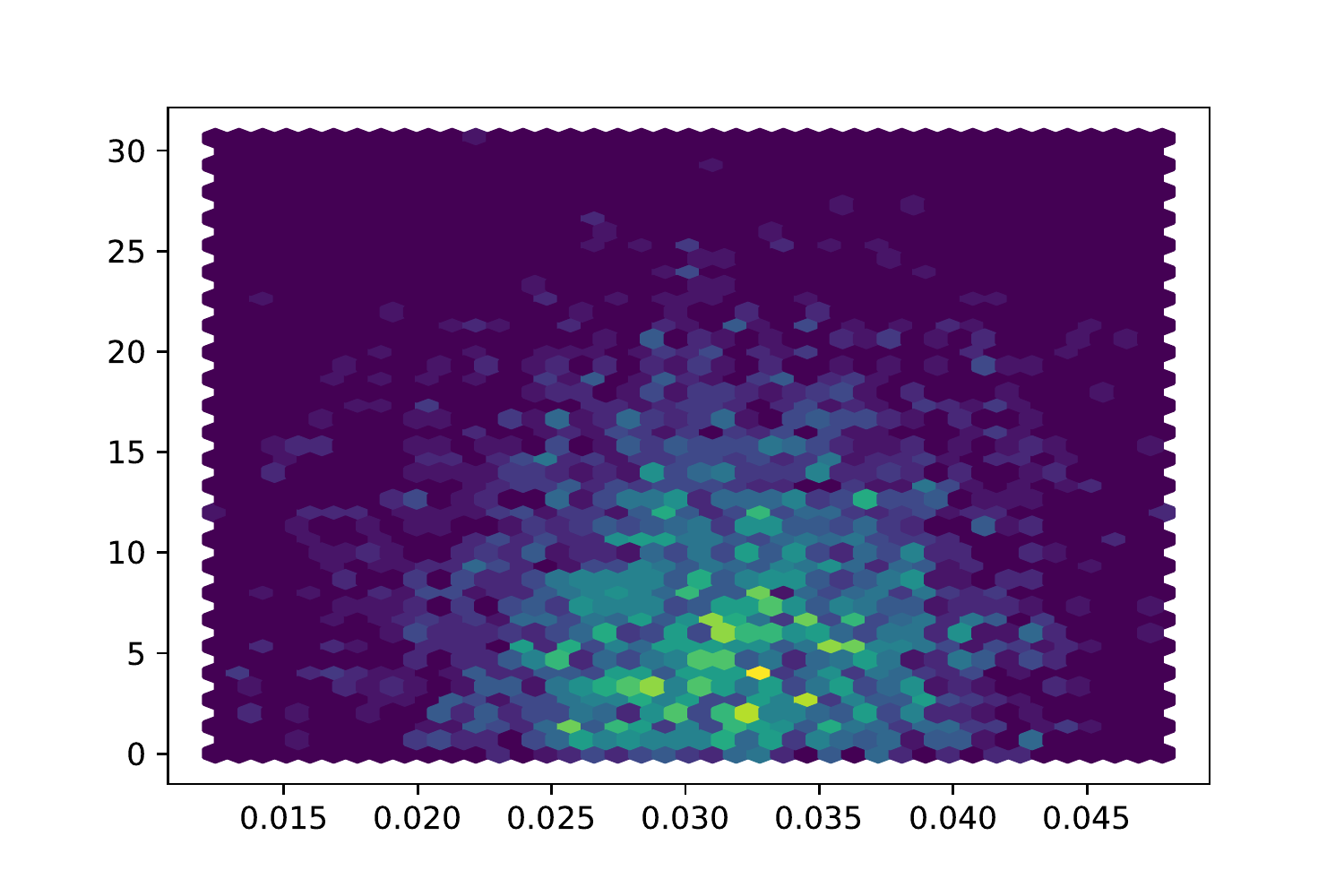} & \includegraphics[scale=0.11]{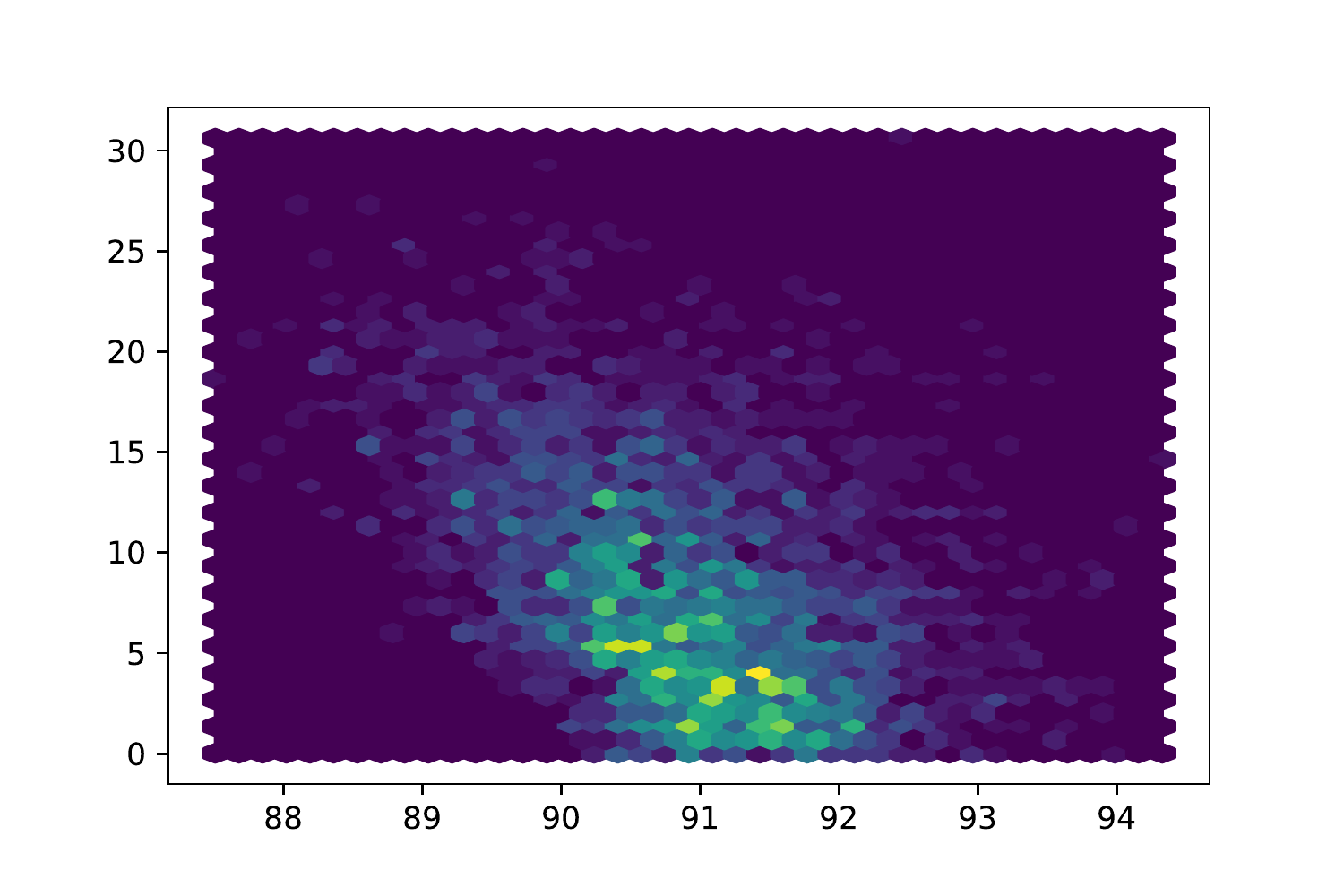} & \includegraphics[scale=0.11]{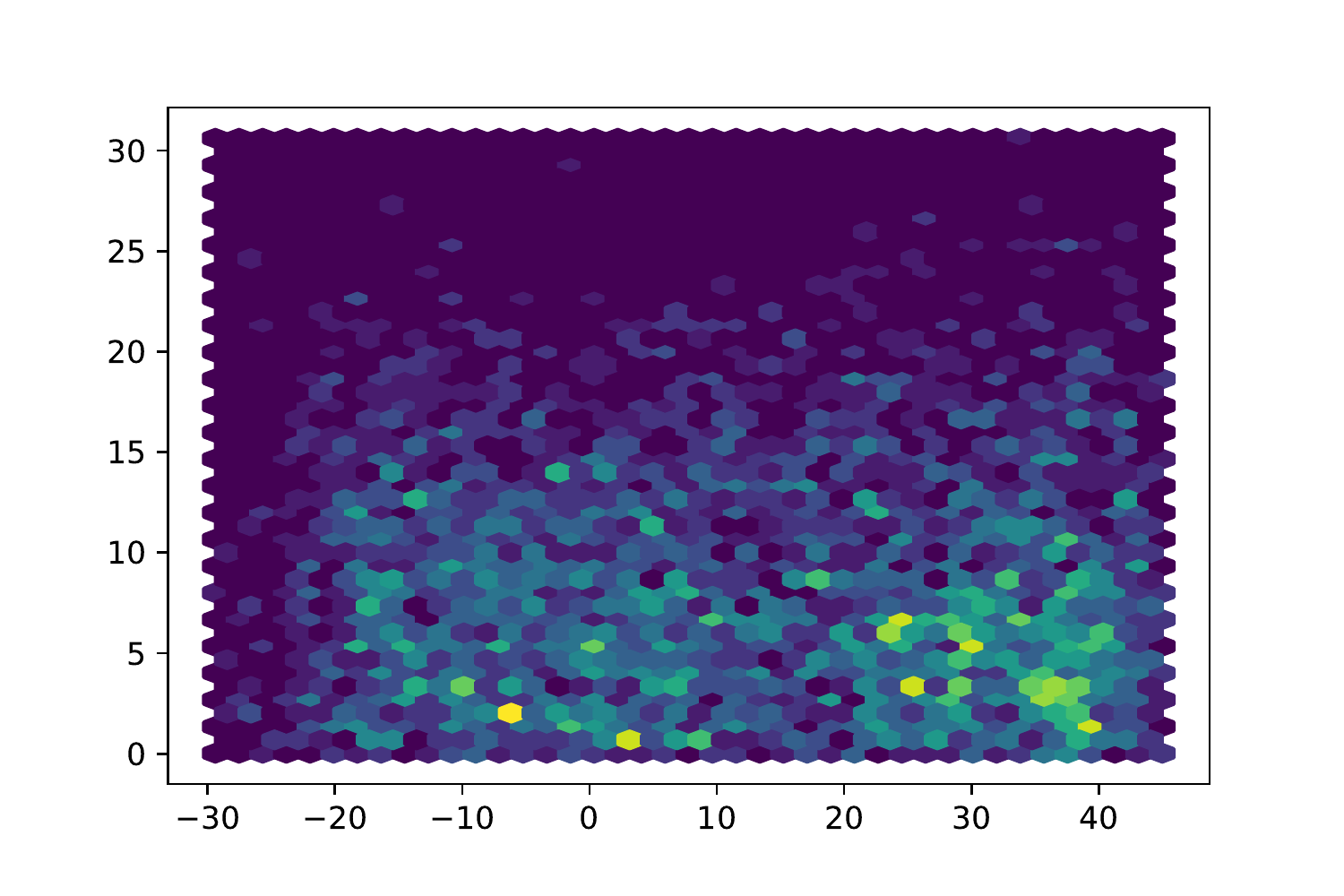} & & & & & \\
    d mean anom & \includegraphics[scale=0.11]{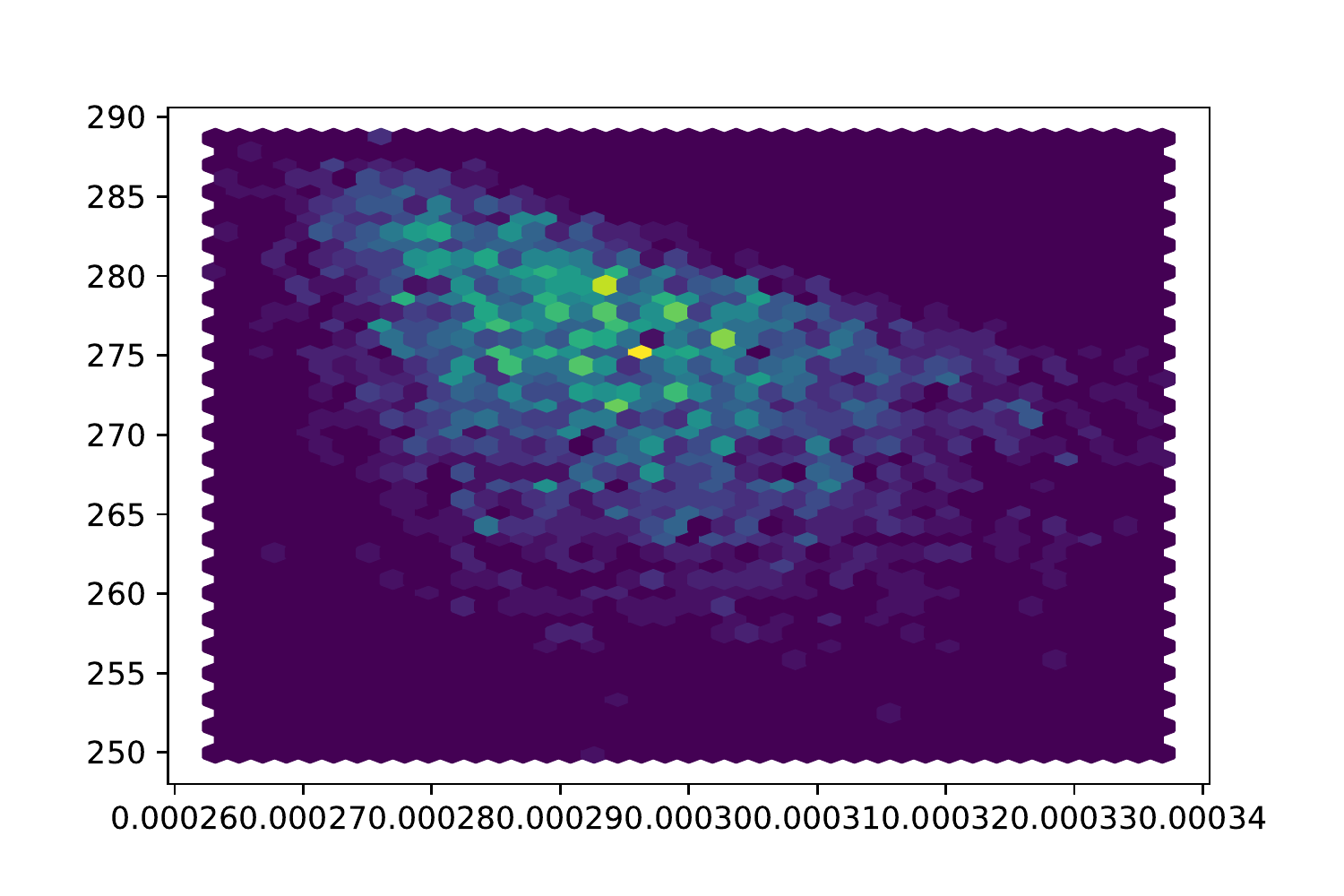} & \includegraphics[scale=0.11]{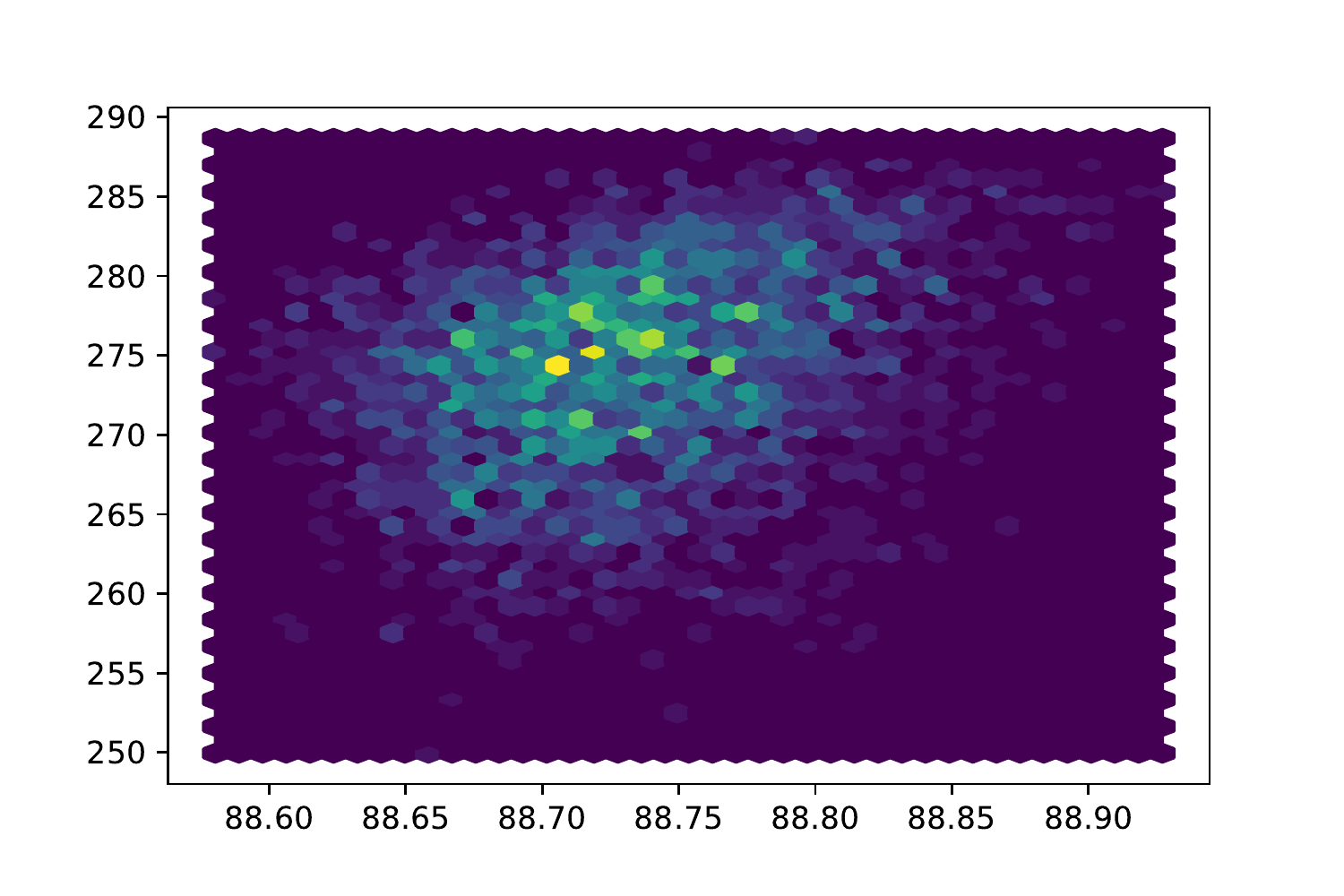} & \includegraphics[scale=0.11]{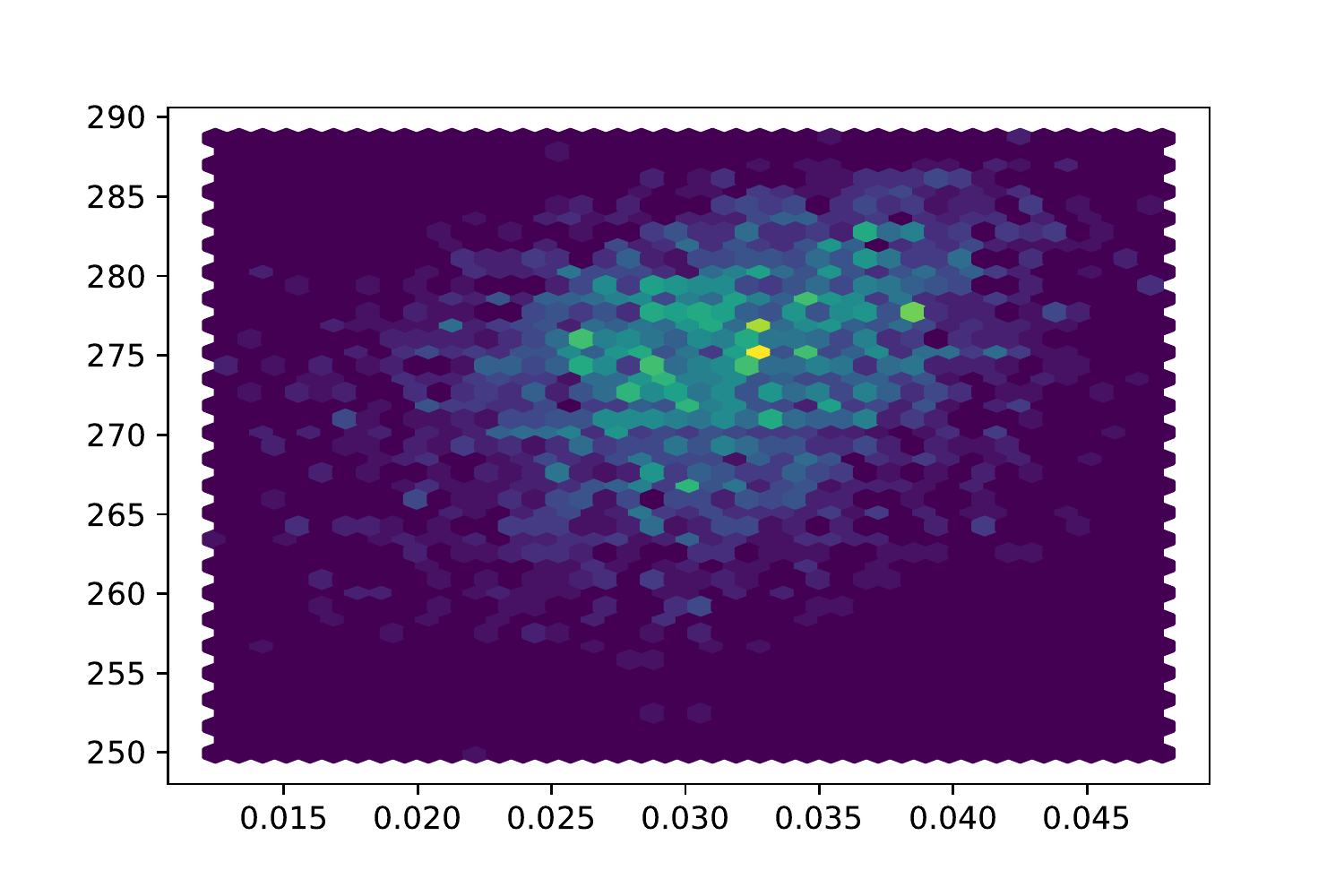} & \includegraphics[scale=0.11]{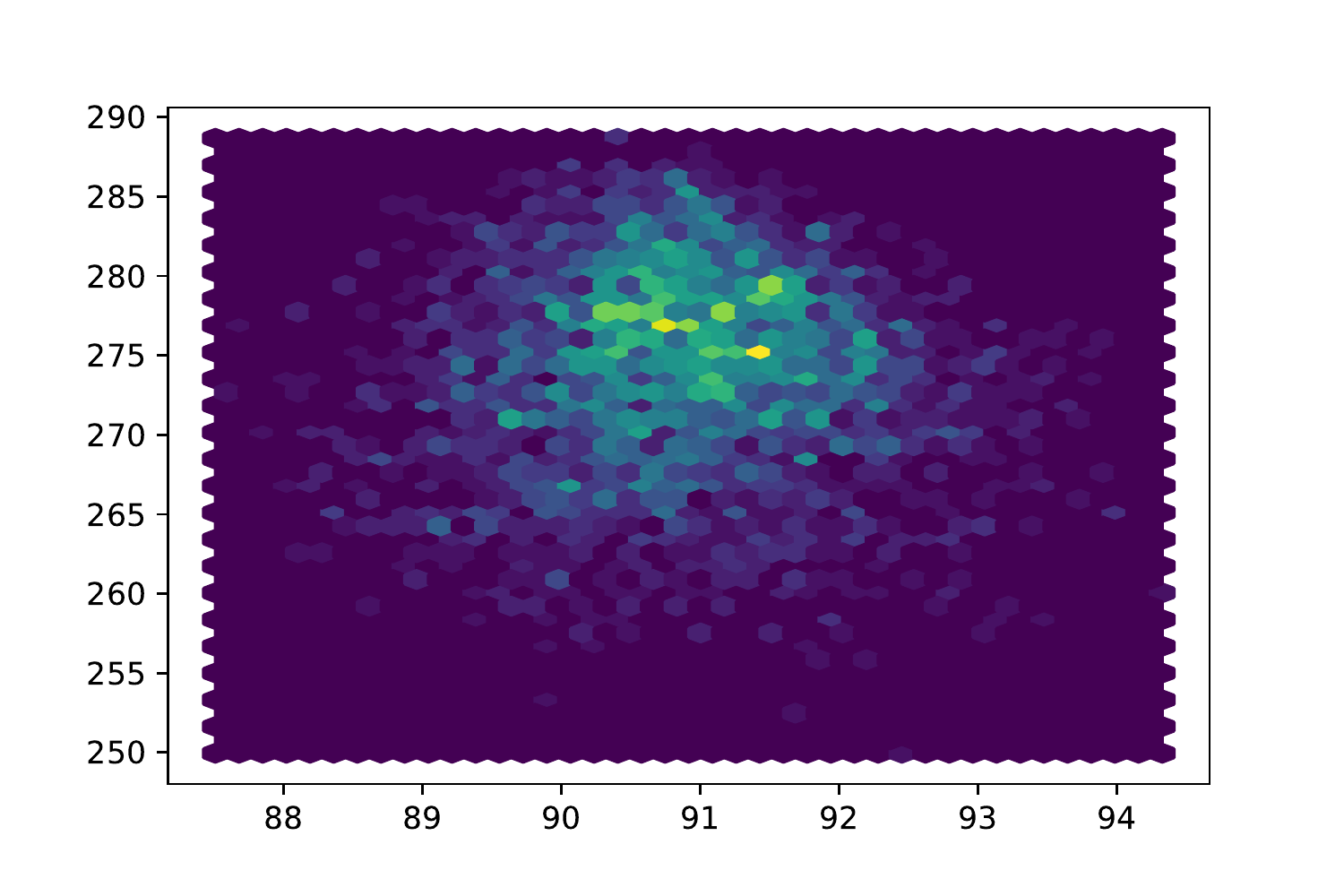} & \includegraphics[scale=0.11]{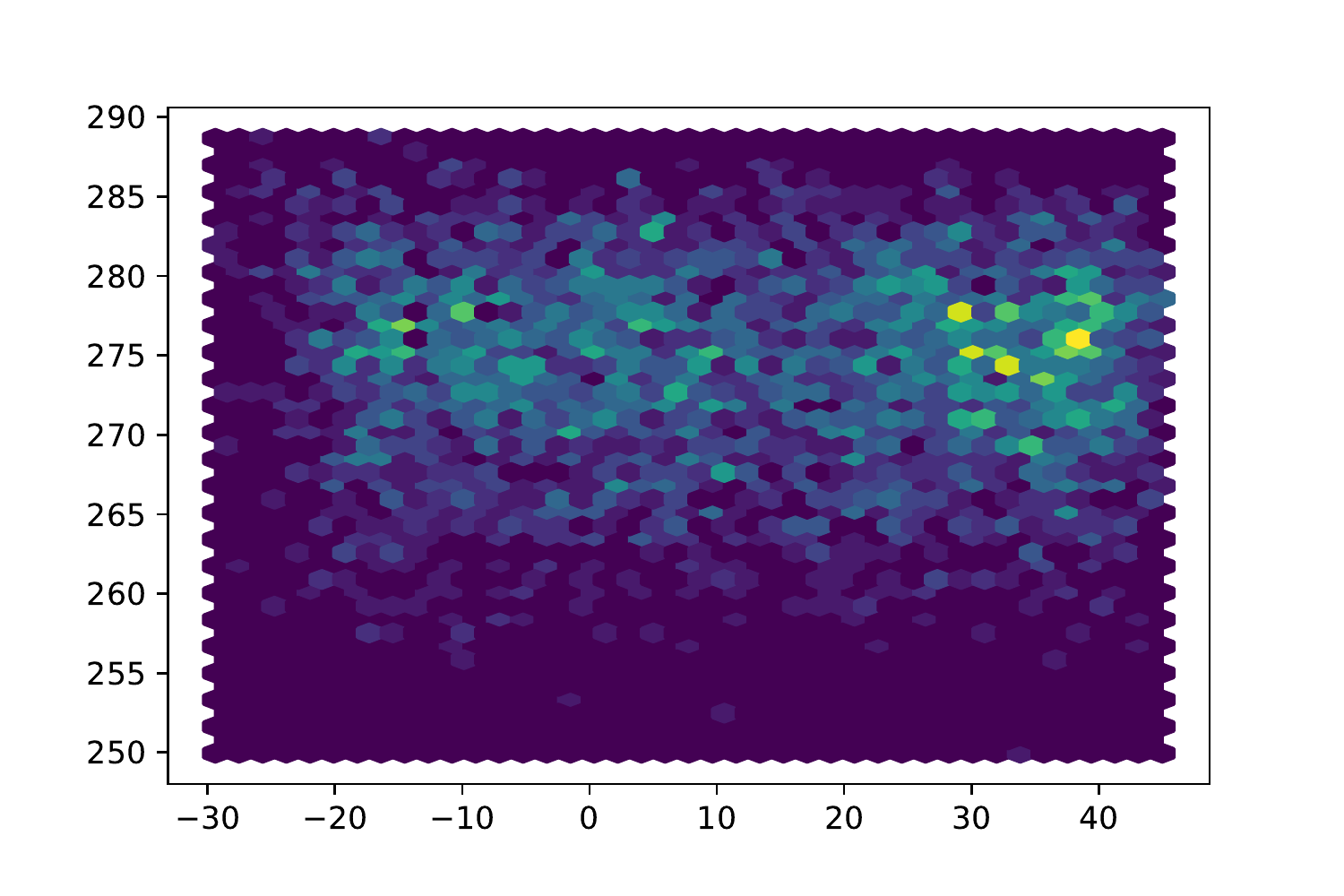} & \includegraphics[scale=0.11]{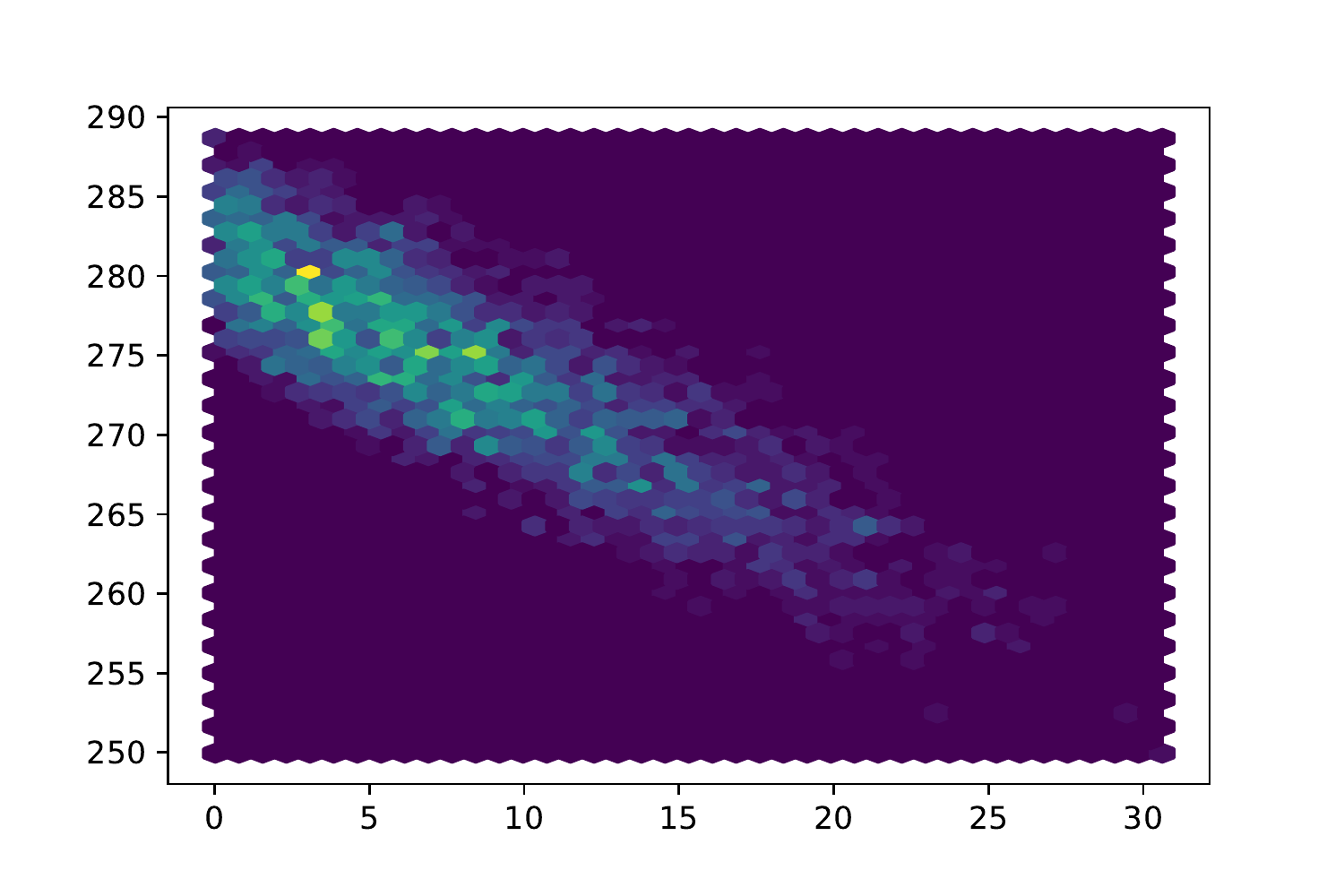} & & & & \\
    c mass & \includegraphics[scale=0.11]{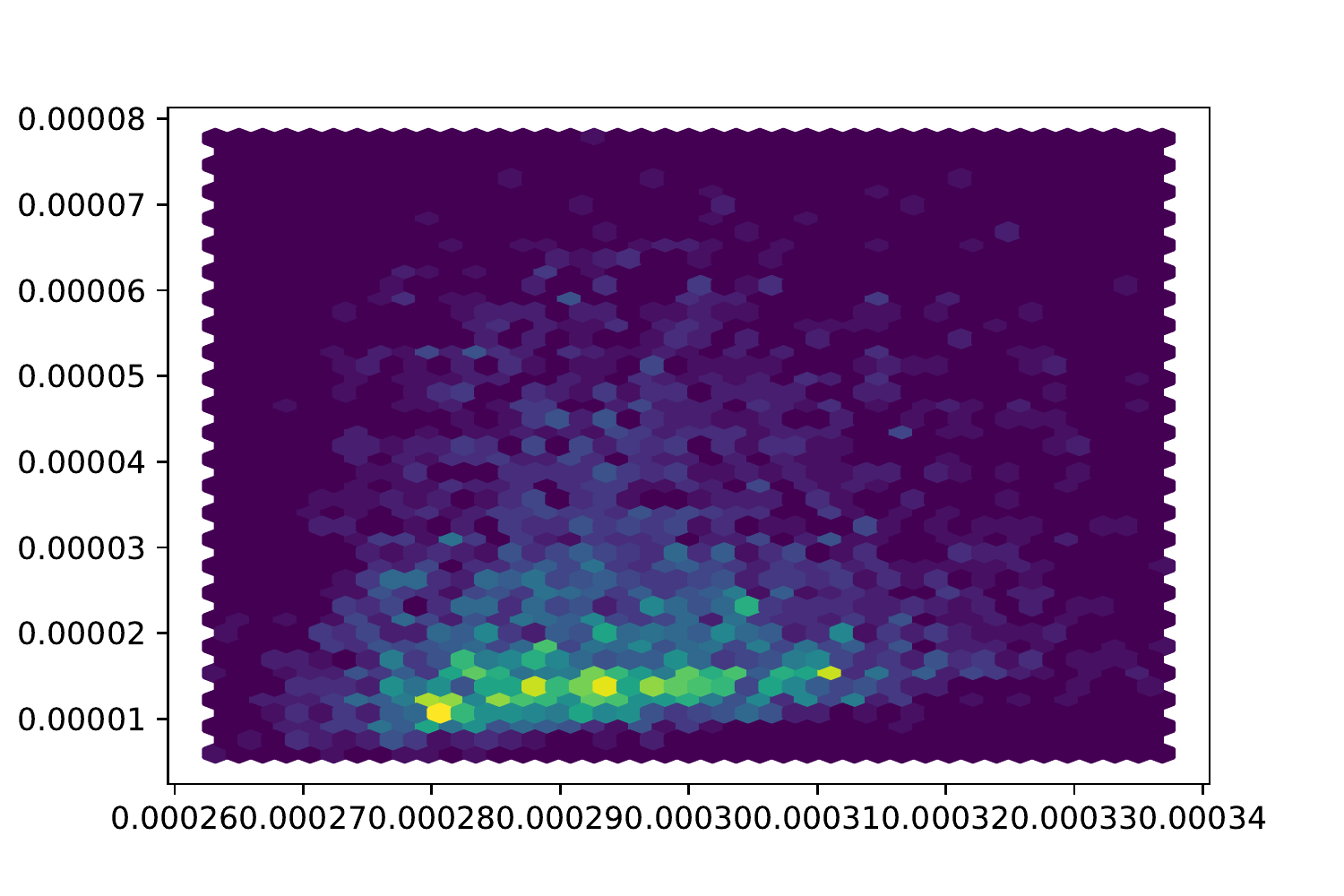} & \includegraphics[scale=0.11]{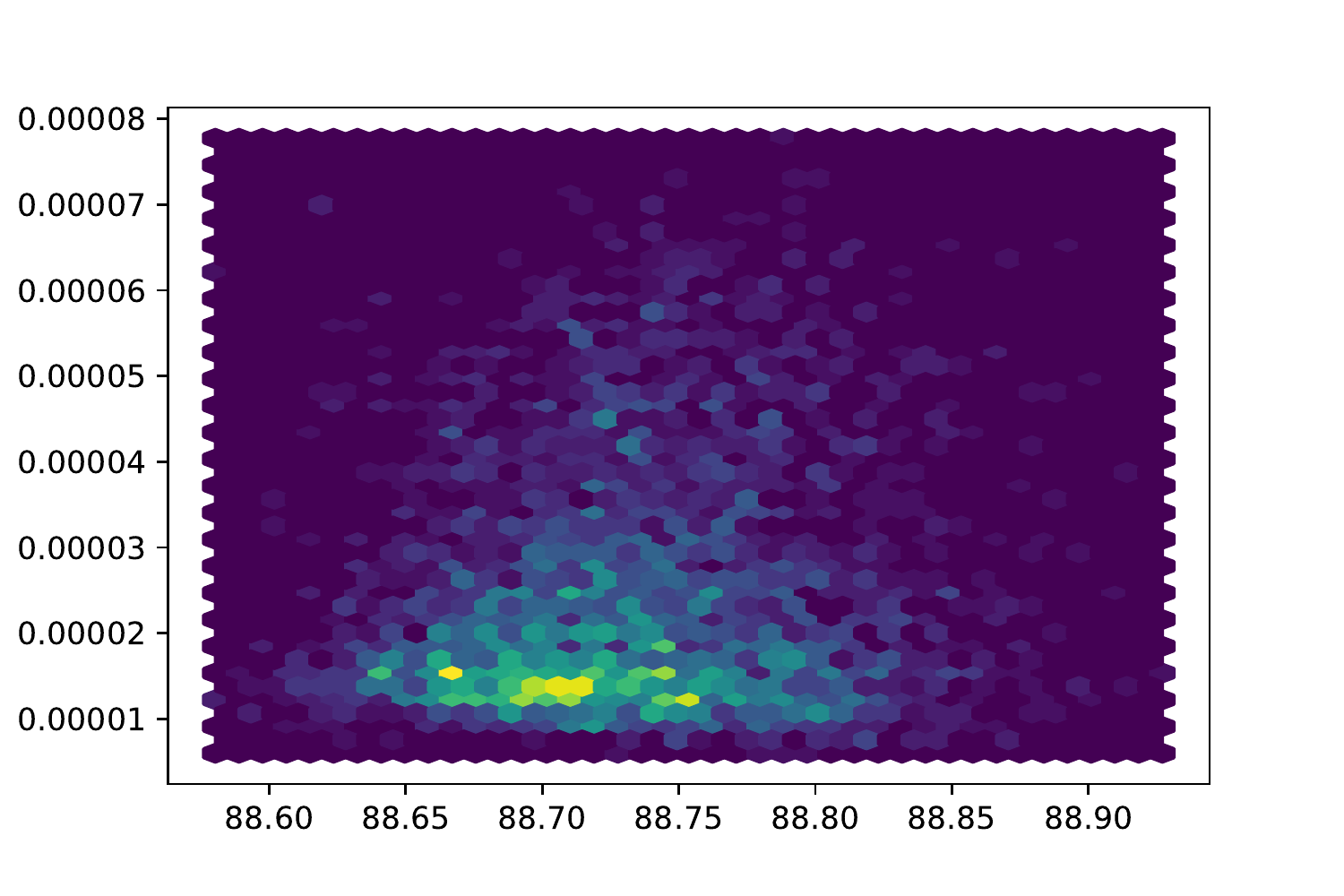} & \includegraphics[scale=0.11]{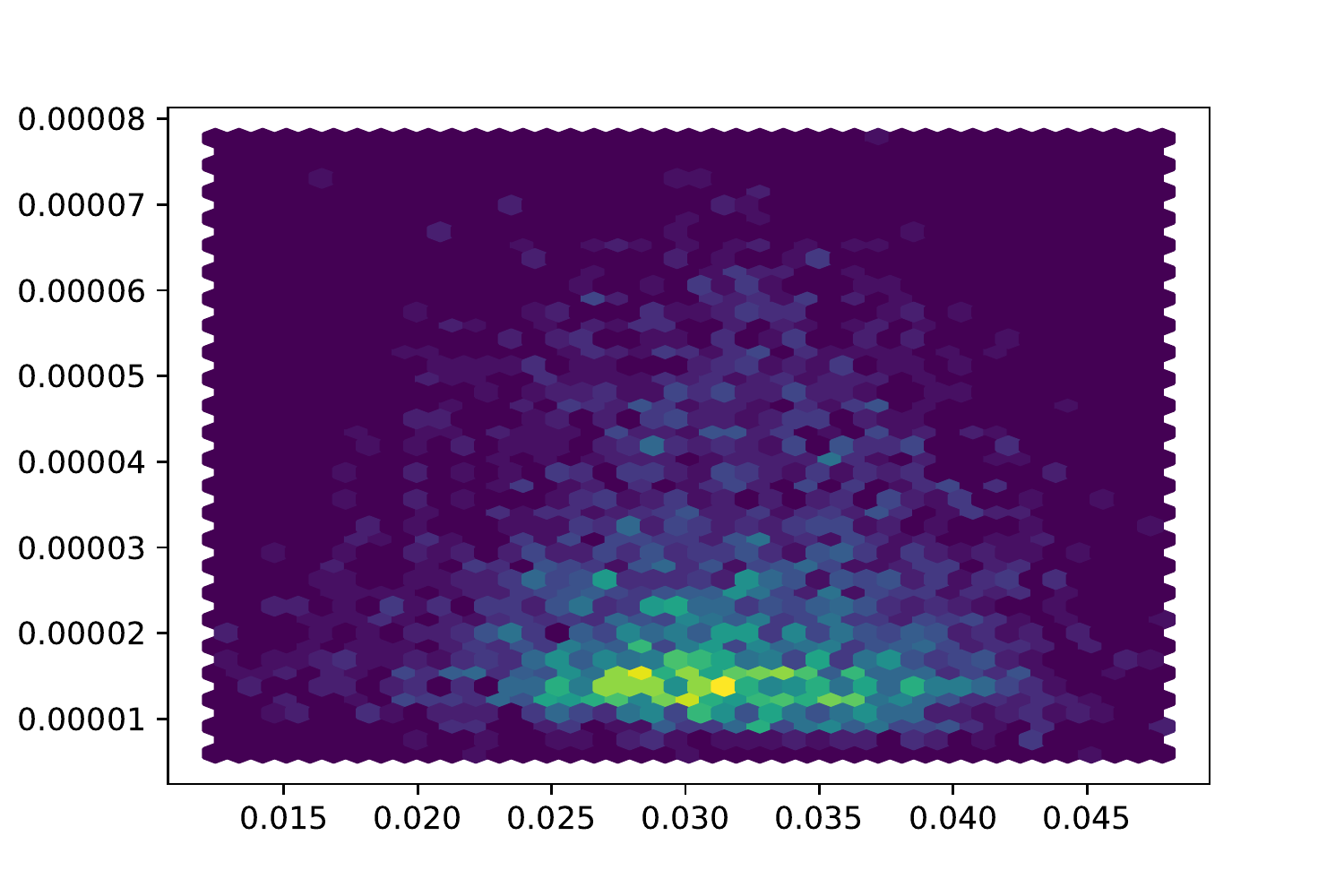} & \includegraphics[scale=0.11]{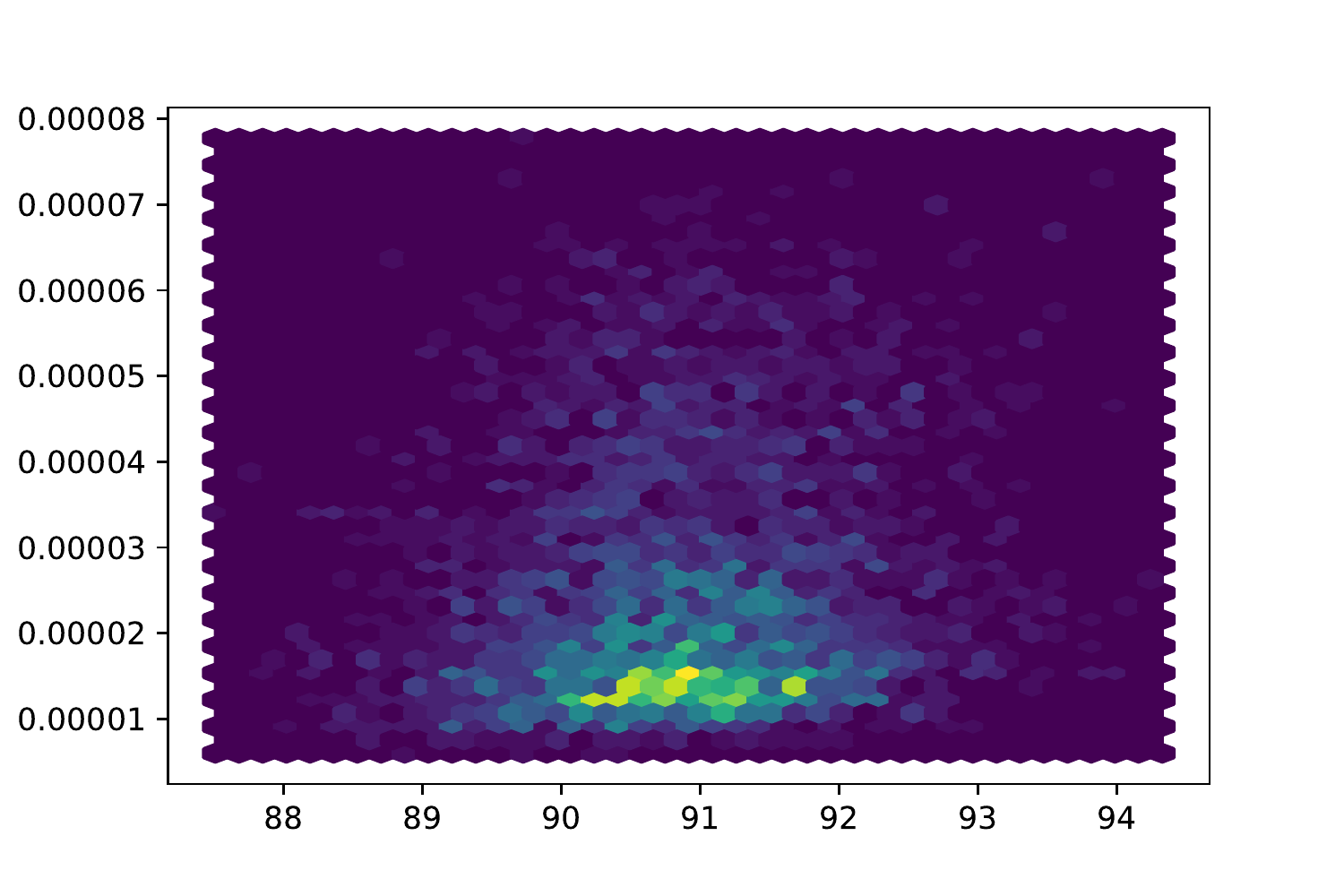} & \includegraphics[scale=0.11]{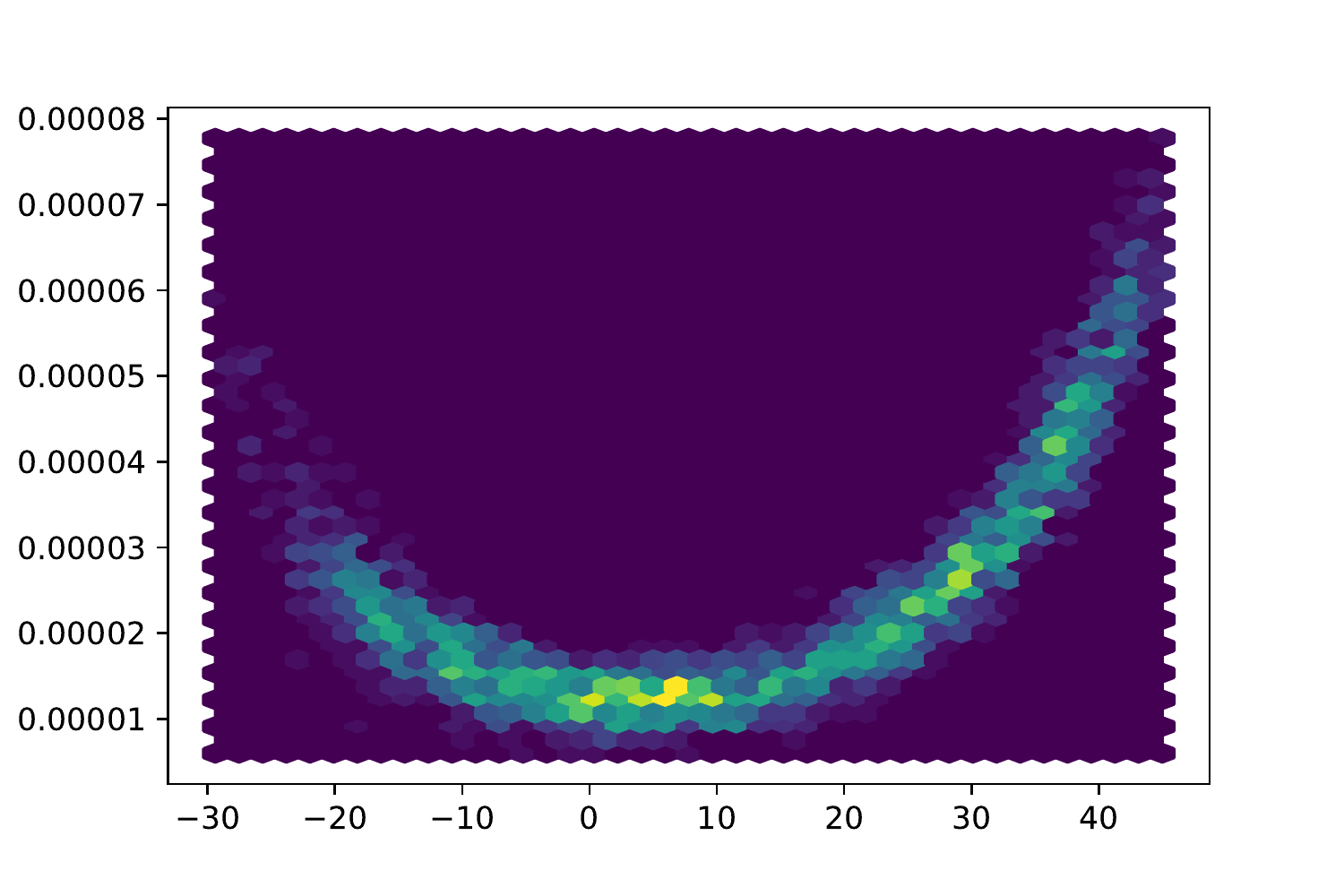} & \includegraphics[scale=0.11]{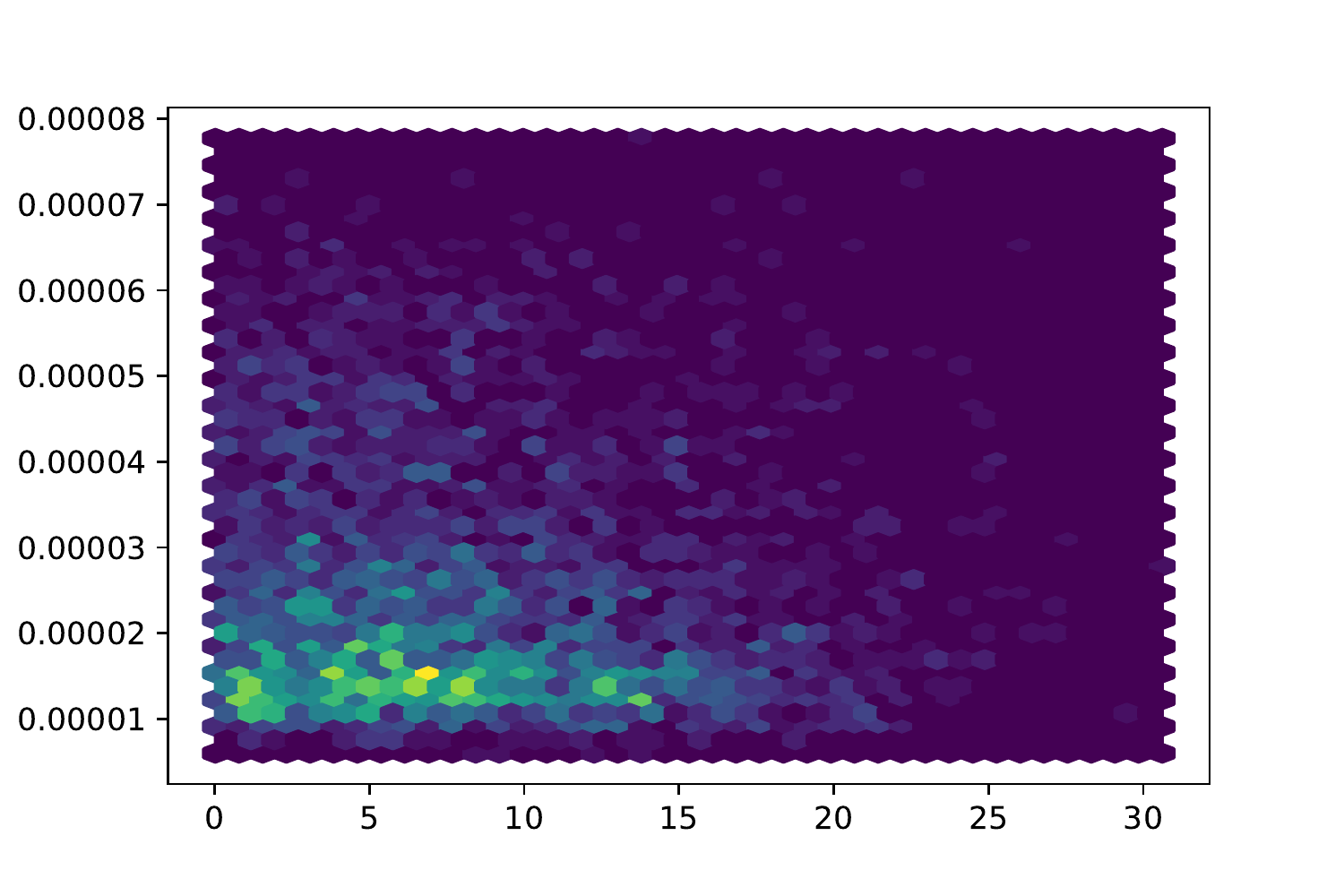} & \includegraphics[scale=0.11]{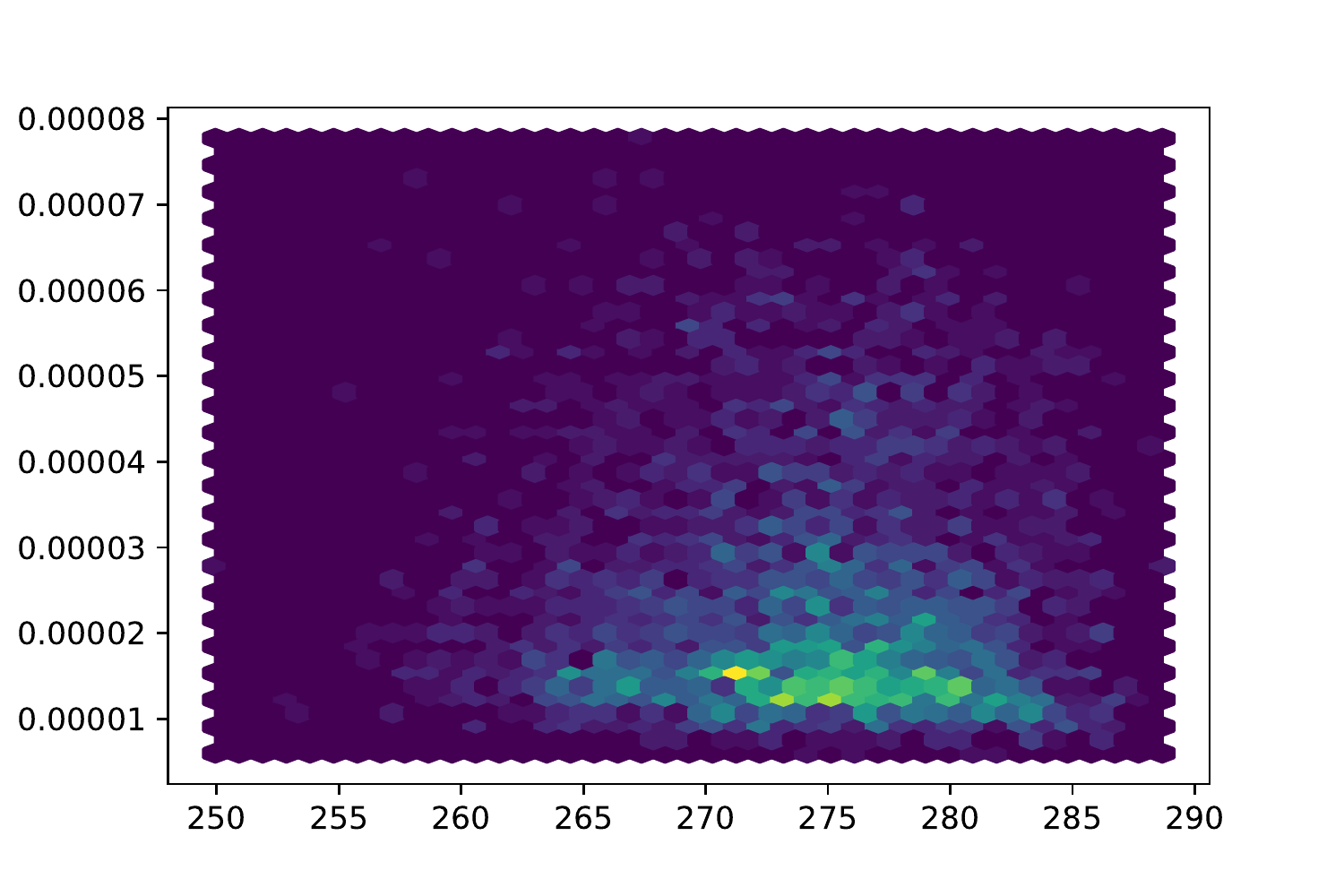} & & & \\
    c ecc & \includegraphics[scale=0.11]{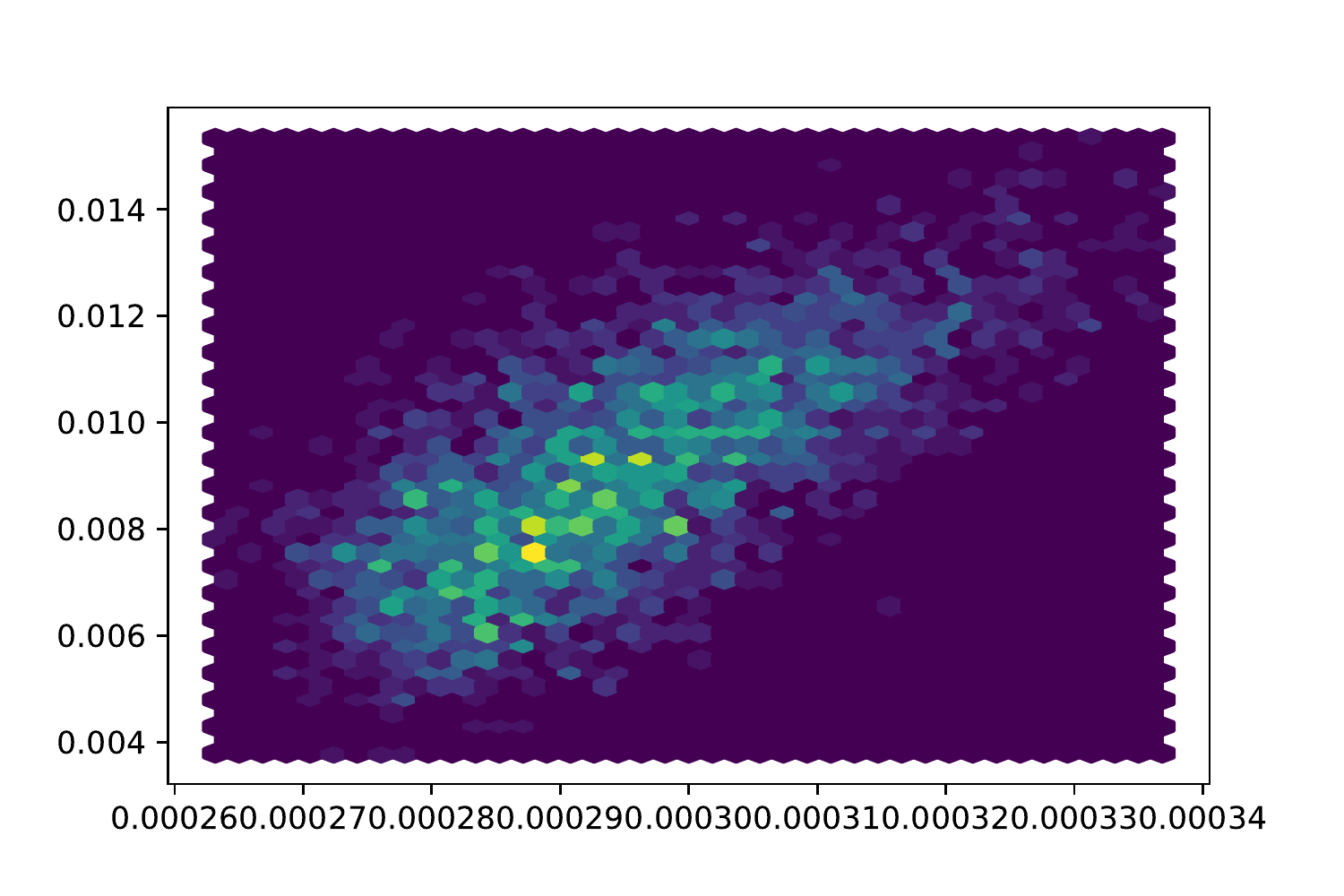} & \includegraphics[scale=0.11]{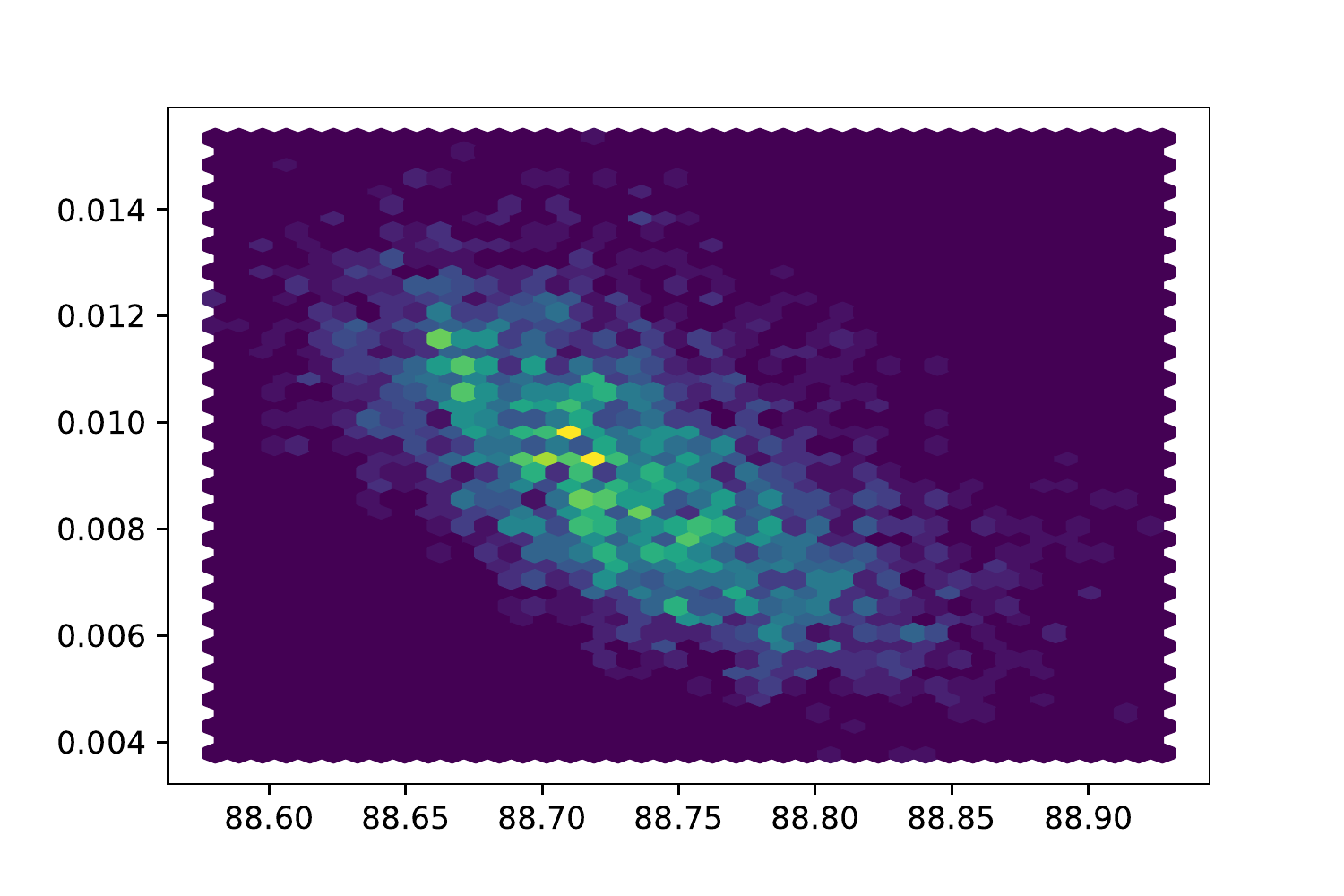} & \includegraphics[scale=0.11]{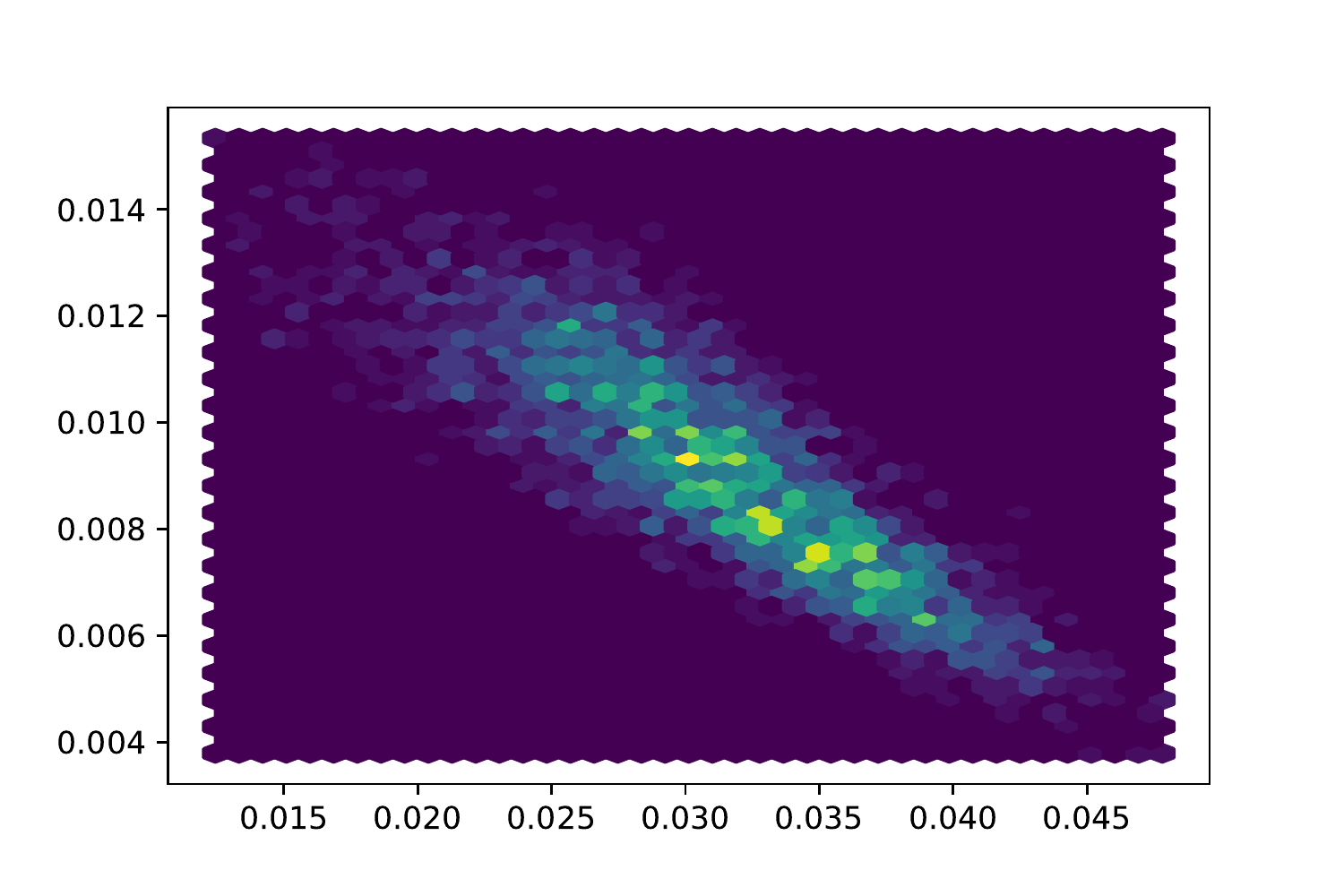} & \includegraphics[scale=0.11]{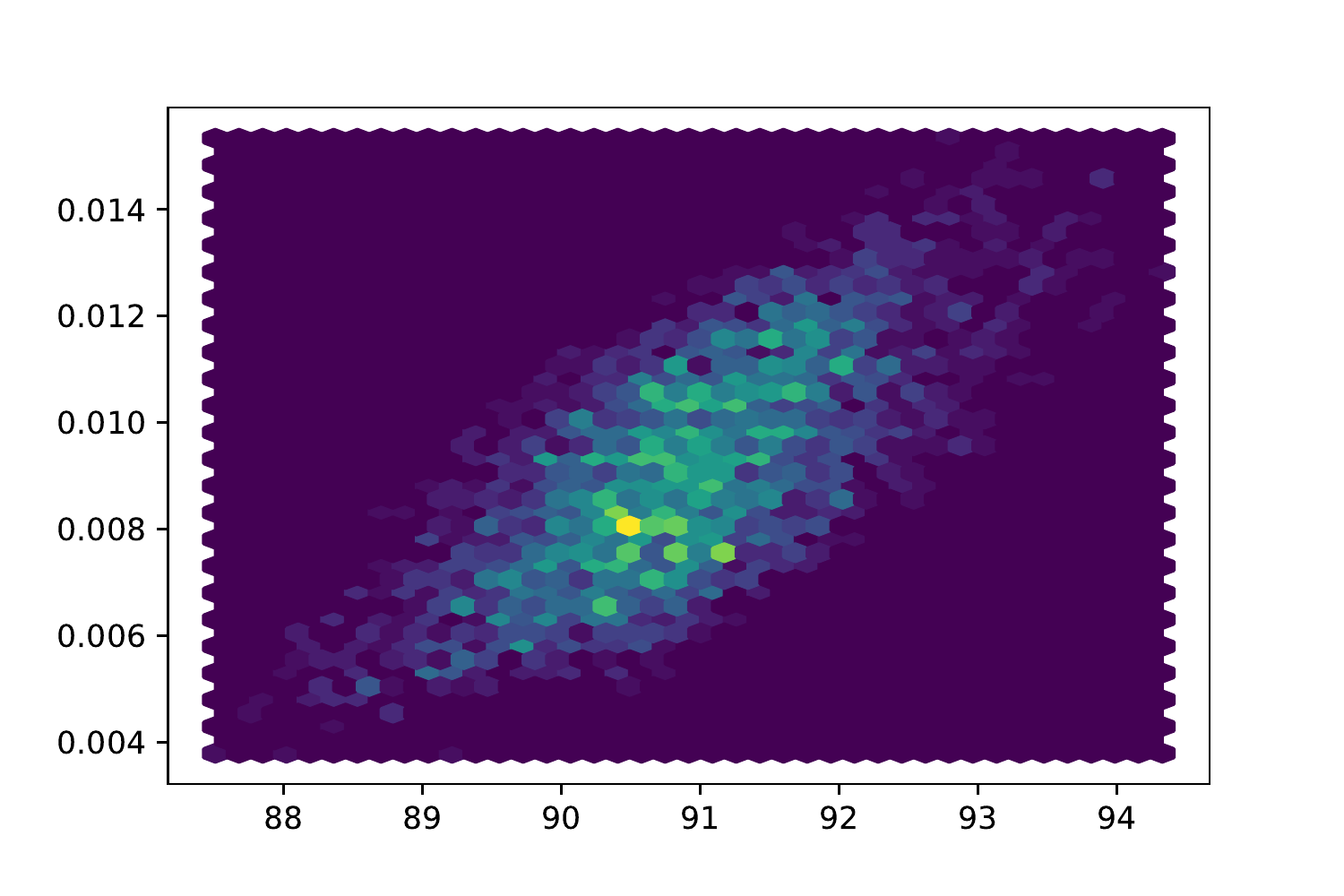} & \includegraphics[scale=0.11]{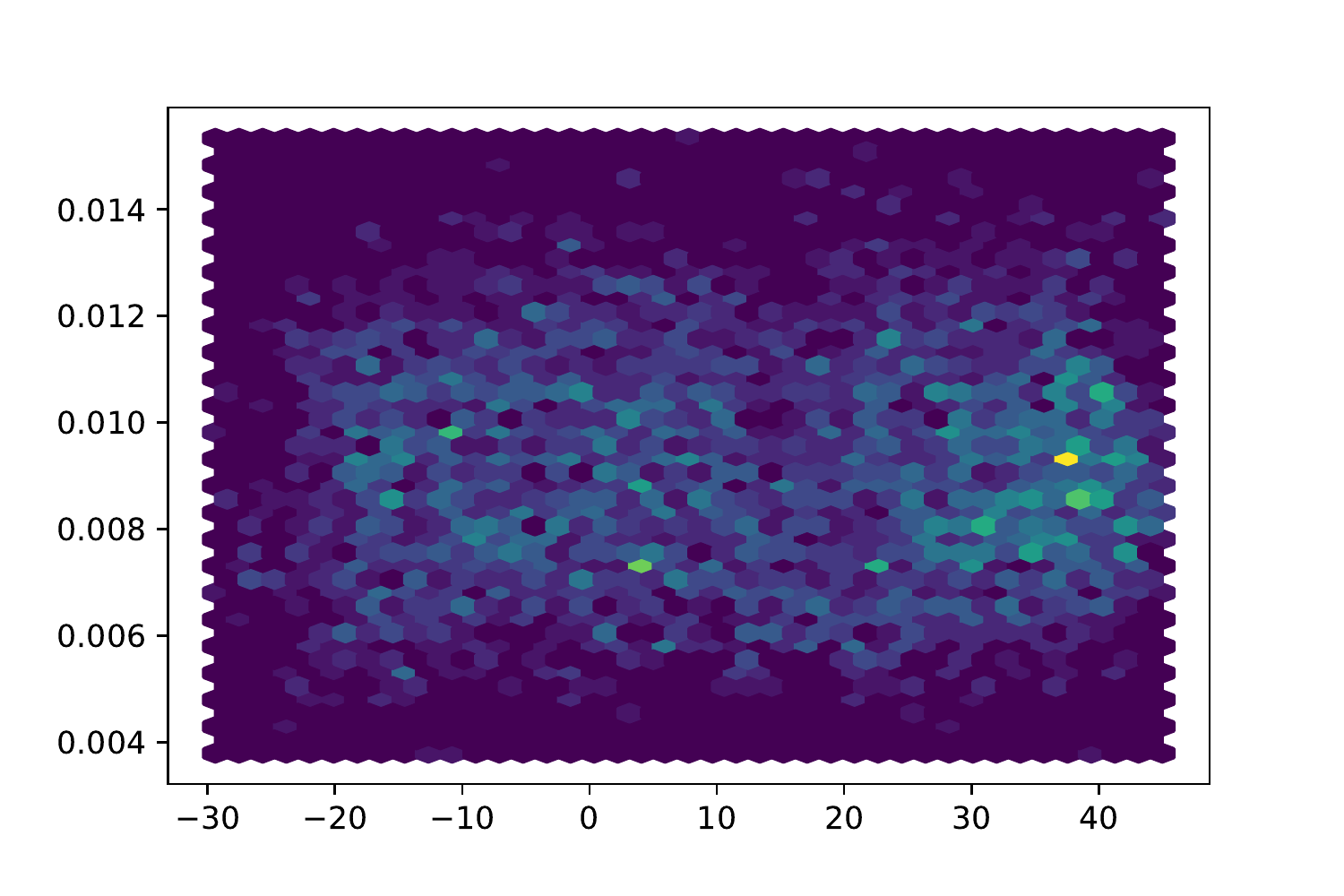} & \includegraphics[scale=0.11]{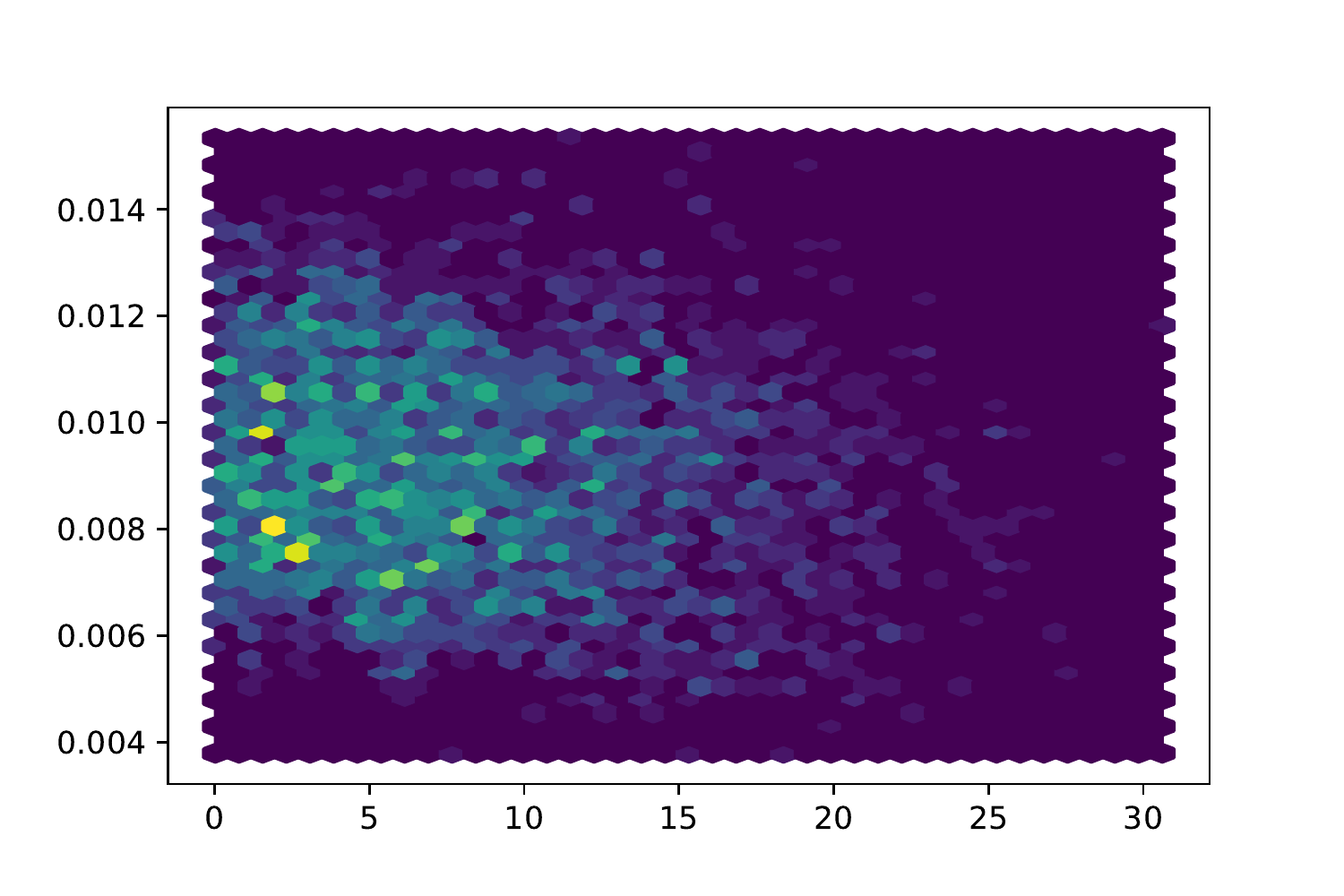} & \includegraphics[scale=0.11]{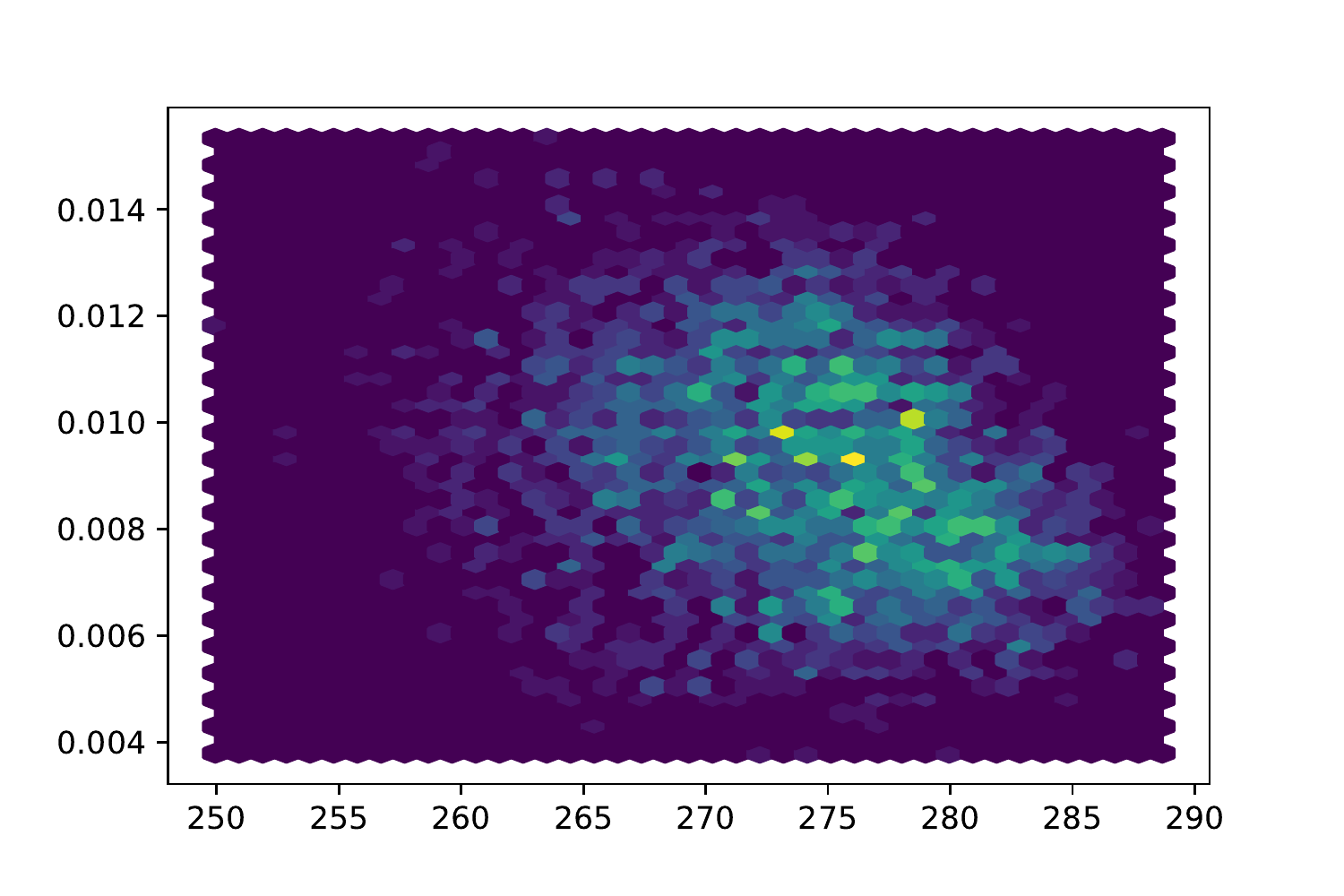} & \includegraphics[scale=0.11]{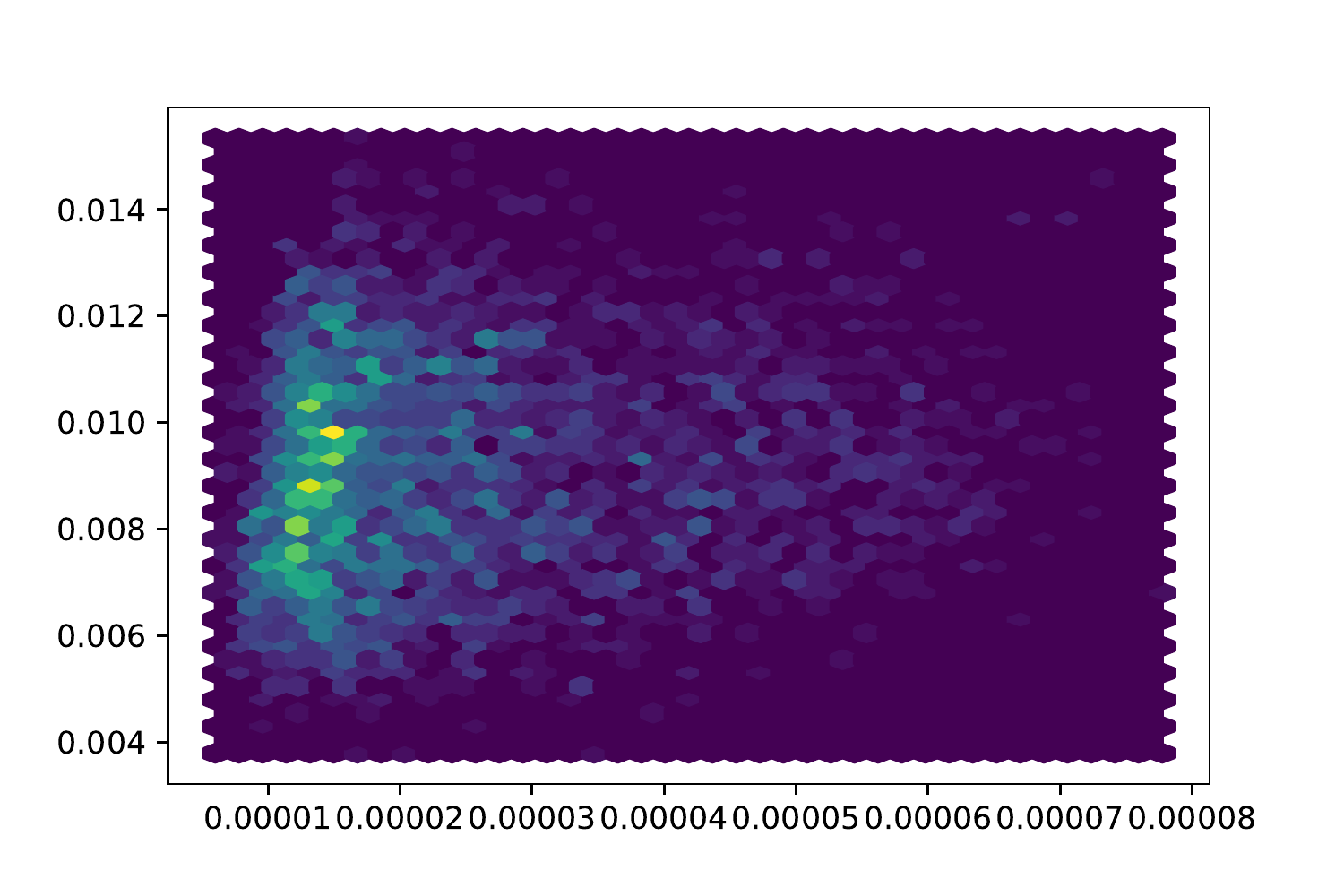} & & \\    
    c asc & \includegraphics[scale=0.11]{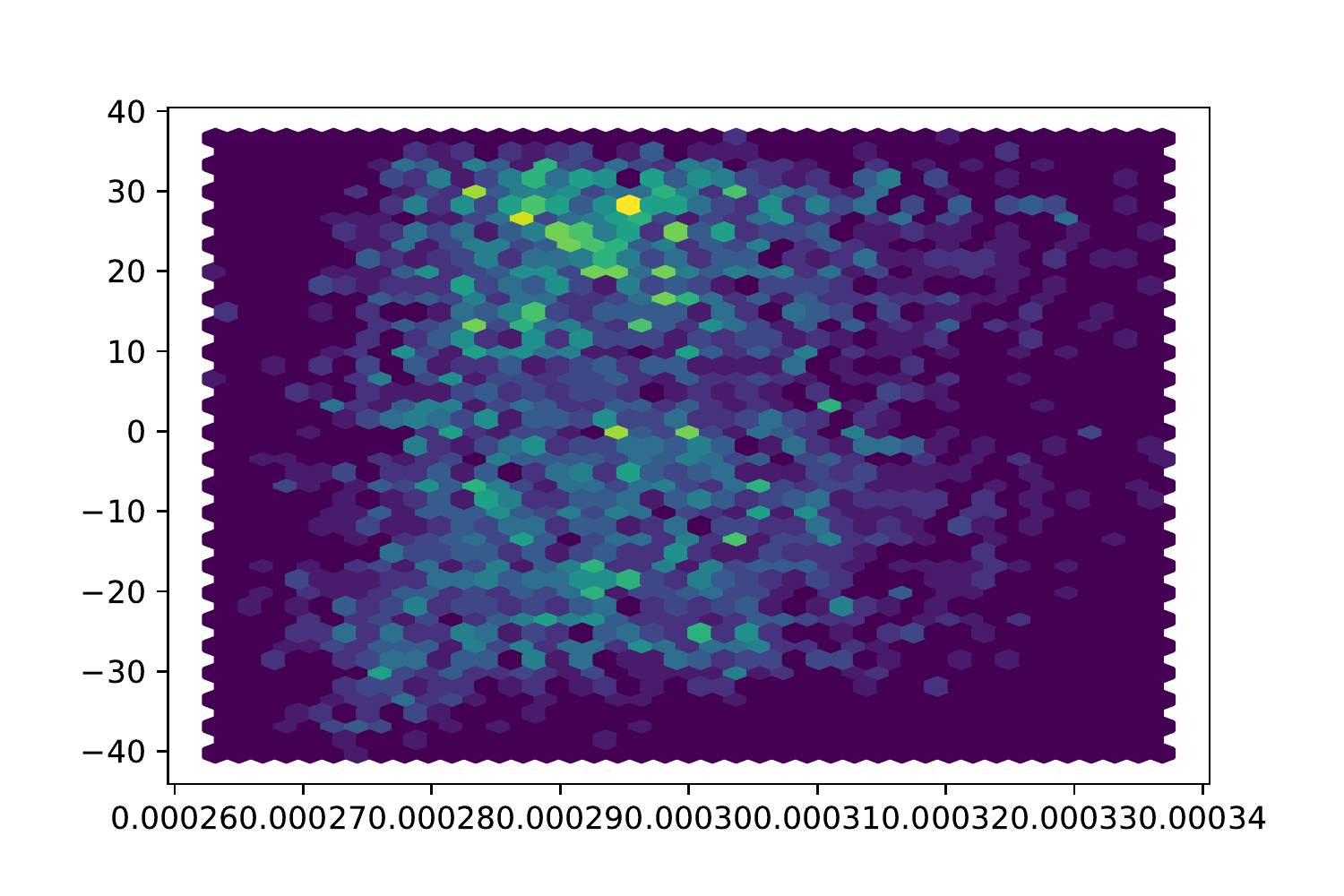} & \includegraphics[scale=0.11]{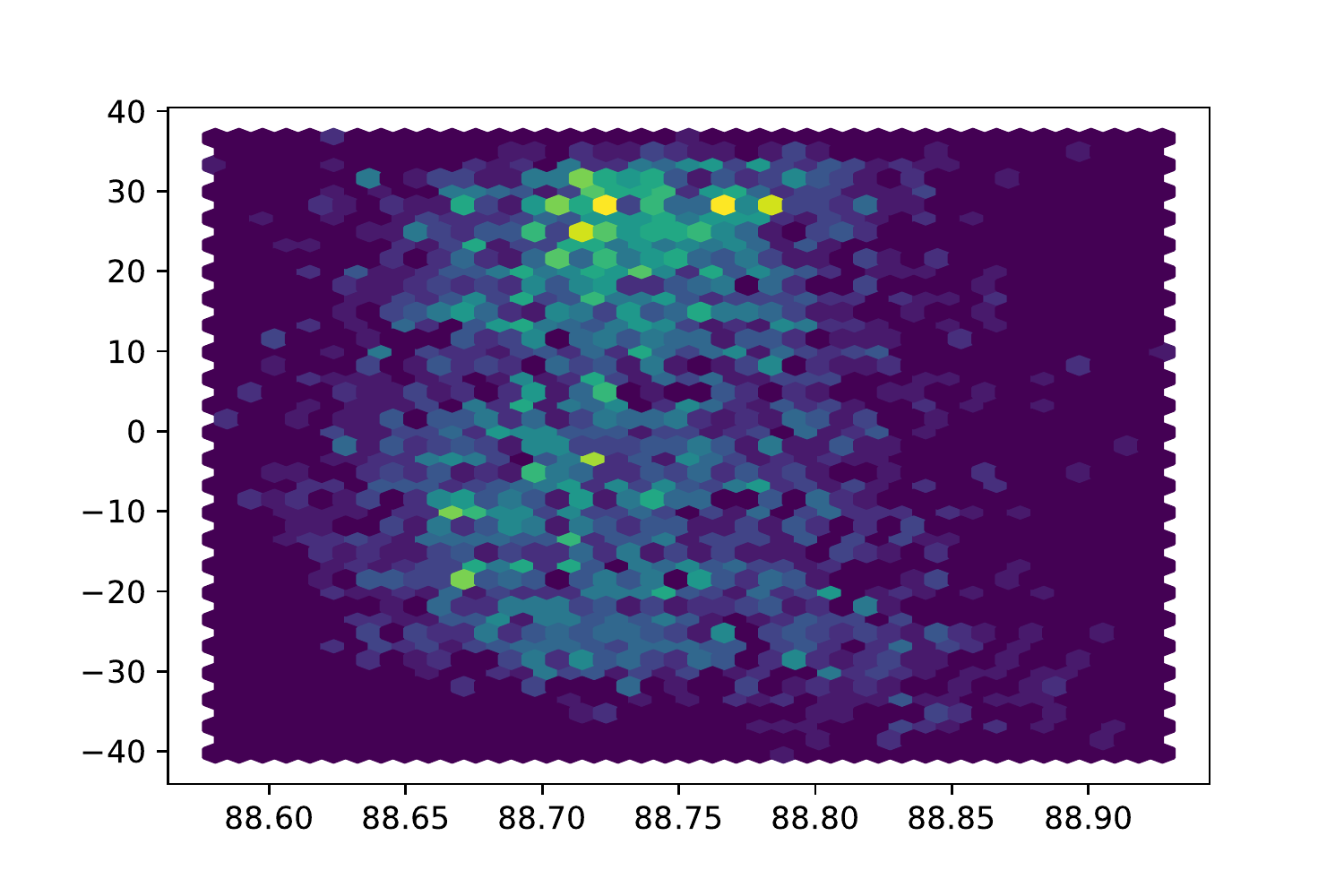} & \includegraphics[scale=0.11]{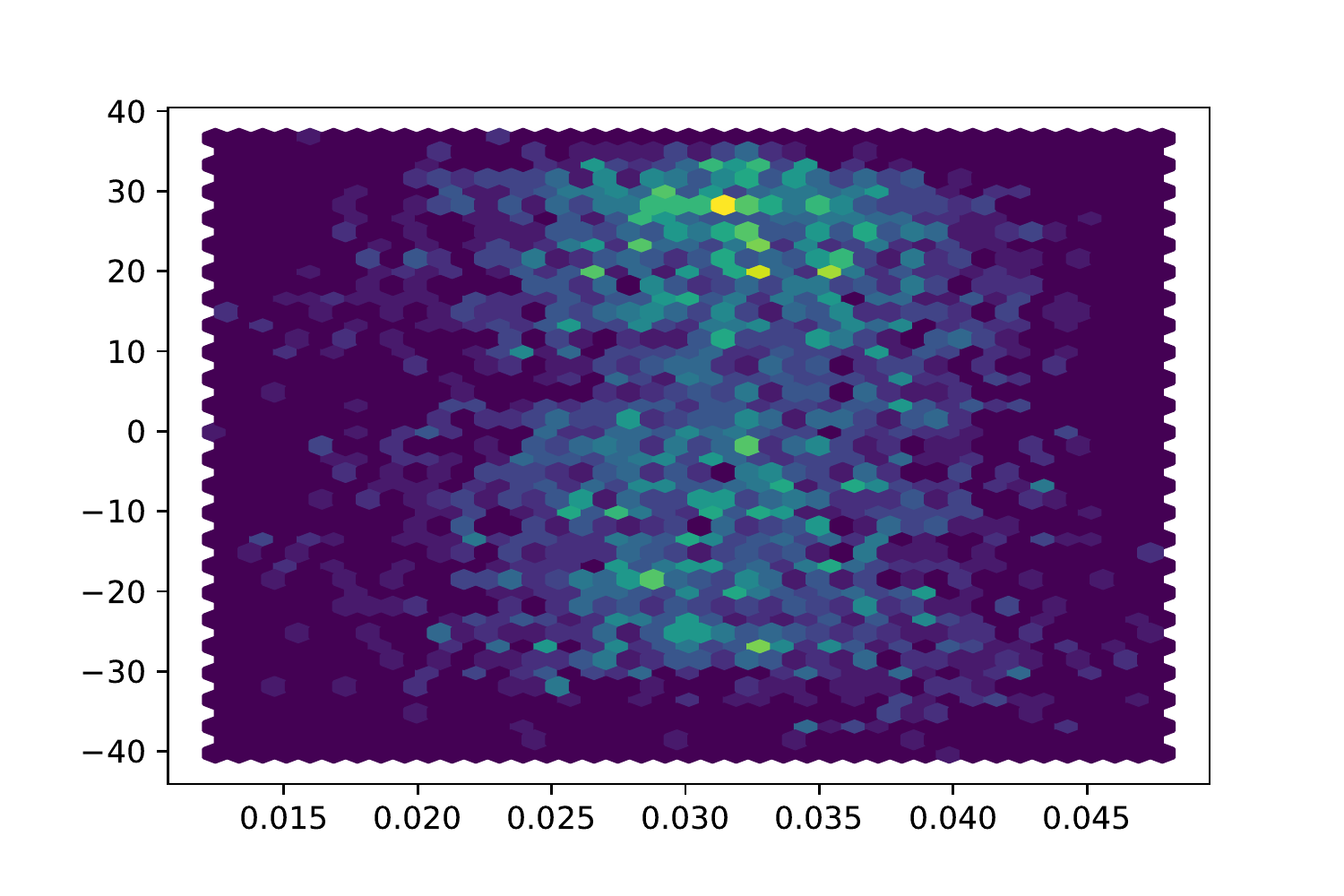} & \includegraphics[scale=0.11]{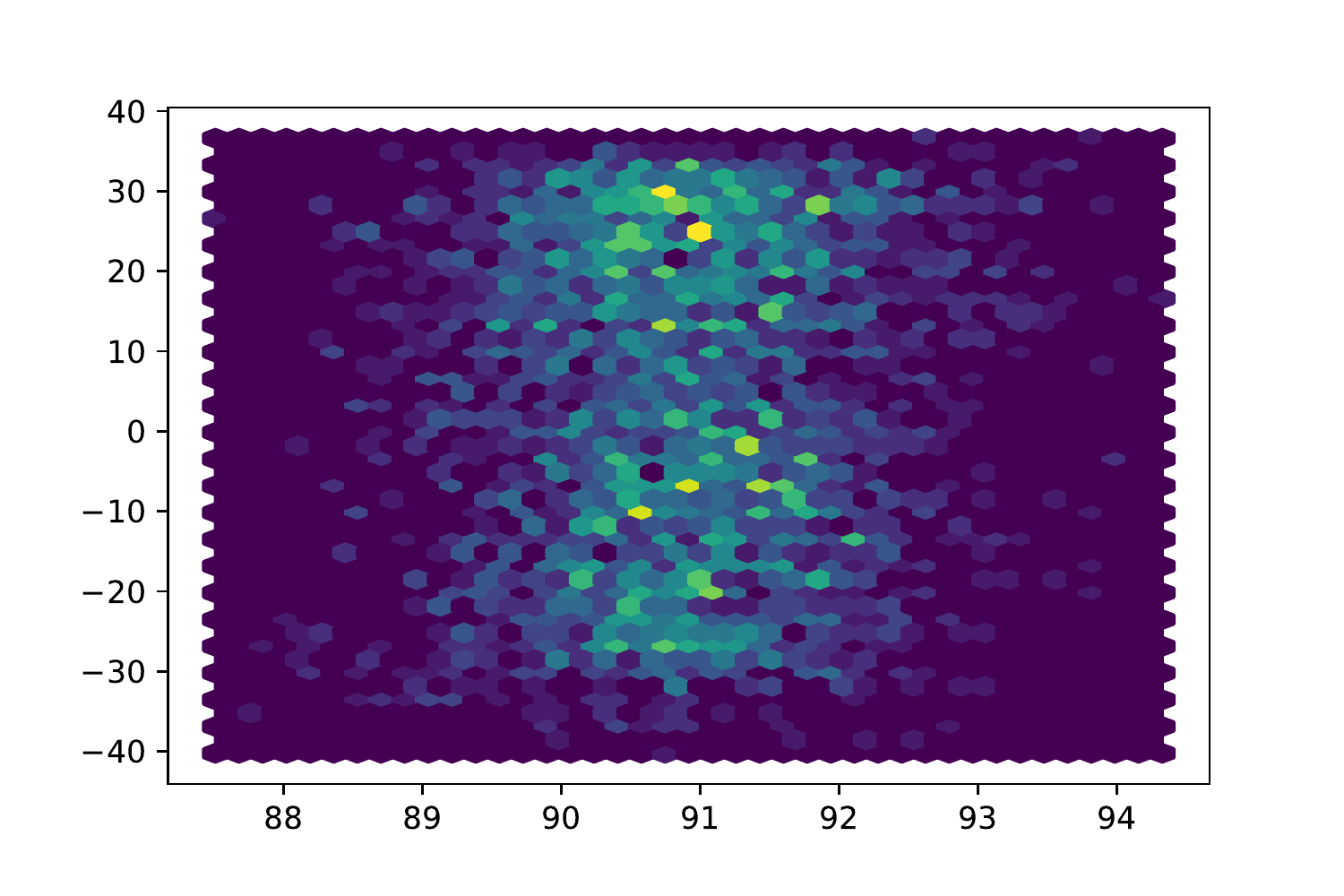} & \includegraphics[scale=0.11]{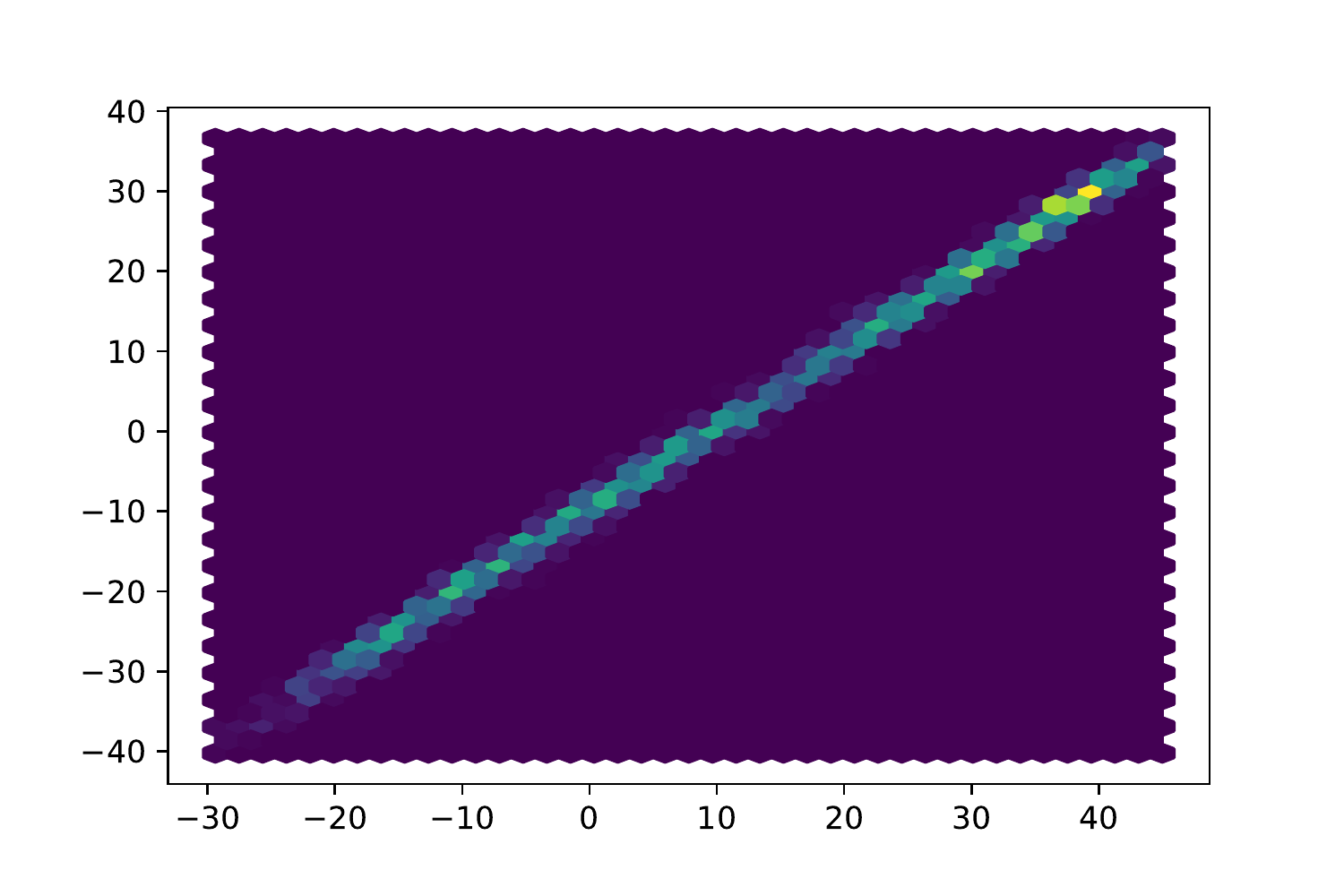} & \includegraphics[scale=0.11]{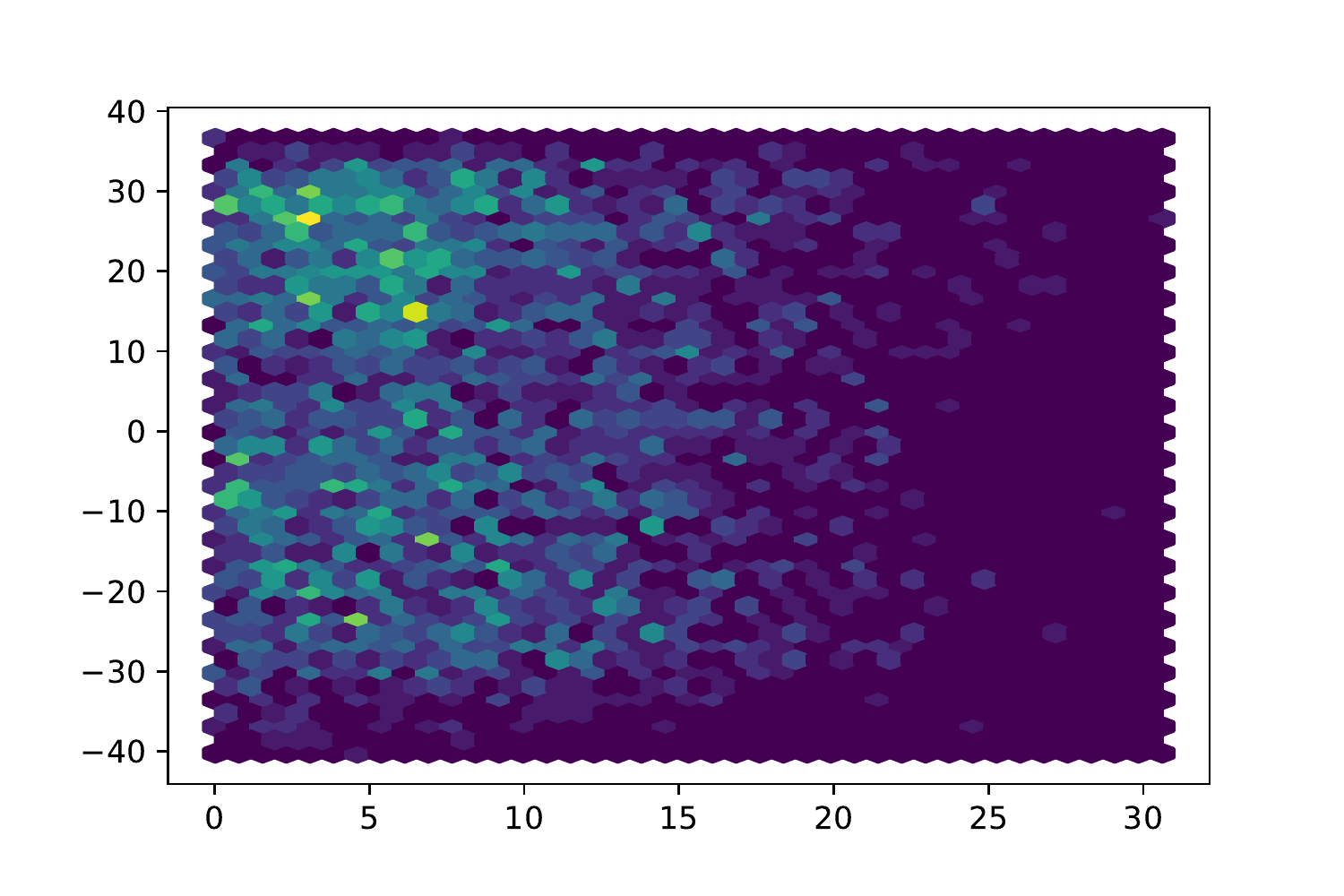} & \includegraphics[scale=0.11]{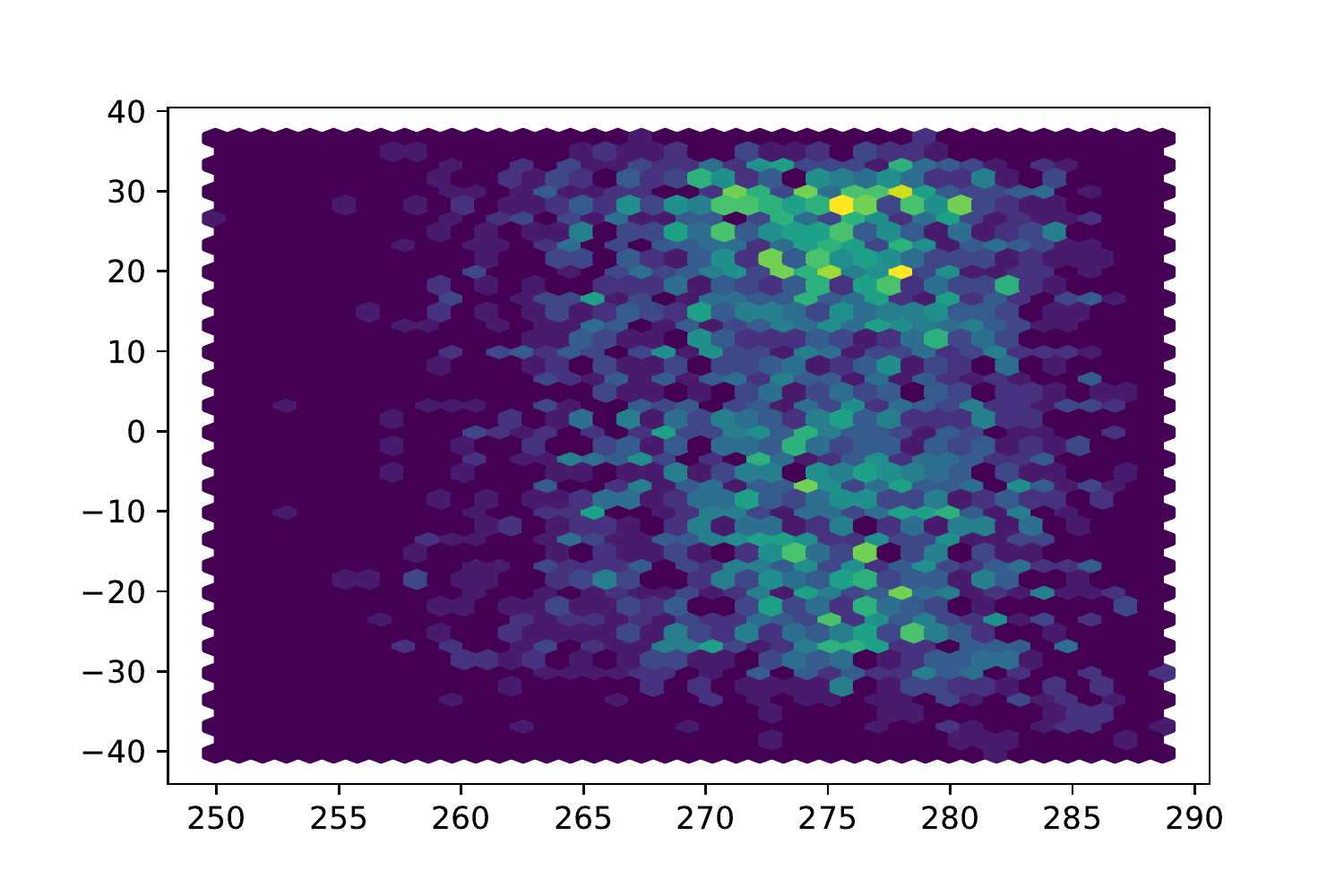} & \includegraphics[scale=0.11]{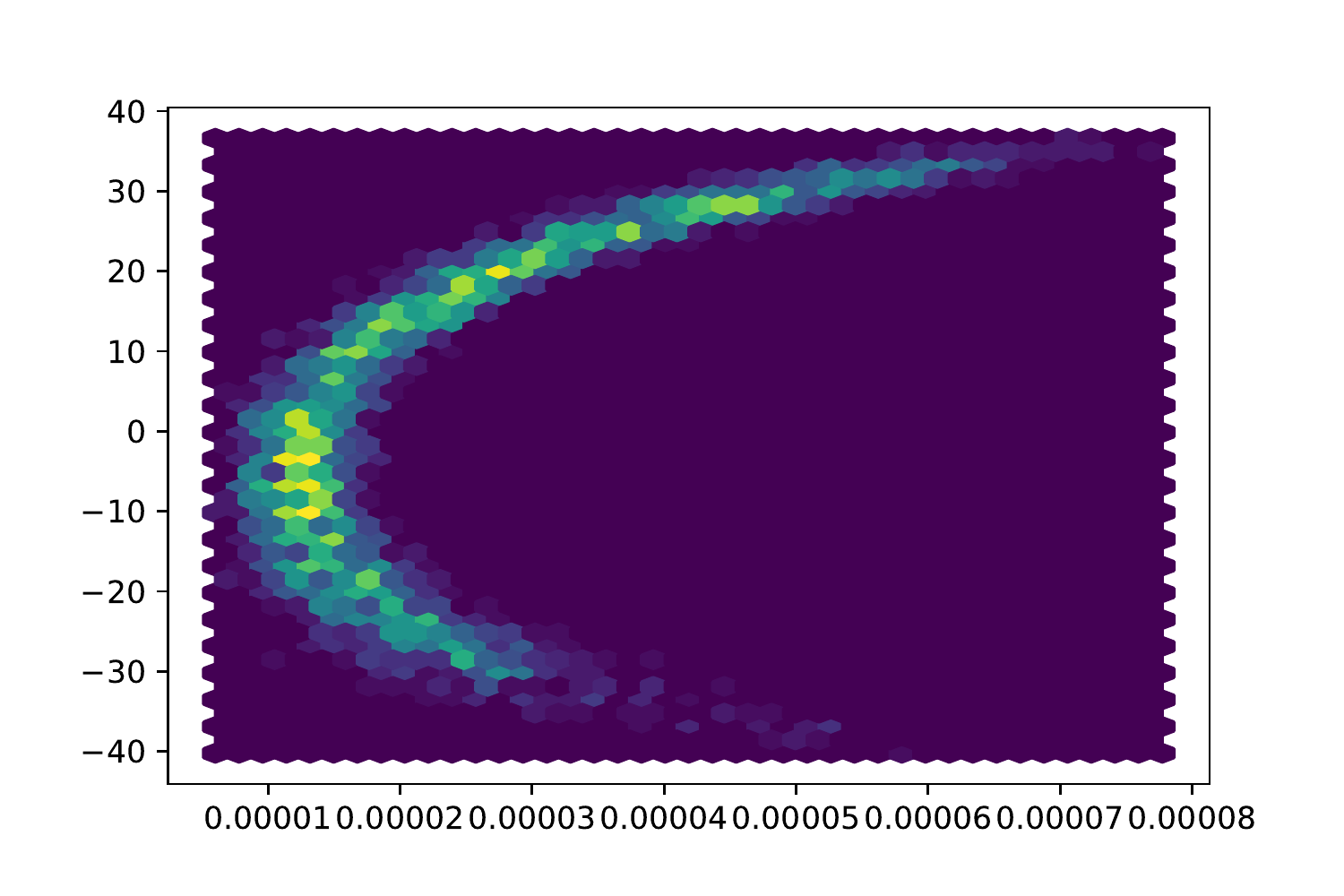} & \includegraphics[scale=0.11]{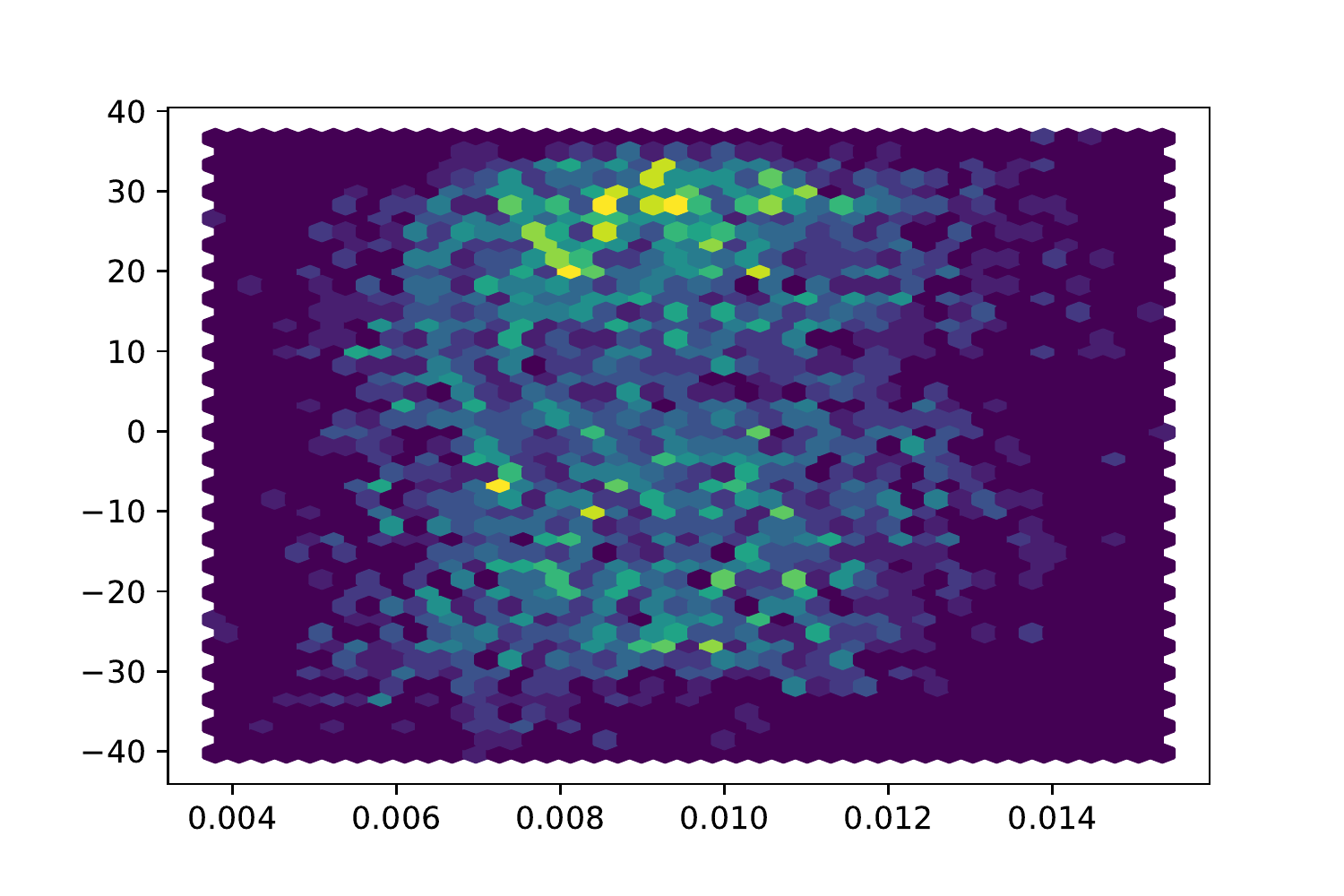} & \\   
    c arg peri & \includegraphics[scale=0.11]{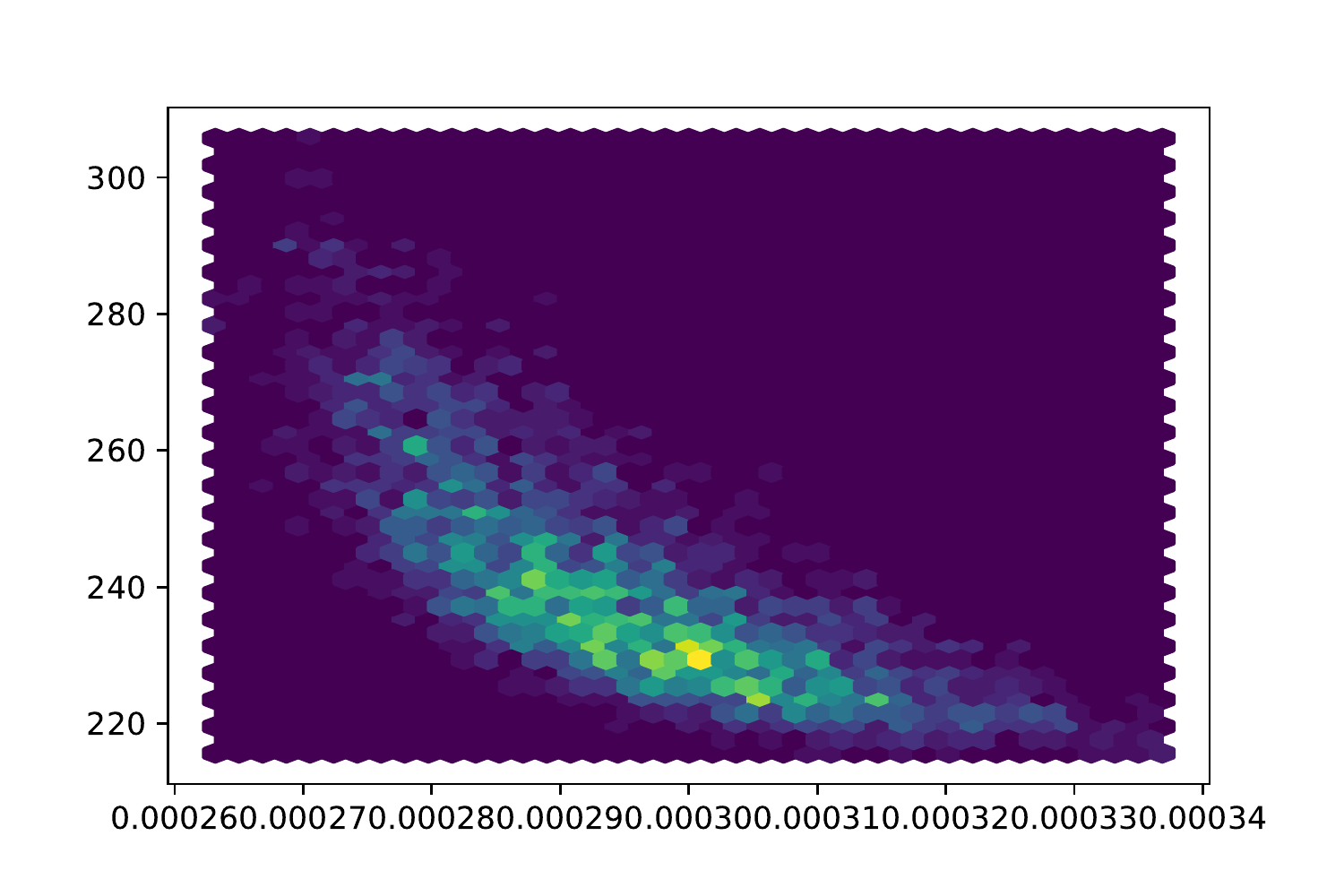} & \includegraphics[scale=0.11]{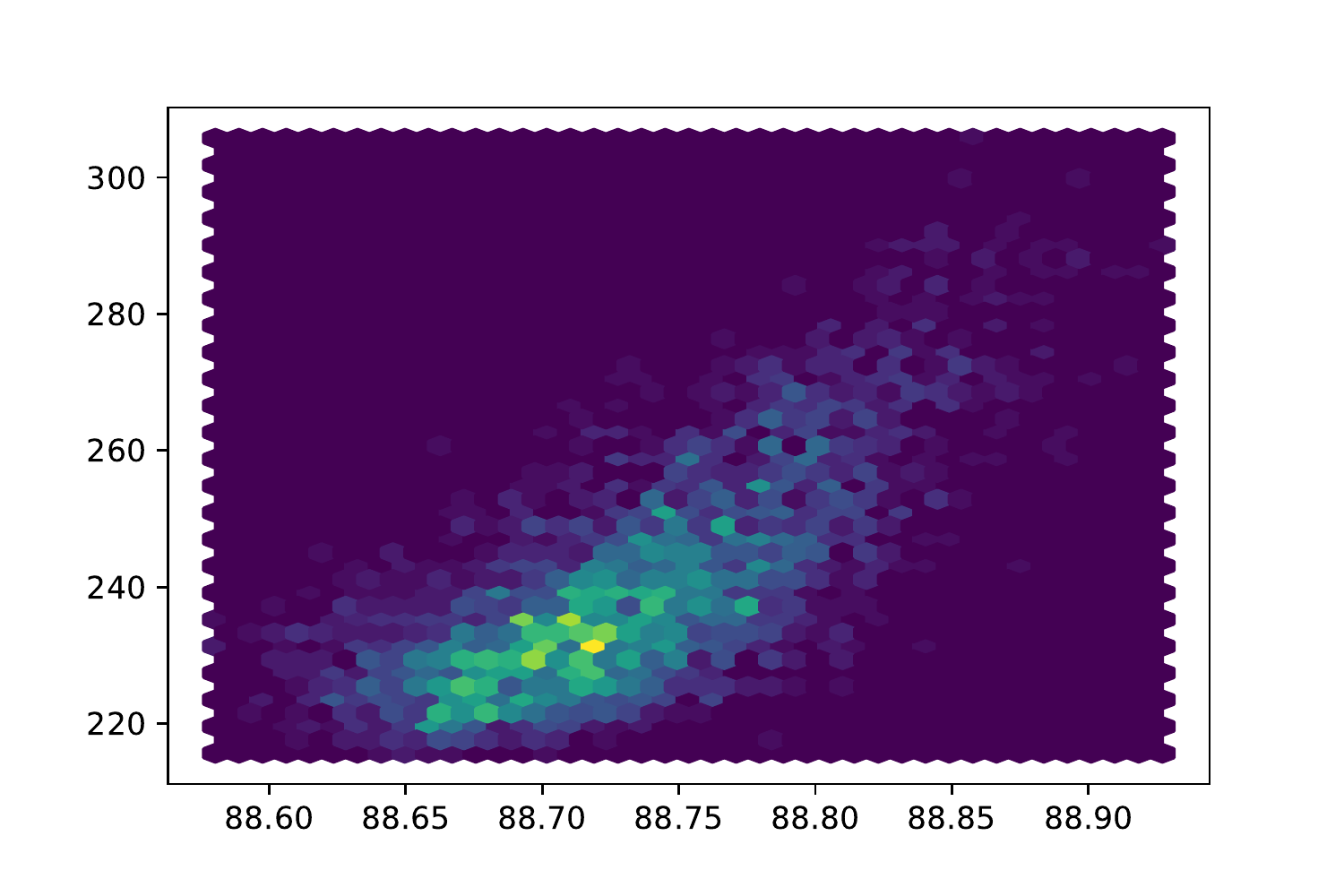} & \includegraphics[scale=0.11]{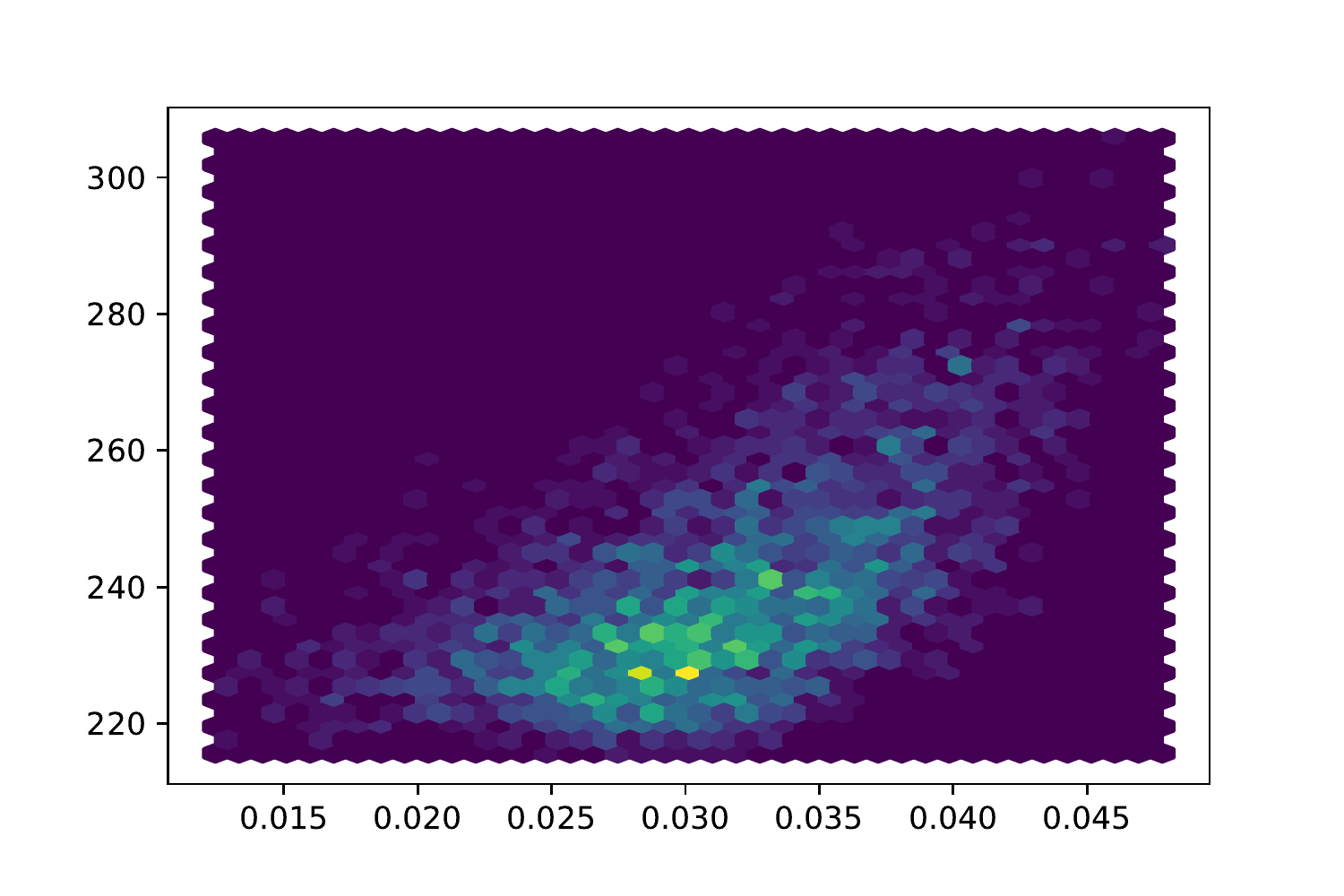} & \includegraphics[scale=0.11]{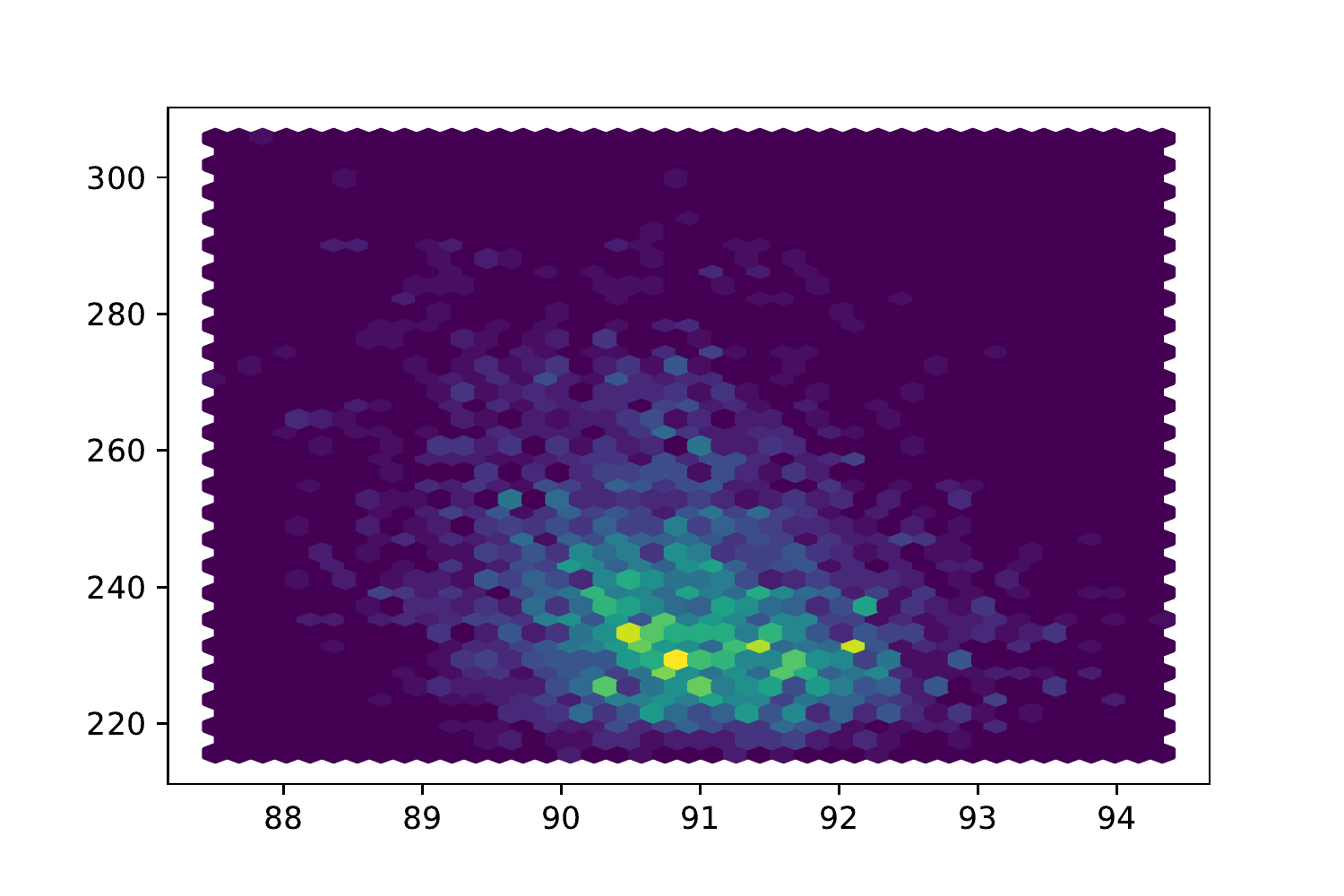} & \includegraphics[scale=0.11]{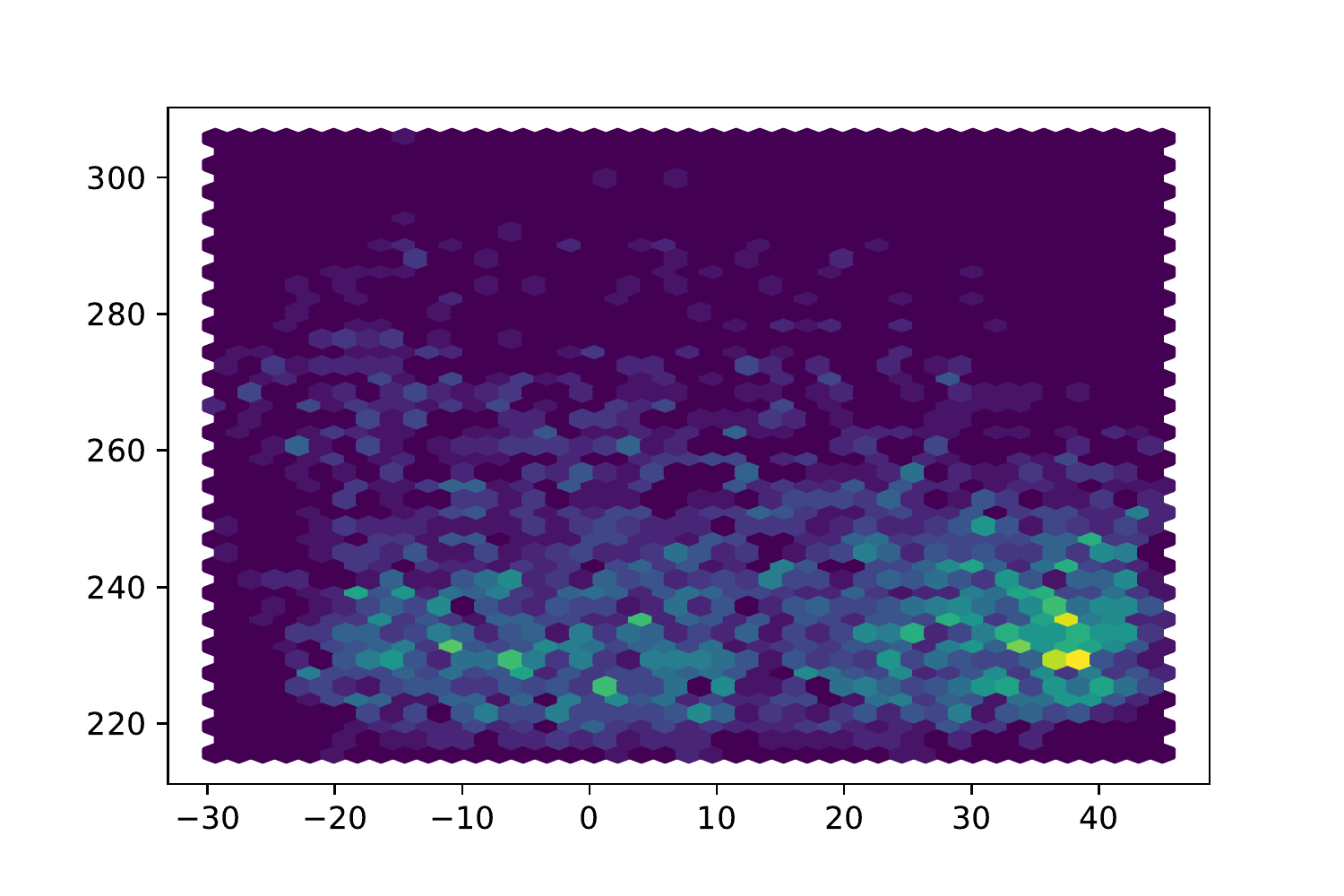} & \includegraphics[scale=0.11]{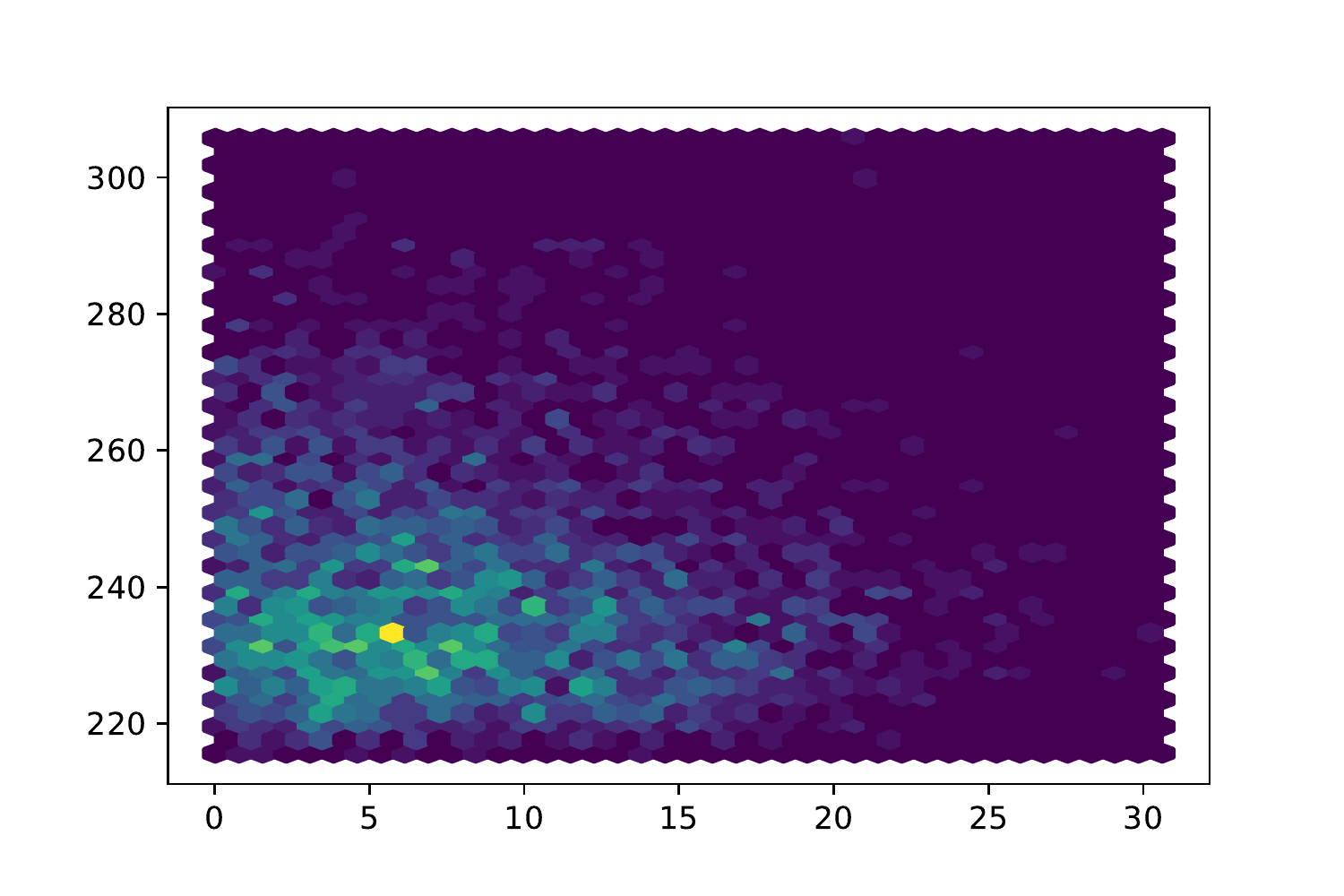} & \includegraphics[scale=0.11]{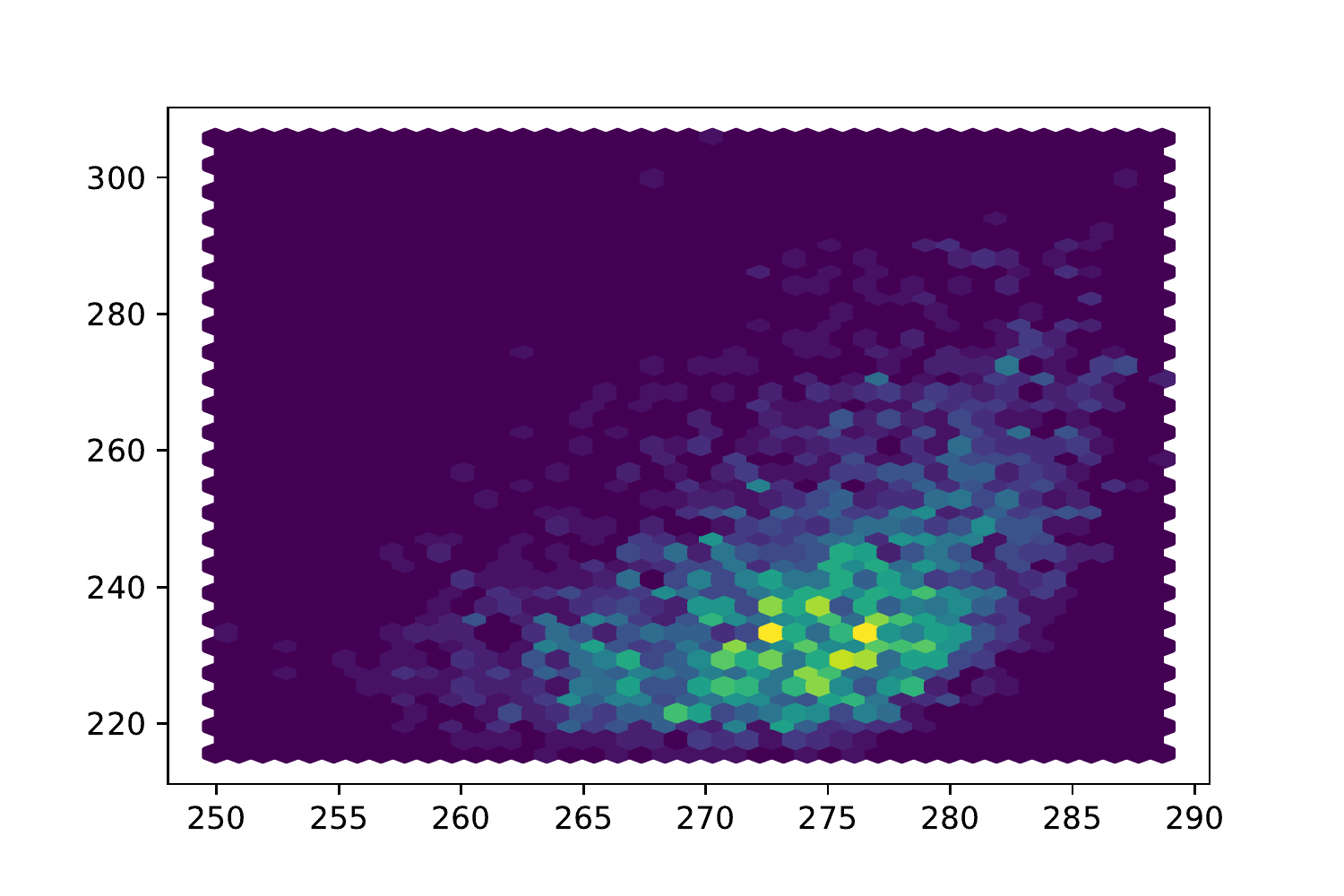} & \includegraphics[scale=0.11]{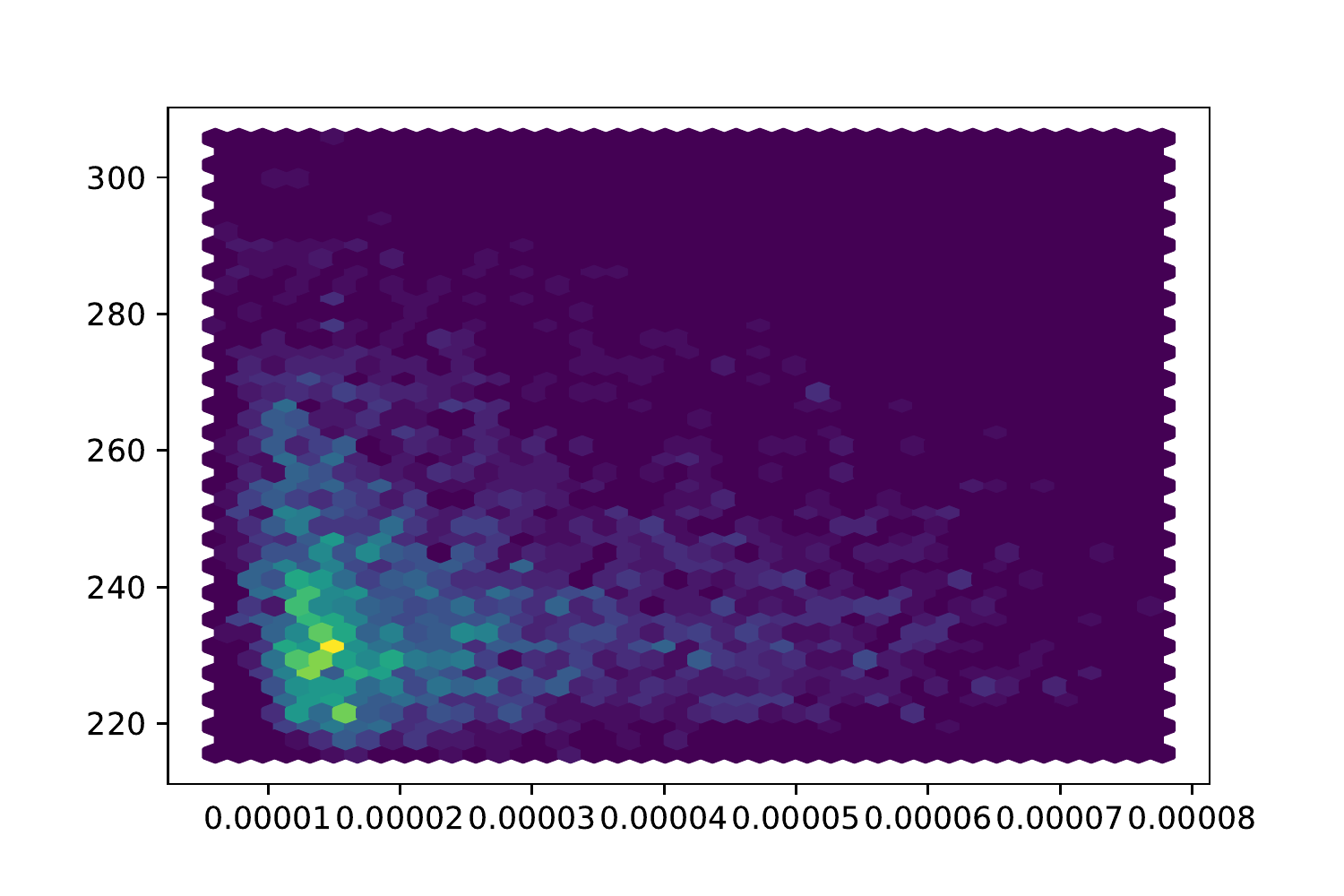} & \includegraphics[scale=0.11]{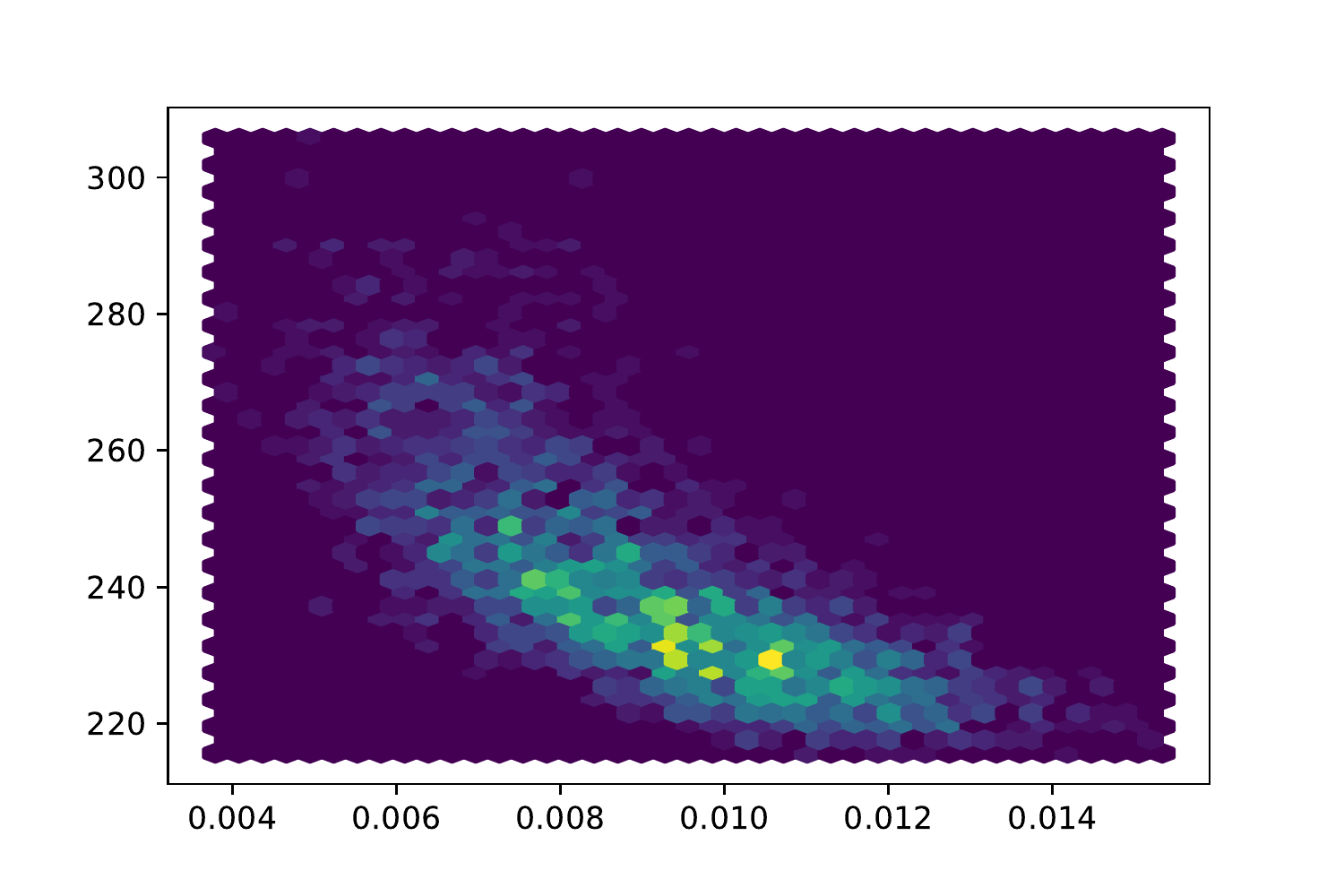} & \includegraphics[scale=0.11]{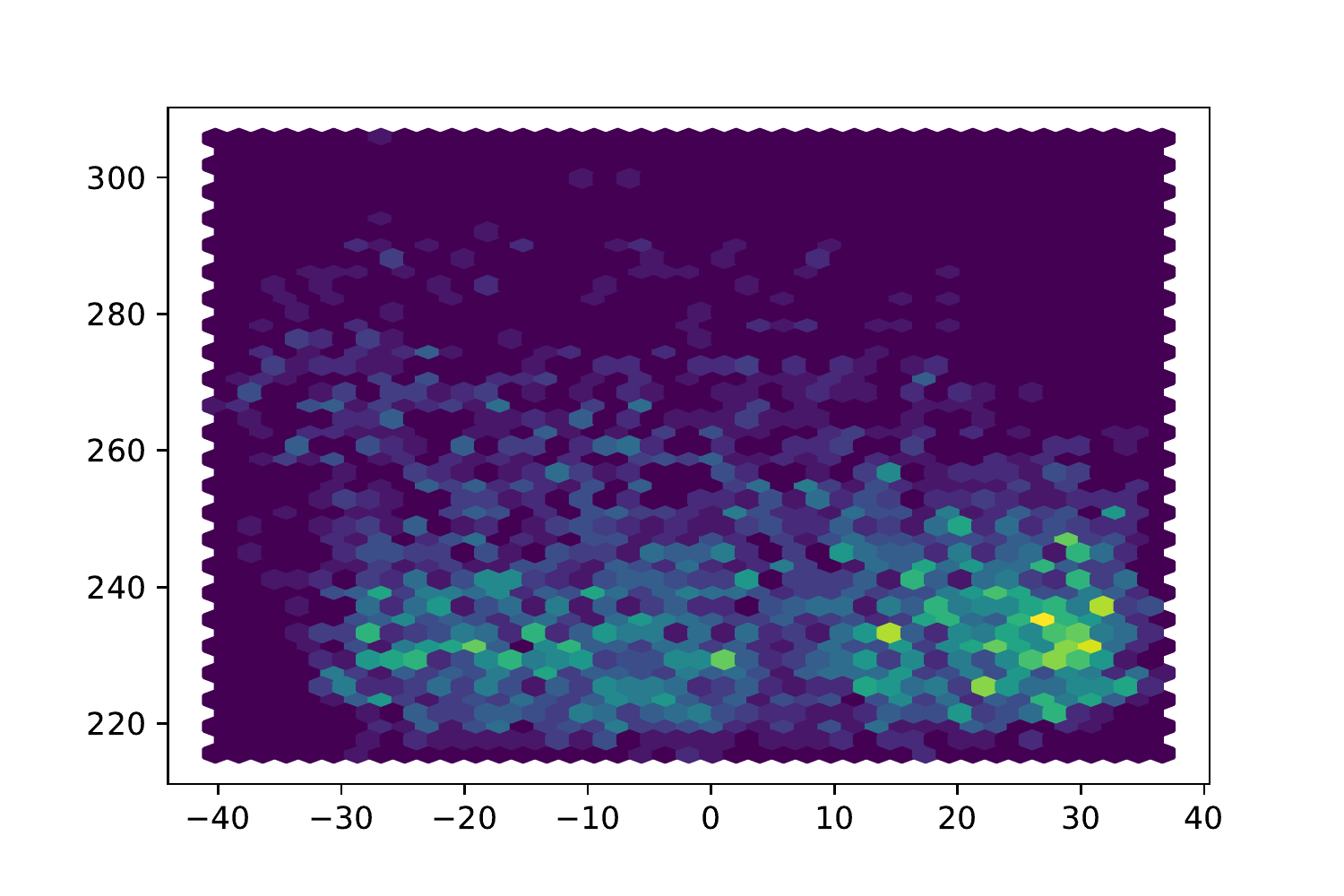} \\   
  \end{tabular}
 \end{table}
\end{landscape}


\bsp	
\label{lastpage}
\end{document}